\documentclass[format=acmsmall, review=false, screen=true]{acmart}

\acmJournal{TOIS}
\setcopyright{acmlicensed}
\acmDOI{}

\usepackage{amsfonts}
\usepackage{siunitx}
\usepackage{multirow}
\usepackage{enumerate}
\usepackage{booktabs}
\usepackage{amssymb}
\usepackage{calc}
\usepackage{paralist}
\usepackage{subcaption}
\usepackage[labelfont=bf,textfont=bf,singlelinecheck=on]{subcaption}
\usepackage{tikz}
\usetikzlibrary{fit,positioning}
\usepackage{graphicx}
\usepackage[justification=justified]{caption}
\usepackage{diagbox}
\usepackage{color}
\usepackage{colortbl}
\usepackage{url}
\usepackage{mathtools}
\usepackage{algorithm}
\usepackage{algorithmic}
\usepackage[normalem]{ulem}
\usepackage{enumitem}
\usepackage{numprint}
\npthousandsep{,}
\npdecimalsign{.}
\usepackage[skip=0pt]{caption}
\usepackage{setspace}
\usepackage{acronym}

\acrodef{J-NCF}{Joint Neural Collaborative Filtering}
\acrodef{DMF}{Deep Matrix Factorization}
\acrodef{NCF}{Neural Collaborative Filtering}
\acrodef{BPR}{Bayesian Personalized Ranking}
\acrodef{RBM}{Restricted Boltzmann Machine}
\acrodef{CF}{Collaborative Filtering}
\acrodef{DL}{Deep Learning}
\acrodef{CDAE}{Collaborative Denoising Auto-Encoder}
\acrodef{DAEs}{Denoising Auto-Encoders}
\acrodef{LFM}{Latent Factor Model}
\acrodef{SVD}{Singular Value Decomposition}
\acrodef{CNN}{Convolutional Neural Network}
\acrodef{CDL}{Collaborative Deep Learning}
\acrodef{BPR}{Bayesian Personalized Ranking}
\acrodef{MF}{Matrix Factorization}

\newcommand{\noop}[1]{}

\newcommand{\todo}[1]{\textcolor{black}{#1}}

\newcommand{\chen}[1]{\textcolor{black}{#1}}

\newcommand{\duwen}{\phantom{$^\blacktriangle$}}

\begin{document}
% Title portion. Note the short title for running heads
\title{Joint Neural Collaborative Filtering for Recommender Systems\footnotemark[2]}

\author{Wanyu Chen}
\orcid{}
\affiliation{%
 \department{Science and Technology on Information Systems Engineering Laboratory}
 \institution{National University of Defense Technology}
 \city{Changsha}
 \postcode{410073}
 \country{China}}
\affiliation{%
  \department{Informatics Institute}
  \institution{University of Amsterdam}
  \city{Amsterdam}
  \country{The Netherlands}
}  
\email{wanyuchen@nudt.edu.cn}

\author{Fei Cai}
\affiliation{%
 \department{Science and Technology on Information Systems Engineering Laboratory}
 \institution{National University of Defense Technology}
 \postcode{410073}
 \city{Changsha}
 \country{China}
}
\email{caifei@nudt.edu.cn}
\authornote{Corresponding author.}

\author{Honghui Chen}
\affiliation{%
 \department{Science and Technology on Information Systems Engineering Laboratory}
 \institution{National University of Defense Technology}
 \postcode{410073}
 \city{Changsha}
 \country{China}}
\email{chenhonghui@nudt.edu.cn}

\author{Maarten de Rijke}
\affiliation{%
  \department{Informatics Institute}
  \institution{University of Amsterdam}
  \city{Amsterdam}
  \country{The Netherlands}
}
\email{derijke@uva.nl}

\begin{abstract}
We propose a \ac{J-NCF} method for recommender systems.
The \ac{J-NCF} model applies a joint neural network that couples deep feature learning and deep interaction modeling with a rating matrix.
Deep feature learning extracts feature representations of users and items with a deep learning architecture based on a user-item rating matrix.
Deep interaction modeling captures non-linear user-item interactions with a deep neural network using the feature representations generated by the deep feature learning process as input.
\ac{J-NCF} enables the deep feature learning and deep interaction modeling processes to optimize each other through joint training, which leads to improved recommendation performance.
In addition, we design a new loss function for optimization, which takes both implicit and explicit feedback, point-wise and pair-wise loss into account.

Experiments on several real-word datasets show significant improvements of \ac{J-NCF} over state-of-the-art methods, with improvements of up to 8.24\% on the MovieLens 100K dataset, 10.81\% on the MovieLens 1M dataset, and 10.21\% on the Amazon Movies dataset in terms of HR@10. NDCG@10 improvements are 12.42\%, 14.24\% and 15.06\%, respectively.
We also conduct experiments to evaluate the scalability and sensitivity of \ac{J-NCF}. \chen{Our experiments show that the \ac{J-NCF} model has a competitive recommendation performance with inactive users and different degrees of data sparsity when compared to state-of-the-art baselines.}
\end{abstract}

% <concept_id>10002951.10003227.10003351.10003269</concept_id>
% <concept_desc>Information systems~Collaborative filtering</concept_desc>
% <concept_id>10002951.10003317.10003347.10003350</concept_id>
% <concept_desc>Information systems~Recommender systems</concept_desc>
% \ccsdesc[500]{Recommender systems}




\footnotetext[2]{To appear in ACM Transactions on Information Systems.}
\maketitle

%\renewcommand{\shortauthors}{G. Zhou et al.}

\section{Introduction}
\label{Intro}
Recommender systems are an effective solution to help people cope with an increasingly complex information landscape.
\ac{CF} approaches have been widely investigated and used for personalized recommendation~\citep{RSsurvey2017,RS-CF2005}.
Many traditional \ac{CF} techniques are based on \ac{MF}~\citep{RSsurvey2017}.
They characterize users and items by latent factors that are extracted from the user-item rating matrix.
In the latent space, traditional \ac{CF} methods, such as the \ac{LFM}~\citep{LFM2009}, often predict a user's preference for an item with a linear kernel, i.e., a dot product of their latent factors, which may not be able to capture the complex structure of user-item {interactions} well.

%The past few years have witnessed the tremendous successes of the \ac{DL} technology in application domains such as computer vision and speech recognition~\citep{DMF2017,CNN2016,DLImage2016}.
%Both academia and industry have been in a race to apply \ac{DL} to a wider range of applications due to its capability in solving many complex tasks.
Recently introduced \ac{DL}-based approaches to recommender systems overcome shortcomings of conventional approaches to recommender systems, such as dynamic user preferences and intricate relationships within the data itself, and are able to achieve high recommendation quality.
Today's \ac{DL}-based approaches to recommender systems mostly use \ac{DL} to explore auxiliary information, e.g., textual descriptions of items or audio features of music, which is then used to model item features~\citep{CNN2016,CTR2011,CDL2015}.
For the user-item rating matrix, recent work mostly continues to use traditional \ac{MF}-based approaches.
\acp{RBM}~\citep{RBM2007} seem to have been the first model to use neural networks to model the user-item rating matrix and obtain competitive results over traditional methods; it is a two-layer network rather than a deep learning structure.
Another recent approach, \ac{CDAE}~\citep{CDAE2016}, is mainly designed for rating prediction with a one-hidden layer neural network.
\ac{NCF}~\citep{NCF2017} uses deep neural networks for learning the interaction function from data with multi-layer perceptrons, yet it does not explore users' and items' features that are known to be helpful in improving \ac{CF} recommendation performance.
\ac{CDAE} and \ac{NCF} only exploit implicit feedback for recommendations instead of explicit rating feedback.
\ac{DMF}~\citep{DMF2017} models the user-item rating matrix with a neural network that maps the users' and items' features into a low-dimensional space with non-linear projections; it uses an inner product to compute interactions between users and items, and applies the same linear kernel (i.e., dot product) as \ac{LFM}~\citep{LFM2009}.

We hypothesize that \ac{DL} should be able to effectively capture both non-linear and non-trivial user-item relationships as well as users' (items') characteristics with multi-layer projections~\citep{RSsurvey2017}.
We propose a \acf{J-NCF} model that enables two processes---feature extraction and user-item interaction modeling---to be trained jointly in a unified \ac{DL} structure.
The \ac{J-NCF} model contains two main networks for recommendation.
The first network uses the rating information of a user (an item) as the network input, and outputs a vector representation for the user (the item).
Then, using the connection of a user's and an item's vectors as input, the second neural network models the user-item interactions and outputs the prediction of the corresponding rating of the user and item.
Thus, these two networks can be coupled tightly and trained jointly in a unified structure.
Interaction modeling can optimize the feature learning process and more accurate feature representations can, in turn, improve the user-item interaction prediction.
We take both implicit and explicit feedback, point-wise and pair-wise loss into account to enhance the prediction performance.
In contrast, previous neural approaches such as \ac{CDAE}, \ac{NCF} and \ac{DMF} are all optimized only with point-wise loss functions and leave dealing with pair-wise loss as future work.

To the best of our knowledge, in the area of recommender systems ours is the first attempt to use a joint neural network to tightly couple feature learning and interaction modeling with the rating matrix.
\ac{J-NCF} allows these two processes to optimize each other through joint training and thereby improve the recommendation performance.

Our experiments on real-world datasets, including the MovieLens dataset and the Amazon Movies dataset, show that \ac{J-NCF} outperforms the {state-of-the-art} baselines in prediction accuracy, with improvements of up to 8.24\% on the MovieLens 100K dataset, 10.81\% on the MovieLens 1M dataset, and 10.21\% on the Amazon Movies dataset in terms of HR@10. NDCG@10 improvements are 12.42\% on the MovieLens 100K dataset, 14.24\% on the MovieLens 1M dataset, and 15.06\% on the Amazon Movies dataset, respectively, over the best baseline model.
In addition, we investigate the scalability and sensitivity of \ac{J-NCF} with different degrees of sparsity and different numbers of users' ratings.
Our experimental results indicate that \ac{J-NCF} achieves competitive recommendation performance when compared to the best state-of-the-art model.

Our contributions in this paper are:
\begin{enumerate}
   \item  We design a Joint Neural Collaborative Filtering model (\ac{J-NCF}) for recommendation, which enables deep feature learning and deep user-item interaction modeling to be coupled tightly and jointly optimized in a single neural network.
   \item  We design a new loss function that explores the information contained in both point-wise and pair-wise loss as well as implicit and explicit feedback.
   \item  We analyse the recommendation performance of \ac{J-NCF} as well as baseline models and find that \ac{J-NCF} consistently yields the best performance.
\chen{\ac{J-NCF} also shows competitive improvements over the best baseline model when applied with inactive users and different degrees of data sparsity.}
\end{enumerate}
We summarize related work in Section~\ref{Relatedwork}. 
Our approach, \ac{J-NCF}, is described in Section~\ref{Approach}. 
Section~\ref{Experiments} presents our experimental setup. 
In Section~\ref{Results}, we report our results to demonstrate the recommendation performance of \ac{J-NCF}. 
\chen{We also investigate the scalability and sensitivity of our model as well as other baselines in Section~\ref{application}.} 
Finally, we conclude our work in Section~\ref{Conclusion}, where we also suggest future research directions.

\section{Related work}
\label{Relatedwork}
We first look back to traditional approaches to recommender systems in Section \ref{RTraditional}, that  focus on modeling the similarity between users (items) for recommendation. Then, as applying deep learning techniques into recommender systems is gaining momentum due to its state-of-the-art performance and high-quality recommendations, we summarize recent work on 
deep learning-based recommender systems in Section~\ref{Rdeep} that can provide a better understanding of user's demands, item's characteristics as well as historical interactions between them by extracting the features of items with auxiliary information, e.g., the content of movies. 

\subsection{Traditional recommender systems}
\label{RTraditional}
In many commercial systems, ``best bet'' recommendations are shown, but the predicted rating values are not. 
This is usually referred to as a top-N recommendation task, where the goal of the recommender system is to find a few specific items that are supposed to be most appealing to the user. 
A similar prediction schema, denoted as Top Popular (Item-pop), recommends the top-N items with the highest popularity (largest number of ratings).

Most top-N recommender systems are based on collaborative filtering~\citep{RS-CF2005}, where recommendations rely on past behavior (ratings) from users, regardless of domain knowledge~\citep{Cf-survey2009}. 
We group these \ac{CF} approaches into two categories, i.e., neighborhood-based methods~\citep{Item2001,neighborRS2003} and latent factor-based models~\citep{LFM2009,FISM2013}. 
Neighborhood-based models share the typical merits of \ac{CF}, which concentrate on exploring the similarity among either users or items. 
For instance, two users are similar because they have rated similarly the same set of items. 
A dual concept of similarity can be defined among items. 
Latent factor-based approaches generally model users and items as vectors in the same ``latent factor'' space by means of a reduced number of hidden factors. 
In such a space, users and items are directly comparable: the rating of a user $u$ on an item $i$ is predicted by the proximity (e.g., inner-product) between the related latent factor vectors.

For neighborhood-based models, algorithms that are centered around user-user similarity typically predict the rating by a user based on the ratings expressed by other users similar to her about such item. 
On the other hand, algorithms centered around item-item similarity compute the user preference to an item based on her own ratings to similar items. 
The similarity between item $i$ and item $j$ is measured as the tendency of users to rate items $i$ and $j$ similarly. 
It is typically based either on the cosine, the adjusted cosine, or (most commonly) the Pearson correlation coefficient~\cite{Item2001}.
The kNN (k-nearest-neighborhood) approach is a representative enhanced neighborhood model~\citep{KNN2016}, which considers only the $k$ items rated by user $u$ that are the most similar to the item $i$ when predicting the rating $r_{ui}$. 
kNN-based approaches discard items that are poorly correlated to the target item, thus decreasing noise for improving the quality of recommendations.
Neighborhood-baesd approaches are similar to the item-item model for user personalization, which is different from our approach based on the user-item model~\cite{Item2001}. 
Thus, we focus on the latent factor modeling approach.

Most research on latent factor modeling is based on factoring the user-item rating matrix, which is known as \ac{SVD}~\citep{LFM2009}. 
\ac{SVD} factorizes the user-item rating matrix to a product of two lower rank matrices, one containing the ``user factors,'' the other containing the ``item-factors.''
Then, with an inner product and biases ($b_{ui}$), the user's preference towards an item can be generated, i.e.,
\begin{equation}
\hat{y}_{ui}=b_{ui}+\mathbf{z_u}\mathbf{{z_i}^\mathrm{T}},
\label{SVD}
\end{equation}
where $\mathbf{z_u}$ and $\mathbf{z_i}$ denote the ``user factors'' and ``item-factors,'' respectively.

Since the conventional \ac{SVD} is undefined in the presence of unknown values, i.e., missing ratings, several solutions have been proposed. 
Earlier work addresses this issue by filling the missing ratings with a baseline estimation~\citep{Application2000}. However, this leads to a very large, dense user rating matrix, where the factorization process becomes computationally infeasible. 
Recent work learns factor vectors directly on known ratings through a suitable objective function that minimizes a prediction error. 
The proposed objective functions are usually regularized in order to avoid overfitting~\citep{improving2007}. 
Typically, gradient descent is applied to minimize the objective function. 
\chen{An advantage of \ac{SVD}-based approaches is that they can provide recommendations for new users after given their ratings towards some items without reconstructing the parameters of the models. Thus for a new user, SVD-based approaches can provide recommendations immediately according to his current ratings.}

Another model based on \ac{SVD}, SVD++~\citep{SVD++2008}, incorporates both explicit and implicit feedback, and shows improved performance over many \ac{MF} models.
This is consistent with our motivation of combining explicit and implicit feedback in \ac{J-NCF}.
\chen{However, applying traditional \ac{MF} methods to sparse ratings matrices can be a non-trivial challenge with high computational costs for decomposing the rating matrix.}

Many traditional recommender systems apply a linear kernel with an inner product of user and item vectors to model user-item interactions.
Linear functions may not be able to give an accurate description of the characteristics of users (items) and user-item interactions: previous work has pointed out that non-linearities have potential advantages for improving the performance of recommender systems with extensive experiments~\citep{DAE2015,CDAE2016,Autorec2016}.

\subsection{Deep learning-based recommender system}
\label{Rdeep}
\ac{DL}-based recommender systems can be divided into two categories, i.e., single neural network models and deep integration models, depending on whether they rely solely on deep learning techniques or integrate traditional recommendation models with deep learning~\citep{RSsurvey2017,Cf-survey2009,DL-RS1,DL-RS2,NAuto2016,Huang:2013,onal-neural-2018,NFM2017,NGCF19}.

For the first category, \ac{RBM}~\citep{RBM2007,ORBM2009,CRBM2105} is an early neural recommender system.
It uses a two-layer undirected graph to model tabular data, such as users' explicit ratings of movies.
\ac{RBM} targets rating prediction, not top-N recommendation, and its loss function considers only the observed ratings.
It is technically challenging to incorporate negative sampling into the training of \acp{RBM}~\citep{CDAE2016}, which would be required for top-N recommendation.
AutoRec~\citep{Autorec2016} uses an Auto-Encoder for rating prediction.
It only considers the observed ratings in the loss function, which does not guarantee good performance for top-N recommendation.
To prevent the Auto-Encoder from learning an identity function and failing to generalize to unseen data, \ac{DAEs}~\citep{DAE2015} have been applied to learn from intentionally corrupted inputs.
Most of the publications listed so far focus on explicit feedback and, hence, fail to learn users' preference from implicit feedback.
\chen{\ac{CDAE}~\citep{CDAE2016} extends \ac{DAEs}; its input is a user's partially observed implicit feedback. Unlike our work, both \ac{DAEs} and \ac{CDAE} use an item-item model for personalization that represents a user with their rated items~\cite{Item2001} and the outputs are the item scores decoded from the learned user's representation.}
\chen{Our work is a kind of user-item model, which learns users' as well as items' representations first and then calculates the relevance between them.} 
The proposed \ac{J-NCF} model is a user-item model that personalizes by modeling user-item interactions.
Also, \ac{CDAE} applies a linear kernel to model the relationship between users and items, whereas \ac{J-NCF} applies a non-linear kernel.

Several \ac{CNN}-based recommendation models have been proposed~\citep{CNN2016,CTR2011,DLcontentRS}.
They primarily use \acp{CNN} to extract item features with auxiliary information, e.g., review text or contextual information, which we will incorporate in our future work.
As for Recurrent Neural Networks, they are used in recommender systems that address the temporal dynamics of ratings and sequential features~\citep{2016session-based,RNN2016}.

Most closely related to our model is \acf{NCF}~\citep{NCF2017}.
It uses multi-layer perceptrons to model the two-way interaction between users and items, which is meant to capture the non-linear relationship between users and items. Let $v_u^\mathit{user}$ and $v_u^\mathit{item}$ denote the side information (e.g., the feature information), then, the prediction rule of NCF is formulated as follows:
\begin{equation}
\label{NCF}
\hat{y}_{ui}=f(U^\mathrm{T}\cdot v_u^\mathit{user},V^\mathrm{T}\cdot v_u^\mathit{item}\mid U,V,\theta),
%\nonumber
\end{equation}
where the function $f(\cdot)$ defines the multilayer perceptron, and $\theta$ are the parameters of the network.
However, \ac{NCF} randomly initializes the representation of users and items, with just a one-hot identifier of user $u$ and item $i$ respectively, which only explores the users' and items' features in a limited manner.
\ac{J-NCF} adopts a joint neural network structure to capture both user and item features, and user-item relationships, as we hypothesize that the two parts can be optimized through tight coupling and joint training.
In addition, \ac{NCF} only exploits implicit feedback for item recommendations and ignores explicit feedback.

An extension based on \ac{NCF} is CCCFNet (Cross-domain Content-boosted Collaborative Filtering neural Network)~\citep{CCCFNet2017}. 
The basic building block of CCCFNet is also a dual network (for users and items, respectively). 
It models the user-item interactions in the last layer with the dot product. 
Unlike our work, it applies content information with a neural network to capture the user's preferences and item features. 
In addition, DeepFM (Deep Factorization Machine)~\citep{DeepFM2017} is an end-to-end model that seamlessly integrates factorization machine and MLP. 
However, it also applies content information and thus models higher-order feature interactions via a deep neural network and low-order interactions via a factorization machine. 
In contrast, \ac{J-NCF} adopts the rating information to explore both user and item features, which are  easier to collect.

As to deep integration models, \ac{CDL}~\citep{CDL2015} is a hierarchical Bayesian model that integrates stacked \ac{DAEs} into traditional probabilistic \ac{MF}.
It differs from our work in two ways: \begin{inparaenum}
\item it extracts deep feature representations of items from the content information which we do not explore, and
\item it uses a linear kernel to model relations between users and items with the dot product of user and item vectors\end{inparaenum}.

A well-known integration model is DeepCoNN (Deep Cooperative Neural Network)~\citep{Jointreview2017}, which adopts two parallel convolutional neural networks to model user behavior and item properties from review texts. 
In the final layer, a factorization machine is applied to capture their interactions from rating predictions. 
It alleviates the sparsity problem and enhances model interpretability by exploiting a rich semantic representation of the reviews, which could be investigated in \ac{J-NCF} as future work.

\chen{Wide \& Deep learning~\citep{Wide2017} and DeepFM~\citep{DeepFM2017} are two state-of-the-art recommendation works with deep learning techniques. 
While they focus on incorporating various features of users and items, we aim at exploring deep learning methods for pure collaborative filtering systems. }
Another integration model that is directly relevant to our work is \acf{DMF}~\citep{DMF2017}.
It uses a deep \ac{MF} model with a neural network that maps users and items into a common low-dimensional space.
It follows the \ac{LFM}, which uses the inner product to compute interactions between users and items. This may partially explain why using deep layers does not help to improve the performance of \ac{DMF} (see~\citep[Section 4.4]{DMF2017}).
Unlike \ac{DMF}, we apply multi-layer perceptrons to model user-item interactions using a combination of user and item feature vectors as input.
This does not only help our model to be more expressive in modeling user-item interactions than linear products, but it also helps to improve the accuracy of user and item feature extraction.

\smallskip\noindent%
\todo{On top of the previous work discussed above, our proposed model \ac{J-NCF} combines feature learning and interaction modeling into an end-to-end trainable neural network, which enables the two processes to be optimized jointly. 
Besides this, we design a new loss function that combines point-wise and pair-wise losses to explore the integration of different types of information, i.e., both implicit and explicit feedback.}

\section{Approach}
\label{Approach}
The proposed model, \ac{J-NCF}, has a joint structure with a layer used for modeling users' and items' features (the DF network) and a higher layer used for modeling user-item interactions (the DI network).
These two layers can be trained in a joint manner to give a predicted score of a user's interactions with an item with minimum prediction error.
We first describe the notation used and then detail \ac{J-NCF}.
We also describe the loss function that we use for optimization.

\subsection{Problem formulation and notation}
\label{notations}
First we describe the task of top-N recommendation that we study in this paper.
Suppose that there are $M$ users and $N$ items, denoted as $U=\{\mathit{user}_1,\ldots,\mathit{user}_M\}$  and $I=\{\mathit{item}_1,\ldots,\mathit{item}_N\}$. $R\in \mathbb{R}^{M\times N}$ denotes the rating information, where $R_{ui}$ is the rating given by user $\mathit{user}_u$ to item $\mathit{item}_i$.
The task for top-N recommendation is to return a list containing a set of items for an individual user to maximize the user's satisfaction.

The main notation we use in this paper is listed in Table~\ref{notation}.

\begin{table}[t]
%\small
\centering
\caption{Main notation used in the paper.}
\label{notation}
%\tbl{Main notation used in the paper.}
    {\begin{tabular}{m{4em}<{\raggedright} m{32em}<{\raggedright}}
    \toprule
 \bf Notation  & \bf Description \\
 \midrule
     $U$ & the set of users\\
     $I$ & the set of items\\
     $R_{ui}$ & an explicit rating of user $u$ to item $i$\\
     $\mathbf{v_u}$ & a vector containing a user's ratings; serves as input to Net$_{\mathit{user}}$\\
     $\mathbf{v_i}$ & a vector containing an item's ratings; serves as input to Net$_{\mathit{item}}$\\
     $M$ & the number of unique users\\
     $N$ & the number of unique items\\
     $\mathbf{W^x_{u}}$& the weight matrix for the $x$-th layer in Net$_{\mathit{user}}$ \\
     $\mathbf{b^x_{u}}$& the bias for the $x$-th layer in Net$_{\mathit{user}}$ \\
     $f^x_{u}$& the activation function for the $x$-th layer in Net$_{\mathit{user}}$ \\
     $X$ & the number of layers in DF network\\
     $\mathbf{W^y_{ui}}$& the weight matrix for the $y$-th layer in the DI network\\
     $\mathbf{a_{ui}}$ & a combination of user and item vectors; serves as input to the DI network\\
     $\mathbf{b^y_{ui}}$& the bias for the $y$-th layer in the DI network\\
     $f^y_{ui}$& the activation function for the $y$-th layer in the DI network\\
     $Y$ & the number of layers in the DI network\\
     $\hat{y}_{ui}$ &  the predicted score of the interaction between user $u$ and item $i$\\
     $V^{+}$ & the set of items that a user rates\\
     $V^{-}$ & the set of items that are not rated by a user \\
     $\alpha$ & a tradeoff parameter controlling the contributions of the point-wise loss and pair-wise loss\\
     \bottomrule
    \end{tabular}}
\end{table}

\subsection{Joint Neural Collaborative Filtering}
\begin{figure}[t]
  \centering
   \includegraphics[clip,trim=0mm 0mm 0mm 0mm,width=0.65\columnwidth]{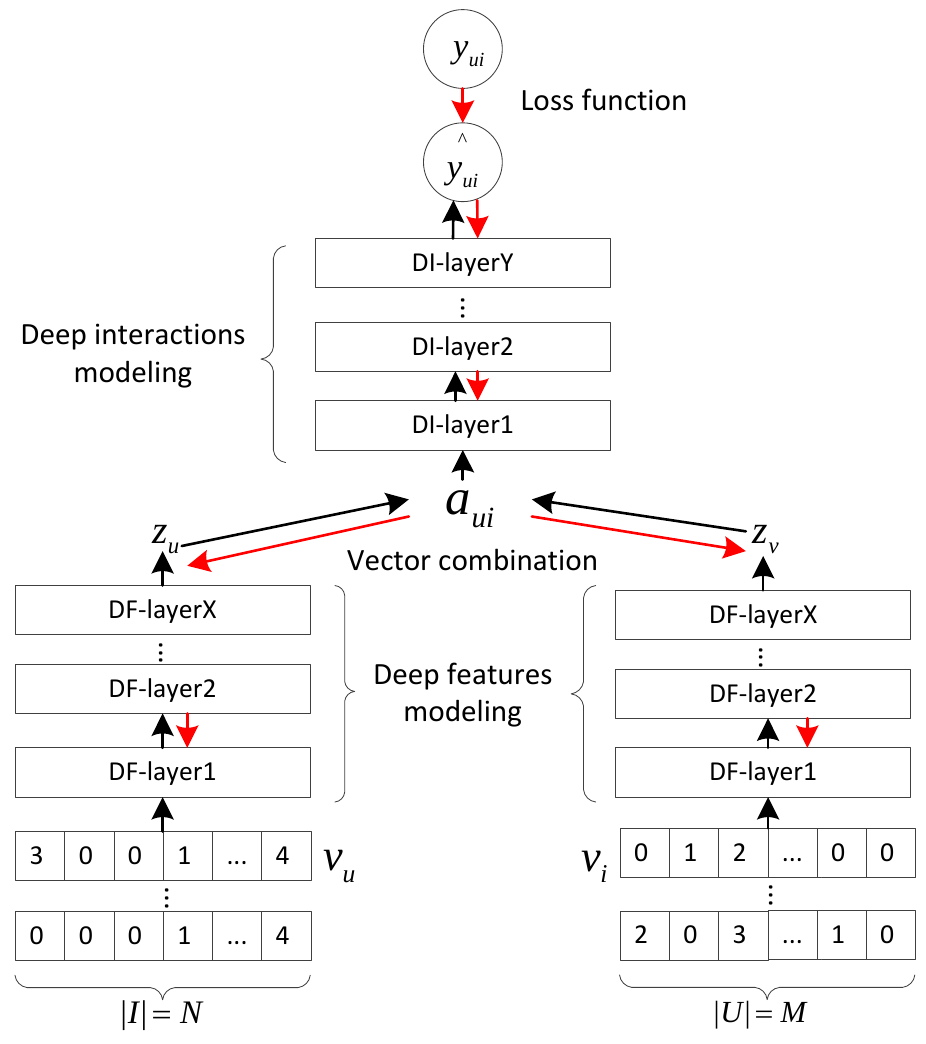}
   \caption{Structure of the \ac{J-NCF} model. Black arrows indicate the forward propagation for calculating the predictions. Red arrows indicate the back propagation for optimizing the parameters. (Best viewed in color.)}
\label{J-NCF}
\vspace{-3mm}
\end{figure}

%\label{J-NCF}
The  joint architecture of the proposed \ac{J-NCF} model is shown in Fig.~\ref{J-NCF}.
The model contains two main networks: a DF network for modeling features and a DI network for modeling interactions between items and users, where the output of the first network serves as the input of the second.

The DF network is used for modeling users' and items' features.
It contains two parallel neural networks coupled in the last layer, one network for users (Net$_{\mathit{user}}$) and another for items (Net$_{\mathit{item}}$).
We give the ratings of a user and an item as inputs to Net$_{\mathit{user}}$ and Net$_{\mathit{item}}$, respectively, which are defined as $\mathbf{v_u}=\langle y_{u1},\ldots,y_{uN}\rangle$ and $\mathbf{v_i}= \langle y_{1i},\ldots,y_{Mi}\rangle$, where
\begin{equation}
\label{equation1}
y_{ui}=\left\{
\begin{array}{ll}
0, & \text{for unknown ratings,}\\
R_{ui}, & \text{when explicit feedback is available.}
\end{array}
\right.
\end{equation}
We think of ratings as non-trivial explicit feedback from users as different ratings indicate different levels of users preference towards items. 
Obviously, there are many unknown ratings between users and items indicating non-preference of a user towards an item. 
Following~\citep{NCF2017,DMF2017}, we regard these unknown ratings as a kind of implicit feedback and mark them as zeroes. 
When pursuing a top-N recommendation task, we are interested only in a correct item ranking and care less about the exact rating scores. 
This grants us some flexibility, like considering all missing values in the user rating matrix as
zeros~\citep{Performance2010}. Thus we can take both explicit and implicit feedback into consideration with Eq.~\eqref{equation1}.

Then, with multi-layer perceptrons (MLP), the initial high-dimen\-sional rating vectors of users and items are mapped to lower-dimensional vectors.
Since Net$_{\mathit{user}}$ and Net$_{\mathit{item}}$ only differ in their inputs, we focus on illustrating the process for Net$_{\mathit{user}}$; the same process is applied for Net$_{\mathit{item}}$ with similar layers.
The MLP model in the DF network is defined as:
\begin{equation}
\label{equation2}
\begin{aligned}
\mathbf{z^1_u}&=f^1_u(\mathbf{W^1_u}\mathbf{v_u}+\mathbf{b^1_u}) \\
\mathbf{z^2_u}&=f^2_u(\mathbf{W^2_u}\mathbf{z^1_u}+\mathbf{b^2_u}) \\
&\mbox{}\hspace*{1.5mm}\vdots\\
\mathbf{z_u} &=f^X_u(\mathbf{W^X_u}\mathbf{z^{X-1}_u}+\mathbf{b^X_u}),
%\nonumber
\end{aligned}
\end{equation}
where $\mathbf{W^x_u}$, $\mathbf{b^x_u}$ and $f^x_u$ denote the weight matrix, the bias vector and the activation function for the $x$-th layer. 
Here, we use a ReLU as the activation function, as it has been shown to be more expressive than others and can effectively deal with the vanishing gradient problem~\citep{DMF2017,NCF2017}.
$X$ indicates the number of layers used in the DF network.
The output of the final layer $z_u$ is a deep representation of the user features; likewise, $z_i$ is the deep representation for the item features.

As to modeling user-item interactions, traditional \ac{LFM}  methods have been widely used.
Such methods are based on the dot product of user and item vectors, which models a user's preference with a linear kernel.
%Recent work suggests that non-linear functions have advantages over linear kernel methods.
In order to investigate the differences between non-linear and linear functions in modeling user-item interactions, we propose two ways to obtain fused users' and items' feature vectors $\mathbf{a_{ui}}$ as the input of the DI network:
\begin{equation}
\label{equation3}
\mathbf{a_{ui}}=\left\{
\begin{array}{ll}
\begin{bmatrix}
\mathbf{z_u} \\ \mathbf{z_i}
\end{bmatrix}
, & \text{concatenation, or}\\
\mathbf{z_u} \odot \mathbf{z_i}, & \text{multiplication.}
\end{array}
\right.
\end{equation}
The first way is to concatenate the two input vectors $\mathbf{z_u}$ and $\mathbf{z_i}$, which we regard as a non-linear fusion.
The second way is to use the element-wise product of vectors, which uses a linear kernel to generate user-item interactions.
Based on these two ways of fusing the input vectors $\mathbf{z_u}$ and $\mathbf{z_i}$, we propose two versions of \ac{J-NCF}, which we discuss in detail in our experiments.

Generating $\mathbf{a_{ui}}$ is the first step for modeling user-item interactions.
However, it is insufficient for modeling the complex relationship between users and items.
Thus, we adopt intermediate hidden layers to which $\mathbf{a_{ui}}$ is fed so as to obtain a multi-layer non-linear projection of user-item interactions:
\begin{equation}
\label{equation4}
\begin{aligned}
\mathbf{z^1_{ui}}&=f^1_{ui}(\mathbf{W^1_{ui}}\mathbf{a_{ui}}+\mathbf{b^1_{ui}})\\
\mathbf{z^2_{ui}}&=f^2_{ui}(\mathbf{W^2_{ui}}\mathbf{z^1_{ui}}+\mathbf{b^2_{ui}})\\
&\mbox{}\hspace{1.5mm}\vdots\\
\mathbf{z_{ui}}&=f^Y_{ui}(\mathbf{W^Y_{ui}}\mathbf{z^{Y-1}_{ui}}+\mathbf{b^Y_{ui}}),
\end{aligned}
\end{equation}
where $\mathbf{W^y_{ui}}$, $\mathbf{b^y_{ui}}$ and $f^y_{ui}$ denote the weight matrix, the bias vector and the activation function for the $y$-th layer in the DI network. 
A {ReLU is applied again as the activation function.}
$Y$ indicates the number of layers used in the network.
The output of the network is the predicted score of the interaction between user $u$ and item $i$:
\begin{equation}
\label{equation5}
\hat{y}_{ui}=\sigma(\mathbf{h^\mathrm{T}} \mathbf{z_{ui}}),
%\nonumber
\end{equation}
where the sigmoid function $\sigma$ can restrict the output in (0,1). $\mathbf{h}$ can be learnt through the training process with back propagation to control the weight of each dimension in $\mathbf{z_{ui}}$.

\subsection{Loss function}
\label{section:loss}
\chen{Objective functions for training recommender systems can be divided into three groups: point-wise, pair-wise and list-wise.}
Point-wise objectives aim at obtaining accurate ratings, which is more applicable in rating prediction tasks~\citep{FISM2013}.
Pair-wise objectives are usually focused on users' preferences towards pairs of items and are usually considered more suitable for top-N recommendation~\citep{EALS2016,NCF2017,FISM2013,BPR2009}. \chen{List-wise objectives are focused on users' interests towards a list of items, which are also used in some deep learning algorithms.} 
We briefly summarize the three groups of loss functions.

We use $\ell(\cdot)$ to denote a loss function and $\Omega(\theta)$ to represent a regularization term that controls the model complexity and encodes prior information such as sparsity, non-negativity, or graph regularization.

For a \emph{point-wise} loss function, the general calculation is:
\begin{equation}
\label{equation6}
\textit{$L$}=\sum_{u\in U}\sum_{i\in I} \ell_\textit{point-wise}(y_{ui},\hat{y}_{ui}) + \lambda \Omega(\theta),
\end{equation}
There are several types of point-wise loss function. E.g., squared loss is more suitable for explicit feedback than implicit feedback, as it is calculated with:
\begin{equation}
\label{equation7}
\ell_\mathit{squ}=\sum_{u\in U}\sum_{i\in I} w_{ui}(y_{ui}-\hat{y}_{ui})^2,
\end{equation}
where $w_{ui}$ is a hyper-parameter denoting the weight of training instance $(u,i)$.
\chen{The use of squared loss is based on the assumption that observations are generated from a Gaussian distribution, however, it may not tally well with implicit data~\citep{2007PMF}. For implicit feedback, there is a point-wise loss function mainly used for classification tasks~\citep{DMF2017, NCF2017}, named log loss~\cite{FISM2013}, which can perform better with implicit feedback than squared loss:}
\begin{equation}
\label{equation8}
\ell_{\log}=-\sum_{u\in U}\sum_{i\in I} y_{ui}\log\hat{y}_{ui} + (1-y_{ui})\log(1-\hat{y}_{ui}).
\end{equation}
\chen{\emph{Pair-wise} loss considers the relative order of the prediction for pairs of items, which is a more reliable kind of information for top-N recommendation. \citet{2018lossfunction} investigate several popular pair-wise loss functions, i.e., TOP1, BPR-max and TOP1-max. We give a brief introduction of them. }
\chen{TOP1 is the regularized approximation of the relative rank of the relevant item, which can be calculated as:}
\begin{equation}
\label{equation9}
\ell_\textrm{TOP1}=\frac{1}{|N_S|}{\sum_{{j\in {N_S}}}}\sigma(\hat{y}_{uj}-\hat{y}_{ui})+\sigma(\hat{y}^2_{uj}),
%\nonumber
\end{equation}
where $\hat{y}_{uj}$ and $\hat{y}_{ui}$ denote the prediction scores for a negative item $j$ and a positive item $i$, respectively; $N_S$ is the set of negative samples. 
\chen{The first part of TOP1 aims to ensure that the target score is higher than the score of the negative samples, while the second part pushes the score of the negative samples down. }
\chen{As for BPR-max and TOP1-max, they have been proposed by~\citet{2018lossfunction} to overcome the vanishing gradients as the number of negative samples increases. 
The idea is to have the target score compared with the most relevant sample score, which is the maximum score amongst the samples. 
As the maximum operation is non-differentiable, softmax scores are used to preserve differentiability.}
\chen{By summing over the individual losses weighted by the corresponding softmax scores $s_j$, TOP1-max can be calculated as:}
\begin{equation}
\label{equation10}
\ell_\textrm{TOP1-max}={\sum_{{j\in {N_S}}}}s_j(\sigma(\hat{y}_{uj}-\hat{y}_{ui})+\sigma(\hat{y}^2_{uj})).
\end{equation}
\chen{And the BPR-max loss function can be calculated as:}
\begin{equation}
\label{equation11}
\ell_\textrm{BPR-max}=-\log{\sum_{{j\in {N_S}}}}s_j\sigma(\hat{y}_{ui}-\hat{y}_{uj}).
\end{equation}
\chen{For \emph{list-wise} loss, many deep learning-based methods combine cross-entropy loss with softmax, which introduces list-wise properties into the loss. 
We refer to it as softmax+cross-entropy (XE) loss, which can be calculated with the following function:}
\begin{equation}
\label{equation12}
\ell_\textrm{XE}=-\log s_i=-\log\frac{e^{\hat{y}_{ui}}}{{\sum_{{j\in {N_S}}}}e^{\hat{y}_{uj}}}
\end{equation}
Most deep learning-based models only use the point-wise loss function for optimization and leave the pair-wise loss function for future work~\citep{DMF2017,NCF2017}.
%We notice that point-wise and pair-wise loss have their advantages and disadvantages. 
Point-wise loss only uses the rating information and ignores the information contained in the relative order of pairs of items. Pair-wise loss, in contrast, ignores the information of a user's individual preference for a certain item.
Thus, unlike previous work, \ac{NCF} and \ac{DMF}, our proposed \ac{J-NCF} model considers both point-wise and pair-wise loss for the top-N recommendation task and combines them into a new loss function:
\begin{equation}
\label{equation13}
\textit{L}= \alpha L_\textit{pair-wise} + (1-\alpha) L_\textit{point-wise},
%\nonumber
\end{equation}
where $\alpha$ is used to control the weights of the two parts.

For point-wise loss, we adopt the log loss (Eq.~\eqref{equation8}), which can integrate both implicit and explicit feedback. 
\chen{As to pair-wise loss, combining with different pair-wise losses yields different new loss functions, i.e., point-wise+TOP1, point-wise+BPR-max, and point-wise+TOP1-max. 
We analyze the performance of these different combined loss functions with experiments in Section~\ref{Results}.}

Acknowledging that explicit and implicit feedback both contain information about a user's preference towards items, we combine both kinds of feedback in our loss function for optimization and rewrite Eq.~\eqref{equation13} in detail as
\begin{equation}
\label{equation14}
\chen{\textit{L}= \alpha L_\textit{pair-wise} +(1-\alpha)(-Y_{ui}\log\hat{y}_{ui} - (1-Y_{ui})\log(1-\hat{y}_{ui})),}
\end{equation}
where $Y_{ui} =\frac{y_{ui}}{Max(R_u)}$, and $Max(R_u)$ denotes the largest rating score of user $u$ given to items, so that different values of $y_{ui}$ have a different influence on the loss. For example, if the largest rating score of a user $u$ given to items is 4, when he rates an item $i$ with 2, we can generate $Y_{ui} =\frac{y_{ui}}{Max(R_u)}=\frac{2}{4}$. We refer to our loss function Eq.~\eqref{equation14} as a ``\emph{hybrid}'' loss function.

%
%\smallskip\noindent
We have developed the joint neural network structure of the \ac{J-NCF} model.
The training process of \ac{J-NCF} is shown in Algorithm~\ref{algorithm1}.
We first initialize the parameters in the network and modify the rating matrix from step~\ref{step1} to~\ref{step4}.
Then, in step~\ref{step8} and~\ref{step9}, we generate deep feature representations for both users and items with the DF network.
In step~\ref{step10} and~\ref{step11}, we calculate the predicted scores for the user-item interactions with the DI network.
Finally, we use the hybrid loss function in Eq.~\eqref{equation14} and back propagation to optimize the network parameters with step~\ref{step12} and~\ref{step13}.

\begin{algorithm}[t]
\caption{Joint Neural Collaborative Filtering.}
\label{algorithm1}
\begin{algorithmic}[1]
\REQUIRE Epochs: training iterations;\\
         $R$: the original rating matrix;\\
         $U$: user set; \\
         $I$: item set;
\ENSURE $\mathbf{W^{x}_u}$ ($x=1,\ldots,X$): Weight matrix of Net$_{\mathit{user}}$;\\
        $\mathbf{b^{x}_u}$ ($x=1,\ldots,X$): Bias of Net$_{\mathit{user}}$;\\
        $\mathbf{W^{x}_i}$ ($x=1,\ldots,X$): Weight matrix of Net$_{\mathit{item}}$;\\
        $\mathbf{b^{x}_i}$ ($x=1,\ldots,X$): Bias of Net$_{\mathit{item}}$;\\
        $\mathbf{W^{y}_{ui}}$ ($y=1,\ldots,Y$): Weight matrix of DI network;\\
        $\mathbf{b^{y}_{ui}}$ ($y=1,\ldots,Y$): Bias of DI network.
\STATE randomly initialize $\mathbf{W_u}$, $\mathbf{W_i}$,$\mathbf{W_{ui}}$, $\mathbf{b_u}$,
      $\mathbf{b_i}$ and $\mathbf{b_{ui}}$\label{step1};
\STATE $y_{ui}$ $\gets$ use Eq.~\eqref{equation1} with $R$;
\STATE $V^+$ $\gets$ all none zero interactions pairs\label{step4};
%\STATE $V^-$ $\gets$ all zero interactions.
\FOR{ epoch in range(Epochs)}
    \STATE random shuffle of $V^+$
    \FOR {$\langle u,i\rangle \in V^+$}
	   \STATE sample the set of negative samples $N_S$ 
	    \FOR{$j \in N_S$}
                  \STATE $\mathbf{v_u}$, $\mathbf{v_i}$, $\mathbf{v_j}$ $\gets$ $y_{ui}$ with Eq.~\eqref{equation1};\label{step8}
                  \STATE $\mathbf{z_{u}}$, $\mathbf{z_{i}}$, $\mathbf{z_{j}}$ $\gets$ use Eq.~\eqref{equation2} with $\mathbf{v_u}$, $\mathbf{v_i}$, $\mathbf{v_j}$ as inputs;\label{step9}
                  \STATE $\mathbf{a_{ui}}$, $\mathbf{a_{uj}}$ $\gets$ use Eq.~\eqref{equation3} with $\mathbf{z_{u}}$, $\mathbf{z_{i}}$, $\mathbf{z_{j}}$;\label{step10}
                  \STATE $\hat{y}_{ui}$, $\hat{y}_{uj}$ $\gets$ use Eq.~\eqref{equation4} and Eq.~\eqref{equation5};\label{step11}
                  \STATE $\textit{L}$ $\gets$ use Eq.~\eqref{equation14} with $y_{ui}$, $\hat{y}_{ui}$ and $\hat{y}_{uj}$ as inputs;\label{step12}
                  \STATE use back propagation to optimize the parameters;\label{step13}
             \ENDFOR
    \ENDFOR
\ENDFOR	
\STATE \textbf{return} $\mathbf{W_u}$, $\mathbf{W_i}$, $\mathbf{W_{ui}}$, $\mathbf{b_u}$,
      $\mathbf{b_i}$ and $\mathbf{b_{ui}}$.
\end{algorithmic}
\end{algorithm}

\section{Experimental setup}
\label{Experiments}
We design experiments on a variety of datasets to examine the effectiveness of \ac{J-NCF}.
We first explain the research questions and the models we use for comparison in Section~\ref{Models}.
The datasets and experiments are described in Section~\ref{Dataset}.

\subsection{Model summary and research questions}
\label{Models}
We conduct experiments with the aim of answering the following research questions:
\begin{enumerate}[nosep, align=left, leftmargin=*]

\item[\bf{RQ1}] Does our proposed \ac{J-NCF} method outperform state-of-art collaborative filtering baselines for recommender systems?

\item[\bf{RQ2}] \chen{How is the performance of \ac{J-NCF} impacted by different choices for the pair-wise loss in Eq.~\eqref{equation14}? }

\item[\bf{RQ3}] Does the hybrid loss function Eq.~\eqref{equation11}, which combines point-wise and pair-wise loss, help to improve the performance of \ac{J-NCF}?

\item[\bf{RQ4}] Are deeper layers of hidden units in the DF network and DI network helpful for the recommendation performance of \ac{J-NCF}?

\item[\bf{RQ5}] Does the combination of explicit and implicit feedback help to improve the performance of \ac{J-NCF}?

\item[\bf{RQ6}] How does the performance of \ac{J-NCF} vary across users with different numbers of interactions?

\item[\bf{RQ7}] Is \ac{J-NCF} sensitive to different degrees of data sparsity?

\item[\bf{RQ8}] \chen{How does \ac{J-NCF} perform on a large and sparse dataset?}

\item[\bf{RQ9}] \chen{How do the training and inference times of \ac{J-NCF} compare against those of other neural models?}

\end{enumerate}

\noindent%
We compare \ac{J-NCF} against a number of traditional collaborative filtering baselines and against state-of-the-art deep learning based models:
\begin{enumerate}[nosep, align=left, leftmargin=*]

\item[\bf{Item-pop}]  This method ranks items based on the number of interactions, which is a non-personalized approach to determine recommendation scores~\citep{RS-CF2005}.

\item[\bf{BPR}] This method uses a pairwise loss function to optimize a \ac{MF} model based on implicit feedback. 
We use it as a strong baseline for traditional collaborative filtering method~\citep{BPR2009}.

\item[\bf{NCF}] This is a state-of-the-art neural network-based method for recommender systems. 
It aims to capture the non-linear relationship between users and items. 
Unlike \ac{J-NCF}, it simply uses one-hot vectors representing users and items as the input for modeling user-item interactions.
And it only uses implicit feedback and a point-wise loss function~\citep{NCF2017}.

\item[\bf{DMF}] This method uses multi-layer perceptrons for rating matrix factorization. 
Unlike our work, after projecting users and items into low dimensional vectors, it applies an inner product to calculate interactions between users and items, which is a linear kernel. 
It uses a point-wise loss function for optimization~\citep{DMF2017}.
\end{enumerate}

\noindent%
In addition, following the choices that we identified in Eq.~\eqref{equation3}, we consider two versions of \ac{J-NCF}:
\begin{enumerate}[nosep, align=left, leftmargin=*]
\item[\bf{\ac{J-NCF}$_{\mathit{m}}$}]  This is \ac{J-NCF} using element-wise multiplication for combining a user and an item feature vector as the input for the DI layer, which has a linear kernel inside.
\item[\bf{\ac{J-NCF}$_{\mathit{c}}$}] This is \ac{J-NCF} using concatenation for combining a user and an item feature vector as the input for the DI layer, which is a non-linear way.
\end{enumerate}

\noindent%
We list all the models to be discussed in Table~\ref{models}.
\begin{table}[!t]
%\small
  \centering
\caption{An overview of the models discussed in the paper.}
    \begin{tabular}{lp{9.5cm}l}
    \toprule
Model & Description & Source \\
 \midrule
Item-pop & \raggedright A typical recommendation approach, which ranks items based on the number of interactions. & \citep{RS-CF2005} \\
BPR & \raggedright A recommendation method using a pairwise loss function to optimize an \ac{MF} model based on implicit feedback. & \citep{BPR2009}\\
NCF & \raggedright A state-of-the-art neural based method for recommender systems. & \citep{NCF2017} \\
DMF & \raggedright A method using multi-layer perceptrons for rating matrix factorization. & \citep{DMF2017}\\
 \midrule
J-NCF$_{\mathit{m}}$ & A \raggedright J-NCF model using element-wise multiplication for combining a user and an item feature vector as the input for the DI layer. & This paper \\
J-NCF$_{\mathit{c}}$ & A  \raggedright J-NCF model using concatenation for combining a user and an item feature vector as the input for the DI layer.&  This paper \\
J-NCF$_{\mathit{point}}$ & A  \raggedright J-NCF model with only point-wise loss based on Eq.~\eqref{equation8}. &  This paper \\
J-NCF$_{\mathit{pair}}$ & A  \raggedright J-NCF model with only pair-wise loss based in Eq.~\eqref{equation9}. &  This paper \\
J-NCF$_{\mathit{hybrid}}$ & A  \raggedright J-NCF model with our designed loss function in Eq.~\eqref{equation11}. &  This paper \\
J-NCF$_{\mathit{ex}}$ & A  \raggedright J-NCF model with both explicit and implicit feedback in the input and the loss function. &  This paper \\
J-NCF$_{\mathit{im}}$ & A  \raggedright J-NCF model with only implicit feedback in the input and the loss function. &  This paper \\
    \bottomrule
    \end{tabular}
  \label{models}
%  \vspace*{-0.45\baselineskip}
\end{table}

\subsection{Datasets and experimental setup}
\label{Dataset}
\subsubsection{Datesets.}

We use three publicly available datasets to evaluate our models and the baselines:
\begin{enumerate}[nosep]
    \item  \textbf{MovieLens}, which contains several rating datasets from the MovieLens web site. The datasets are collected over various periods of time, depending on the size of the set~\citep{NCF2017,DMF2017}. We use two sets for our experiments, i.e., MovieLens 100K (ML100K) containing 100,000 ratings from 943 users on 1,682 movies, and MovienLens 1M (ML1M) containing more than 1 million ratings from 6,040 users on 3,706 movies.\footnote{\url{https://grouplens.org/datasets/movielens/}}
    \item  \textbf{Amazon Movies (AMovies)}, which contains 4,607,047 ratings for movies from Amazon, which is bigger and sparser than the MovieLens datasets and used widely in the recommender systems literture for evaluation~\citep{RSsurvey2017,DMF2017}.\footnote{\url{http://jmcauley.ucsd.edu/data/amazon/}}
    \item \chen{\textbf{Amazon Electronics (AEle)}, which is a larger and sparser dataset than the other datasets used in our paper. It contains 7,824,482 ratings of users on different electronics. We use it to test the performance of our model when applied on a large and sparse dataset.\footnote{\url{http://jmcauley.ucsd.edu/data/amazon/}}}
\end{enumerate}

\noindent%
\chen{For the two MovieLens datasets, we do not process them because they are already filtered. For the AMovies dataset, following~\citep{DMF2017,NCF2017}, we filter the dataset so that, similar to the MovieLens data, only users with at least 20 interactions and items with at least 5 interactions are retained. For the larger dataset AEle, we only do minor filtering on the data, i.e., filtering the users with less than 2 interactions and items with less than 5 interactions. To answer \textbf{RQ1} to \textbf{RQ7}, we use the ML100K, ML1M, and AMovies datasets to evaluate our models and baselines. As for \textbf{RQ8} to \textbf{RQ9}, we test the models on all of the datasets.}
The characteristics of the datasets after preprocessing are summarized in Table~\ref{dataset}.

\begin{table}[!t]
\centering
\caption{Dataset statistics. ``Density'' is the density of each dataset (i.e., $\#Density = \#Ratings/(\#Uses\times\#Items)$).}
\label{dataset}
\begin{tabular}{lrrrr}
%\begin{tabular}{m{5em}<{\raggedright} m{3em}<{\raggedright}m{3em}<{\raggedright}m{4em}<{\raggedright}m{4em}<{\raggedright}}
%{m{13em}<{\centering} m{5em}<{\centering} m{5em}<{\centering}}\hline%{|p{8em}|c|c|c|c|}%{m{7cm}<{\centering}
\toprule
    Dataset& \#Users& \#Items& \#Ratings& \#Density(\%)\\
    \midrule
 ML100K & 943 & 1,682 &100,000  &6.3047\\
 ML1M & 6,040 & 3,706 &1,000,209  &4.4685\\
 AMovies & 15,067 & 69,629 &877,736  &0.0837\\
 AEle & 1,221,341 & 157,003 & 4,486,501 & 0.00234\\
  \bottomrule
\end{tabular}
\vspace*{-.5\baselineskip}
\end{table}

\begin{figure*}[t]
  \centering
   \includegraphics[width=1\textwidth]{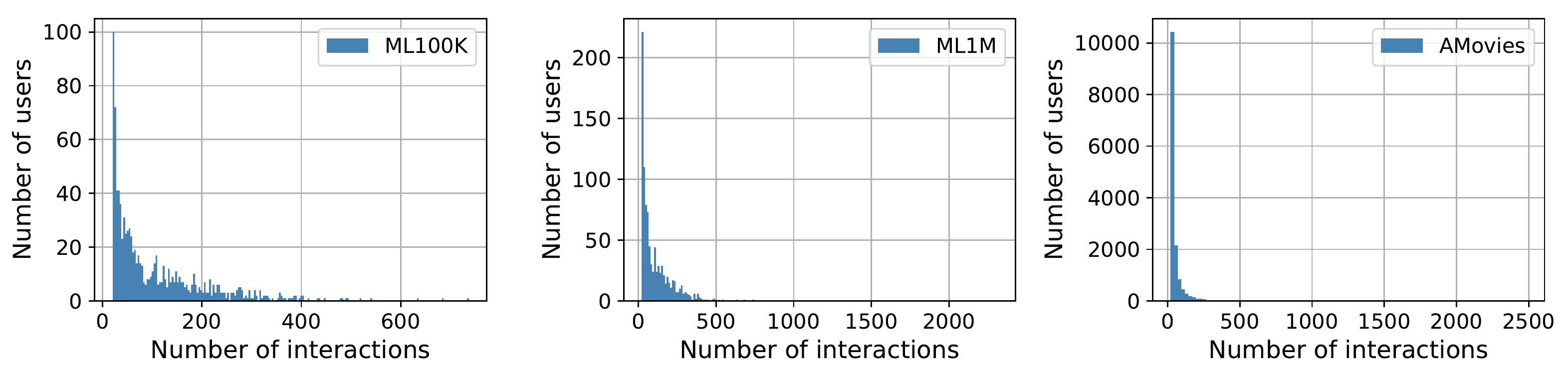}
   \caption{Distribution of users with varying numbers of interactions in the ML100K, ML1M, and AMovies datasets, respectively.}
\label{distribution}
\end{figure*}

In order to answer \textbf{RQ5}, we plot distributions of users with different numbers of interactions in the ML100K, ML1M, and AMovies datasets in Figure~\ref{distribution}.
The x-axis denotes the number of ratings while the y-axis indicates the number of users corresponding to the ratings.
We see that the majority of users in the three datasets only have a few ratings, which we regard as ``inactive users,'' and few ``active users'' have far more ratings.
E.g., in the ML100K dataset, 61.72\% of the users have fewer than 100 ratings, 32.66\% have between 100 and 300 ratings, and only 5.6\% of the users have more than 300 ratings.

As we will see below, the models being considered in this paper achieve different scores when used on datasets with different characteristics, i.e., number of users and number of items (see Section~\ref{Results}).
Thus, for \textbf{RQ6}, in order to evaluate the performance of our model on datasets with different degrees of sparsity, we keep the number of users and items the same.
\chen{Namely, following \citep{FISM2013}, for each of the three datasets, i.e., ML100K, ML1M, and AMovies, we create three versions at different sparsity levels with the the following steps:}
%For each dataset, we start by randomly choosing a subset of users and items from the main dataset.
%This dataset is represented with a `-1' suffix.
%Keeping the same set of users and items, the first sparser version of the dataset with the `-2' suffix is created by randomly removing entries from the first dataset's user-item matrices.
%The second sparser version of the dataset with the `-3' suffix is similarly created by randomly removing entries from the second dataset's user-item matrices.
%The characteristics of the datasets are summarized in Table~\ref{dataset-sparsity}.
\begin{enumerate}[nosep,label=(Step \arabic*),leftmargin=*,align=left]
\item[\chen{Step 1}] \chen{We start by randomly choosing a subset of users and items from the original dataset. This dataset is represented with a `-1' suffix.}
\item[\chen{Step 2}] \chen{We randomly choose a rating record and make a judgment if the numbers of users as well as items are unchanged of the sub-dataset after removing this record. If unchanged, we remove this record; otherwise repeat Step 2.}
\item[\chen{Step 3}] \chen{After several repetitions of Step 2, the first sparser version of the dataset with the `-2' suffix is created.}
\item[\chen{Step 4}] \chen{Repeat Step 2 and Step 3 based on the dataset with a `-2' suffix, the second sparser version of the dataset with the `-3' suffix is created in the same way.}
\end{enumerate}
\chen{The characteristics of the datasets are summarized in Table~\ref{dataset-sparsity}.}

\begin{table}[t]
\centering
\caption{Dataset statistics with different degrees of sparsity.}
\label{dataset-sparsity}
\begin{tabular}{lrrrr}
%\begin{tabular}{m{5em}<{\raggedright} m{3em}<{\raggedright}m{3em}<{\raggedright}m{4em}<{\raggedright}m{4em}<{\raggedright}}
%{m{13em}<{\centering} m{5em}<{\centering} m{5em}<{\centering}}\hline%{|p{8em}|c|c|c|c|}%{m{7cm}<{\centering}
\toprule
    Dataset& \#Users& \#Items& \#Ratings& \#Density(\%)\\
    \midrule
 ML100K-1 & 943 & 1,682 &69,999  &4.4132\\
 ML100K-2 & 943 & 1,682 &39,999  &2.2522 \\
 ML100K-3 & 943 & 1,682 &9,999  &0.6304 \\
 \hline
 ML1M-1 & 3,706 & 6,040 &850,208  &3.7982\\
 ML1M-2 & 3,706 & 6,040 &350,207  &1.5645 \\
 ML1M-3 & 3,706 & 6,040 &167,870  &0.7499 \\
 \hline
 AMovies-1 & 7,402 & 12,080 &87,807  &0.0982\\
 AMovies-2 & 7,402 & 12,080 &37,823  &0.0423 \\
 AMovies-3 & 7,402 & 12,080 &18,867  &0.0211  \\
  \bottomrule
\end{tabular}
\vspace*{-.5\baselineskip}
\end{table}

\subsubsection{Experimental setup.}
For evaluation, we use a \emph{leave-one-out} strategy, which has been used widely in \ac{DL}-based recommender systems~\citep{DMF2017,NCF2017,EALS2016}.
The training set consists of all but the last interaction of every user; the test set contains the latest interaction of every user.
When testing, it is time-consuming to give ranking predictions to all items for every user. 
\chen{Thus following~\citet{NCF2017,DMF2017}, we randomly sample 100 items with which the user has not interacted and then give the test item ranking predictions among the 100 samples. 
Although using this sampling strategy during evaluation may overestimate the performance of all algorithms, \citet{Precision2011,2018lossfunction} have pointed out that the comparison among algorithms still remains fair.}

The majority of the recommender system literature applies error metrics for evaluation, i.e., \emph{Root Mean Squared Error} (RMSE) and \emph{Mean Absolute Error} (MAE).
Such classical error criteria do not really measure the top-N recommendation performance~\citep{Performance2010}. An extensive evaluation of several state-of-the-art recommender algorithms suggests that algorithms optimized for minimizing RMSE do not necessarily perform as expected in terms of the top-$N$ recommendation task~\citep{Performance2010,TOIS2004}. Experimental results also show that improvements in terms of RMSE often do not translate into accuracy improvements~\citep{TOIS2004}.
Thus, here we choose to use accuracy metrics to examine the recommendation performance~\citep{NCF2017}.
Specifically, we use HR and NDCG to evaluate the performance of our models.
\emph{Hit Ratio} (HR) is used to evaluate the precision of the recommender system, i.e., whether the test item is contained in the top-N list.
The \emph{Normalized Discount Cumulative Gain} (NDCG) measures the ranking accuracy of the recommender system, i.e., whether the test item is ranked at the top of the list.

As for parameters, we optimize the hyperparameters by running 100 experiments at randomly selected points of the parameter space. Optimization is done on a validation set, which is partitioned from the training set with the same procedure as the test set~\citep{QS2018}.
\chen{As for the loss function, we test the parameter $\alpha$ from $0$ to $1$ with step size of $0.1$ in our experiment.}
For the neural networks, we randomly initialize model parameters with a Gaussian distribution (mean of 0 and standard deviation of 0.01), optimizing the model with mini-batch Adam~\citep{Adam2014}.
The batch size and learning rate are set to 256 and 0.0001.
\chen{For the baselines, we set the parameters of \ac{DMF} as well as \ac{NCF} following~\citep{DMF2017,NCF2017}, respectively. For \ac{DMF} and \ac{NCF}, we set the batch size to  256, and the learning rate to 0.0001 and 0.001.} 
\chen{For the DF network in DMF model, we apply two layers and the sizes of them are [128, 64]. 
For the DI network in the NCF model, we employ three hidden layers with size [128, 64, 8]. 
For the DF and DI networks in \ac{J-NCF}, without special mention, we employ three layers in DF network with the size of [256, 128, 64] and two layers in DI network with size of [128, 8]. 
Thus the embedding sizes of users as well as items are same in all baseline models as well as \ac{J-NCF}. 
We also keep the size of the last hidden layer of the DI network in \ac{J-NCF} the same as \ac{NCF}, which may determine the model capability.}
\chen{We also test our model as well as the baseline models with different numbers of layers to see if deep layers are beneficial to the overall performance of these models.}
Unless specified, for all the results presented in this paper, the number of recommendations ($N$) is equal to 10~\citep{DMF2017,NCF2017}.

\section{Results and Discussion}
\label{Results}
\subsection{Overall performance}
\label{overalperformance}
To answer \textbf{RQ1}, we examine the recommendation performance of the baselines and the \ac{J-NCF}$_{\mathit{m}}$ and \ac{J-NCF}$_{\mathit{c}}$ models. See Table~\ref{performance}.

\begin{table*}[t]
\captionsetup{justification=justified}
  \centering
  \caption{Performance of recommendation models. The results produced by the best baseline and the best performer in each column are underlined and boldfaced, respectively. Statistical significance of pairwise differences of \ac{J-NCF}$_{\mathit{m}}$ and \ac{J-NCF}$_{\mathit{c}}$ vs.\ the best baseline) is determined by a $t$-test ($^\blacktriangle$/$^\blacktriangledown$ for $\alpha$ = .01, or $^\vartriangle$/$^\triangledown$ for $\alpha$ = .05).}
\label{performance}
\begin{tabular}{lcccccccc}
\toprule
 &\multicolumn{2}{c}{ML100K} && \multicolumn{2}{c}{ML1M} &&\multicolumn{2}{c}{AMovies} \\
 \cmidrule{2-3}\cmidrule{5-6}\cmidrule{8-9}
Model & {HR@10} & {NDCG@10} && {HR@10} & {NDCG@10}&& {HR@10} & {NDCG@10}  \\
\midrule
Item-pop & .3832\duwen & .2018\duwen && .4513\duwen & .2315\duwen &&.5925\duwen & .3493\duwen  \\
BPR & .5762\duwen & .3021\duwen && .6097\duwen & .3711\duwen && .6288\duwen & .3903\duwen  \\
NCF & .6066\duwen & .3488\duwen && .6498\duwen & .3951\duwen && .6782\duwen & .4135\duwen  \\
DMF & \underline{.6309}\duwen & \underline{.3616}\duwen && \underline{.6748}\duwen & \underline{.4221}\duwen &&\underline{.7151}\duwen & \underline{.4616 }\duwen \\
\midrule
J-NCF$_{\mathit{m}}$ & .6627$^\vartriangle$ &.3877$^\vartriangle$ && .7127$^\blacktriangle$ &.4485$^\blacktriangle$ && .7666$^\blacktriangle$ &.5098$^\blacktriangle$  \\
J-NCF$_{\mathit{c}}$ & \bf{.6829}$^\blacktriangle$ &\bf{.4065}$^\blacktriangle$ && \bf{.7377}$^\blacktriangle$ &\bf{.4822}$^\blacktriangle$ && \bf{.7881}$^\blacktriangle$ &\bf{.5311}$^\blacktriangle$  \\
\bottomrule
\end{tabular}
\end{table*}

Let us first consider the baselines.
From Table~\ref{performance}, we see that \ac{DMF} achieves a better performance than the other baselines in terms of HR@10 and NDCG@10. 
Hence, we only use \ac{DMF} as the best baseline for comparisons in later experiments.
\ac{BPR} clearly shows higher improvements over the Item-pop baseline in terms of NDCG@10 than in terms of HR@10, which shows that pairwise loss has a strong performance for ranking prediction. The \ac{NCF} and \ac{DMF} models both show better performance than the two traditional \ac{CF} models, which indicates the utility of \ac{DL} techniques in improving recommendation performance.

Next, we compare the baselines against the \ac{J-NCF} models.
\ac{NCF} and \ac{DMF} both lose against the \ac{J-NCF} models in terms of HR@10 and NDCG@10. This shows that a joint neural network structure that tightly couples deep feature learning and deep interaction modeling helps to improve the recommendation performance.
Regarding the \ac{J-NCF} models, independent of the choice of combining the users' and items' vectors, \ac{J-NCF} achieves a better performance than the \ac{DMF} baseline, resulting in HR@10 improvements ranging from 5.04\% to 8.24\% on the ML100K dataset, 5.62\% to 10.81\% on the ML1M dataset, and 7.21\% to 10.21\% on the AMovies dataset. 
NDCG@10 improvements range from 7.22\% to 12.42\% on the ML100K dataset, 6.25\% to 14.24\% on the ML1M dataset, and 10.44\% to 15.06\% on the AMovies dataset.
Significant improvements against the baseline in terms of HR@10 and NDCG@10 are observed for both \ac{J-NCF}$_{\mathit{c}}$ and \ac{J-NCF}$_{\mathit{m}}$ at the $\alpha=.01$ level, except for \ac{J-NCF}$_{\mathit{m}}$ on the ML100K dataset, for which we observe significant improvements at the $\alpha=.05$ level in terms of HR@10 and NDCG@10.
The higher improvements in NDCG@10 over HR@10 may be due to the fact that we incorporate pair-wise loss in our loss function, which motivates us to conduct a further investigation to answer \textbf{RQ3}.

Comparing \ac{J-NCF}$_{\mathit{c}}$ and \ac{J-NCF}$_{\mathit{m}}$, we see that  \ac{J-NCF}$_{\mathit{c}}$ achieves the best performance, with improvements of 3.05\%, 3.51\% and 2.81\% in terms of HR@10, and 4.85\%, 7.51\% and 4.18\% in terms of NDCG@10  over \ac{J-NCF}$_{\mathit{m}}$ on the three datasets,  respectively.
The complex relationship between users and items can be described better with a non-linear kernel than linear kernel, which is consistent with the findings in~\citep{CRBM2105,NCF2017}.

\newcommand{\OurModel}{\ac{J-NCF}$_{\mathit{c}}$}

\subsection{Impact of different loss functions}
\label{diff-lossfunction}
\chen{As we have mentioned in Section~\ref{section:loss}, there are several kinds of pair-wise loss functions that can be incorporated in Eq.~\eqref{equation13}. 
When \ac{J-NCF} combines the point-wise loss, i.e., log loss, with TOP1, TOP1-max, and BPR-max pair-wise losses, it gives rise to the J-NCF$_{\mathit{TP}}$, J-NCF$_{\mathit{TMP}}$ and J-NCF$_{\mathit{BMP}}$ models, respectively.  
Additionally, list-wise loss, i.e., softmax+cross-entropy (XE), can also be applied with \ac{J-NCF}, which gives rise to the J-NCF$_{\mathit{XE}}$ model.
In order to  investigate the impact of various loss functions on \ac{J-NCF}, we examine the recommendation performance of J-NCF$_{\mathit{TP}}$, J-NCF$_{\mathit{TMP}}$, J-NCF$_{\mathit{BMP}}$ as well as J-NCF$_{\mathit{XE}}$ models where the parameter $\alpha$ in Eq.~\eqref{equation13} ranges from $0$ to $1$ with a step size of $0.1$. 
Fig.~\ref{diff-lossfunction} shows the results.}

\begin{figure*}[!t]
        \begin{subfigure}[t]{0.49\textwidth}
                \includegraphics[width=\textwidth]{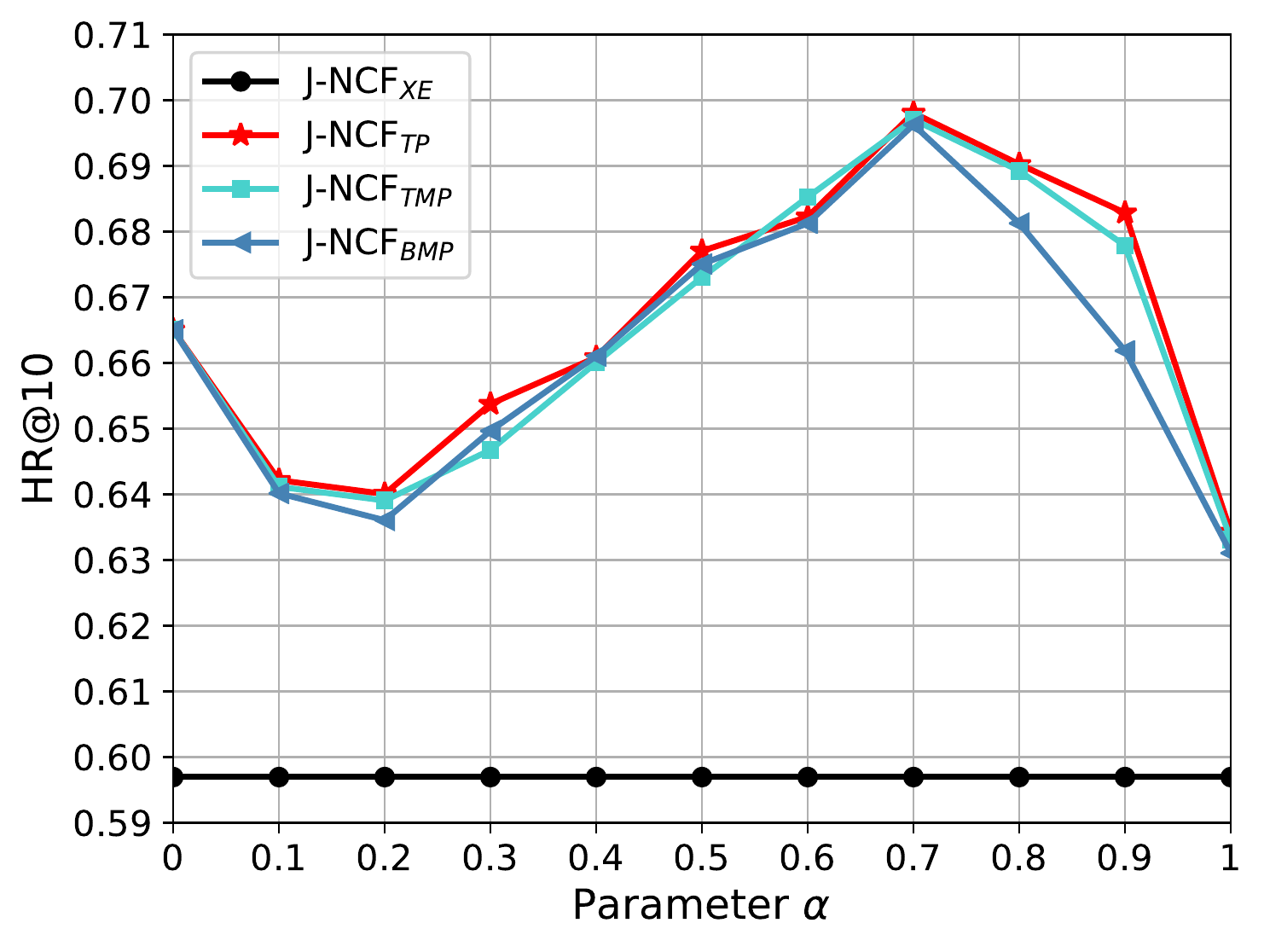}
                \caption{\raggedright Performance in terms of HR@10 on the ML100K dataset.}
                \label{diff-ML100k_hr}
        \end{subfigure}
        ~%space
        \begin{subfigure}[t]{0.5\textwidth}
                \includegraphics[width=\textwidth]{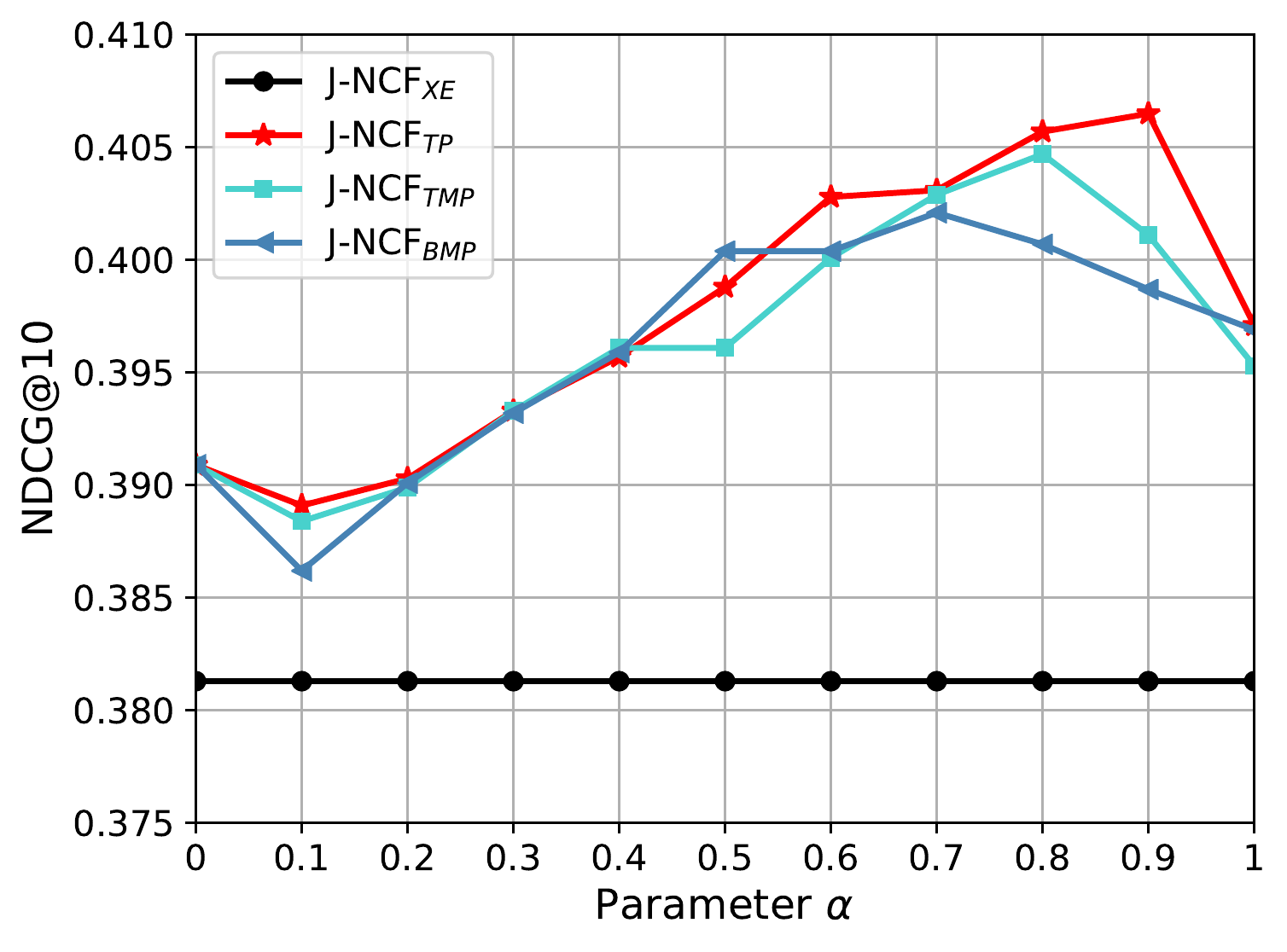}
                \caption{\raggedright Performance in terms of NDCG@10 on the ML100K dataset.}
                \label{diff-ML100k_ndcg}
        \end{subfigure}
        \begin{subfigure}[t]{0.49\textwidth}
                \includegraphics[width=\textwidth]{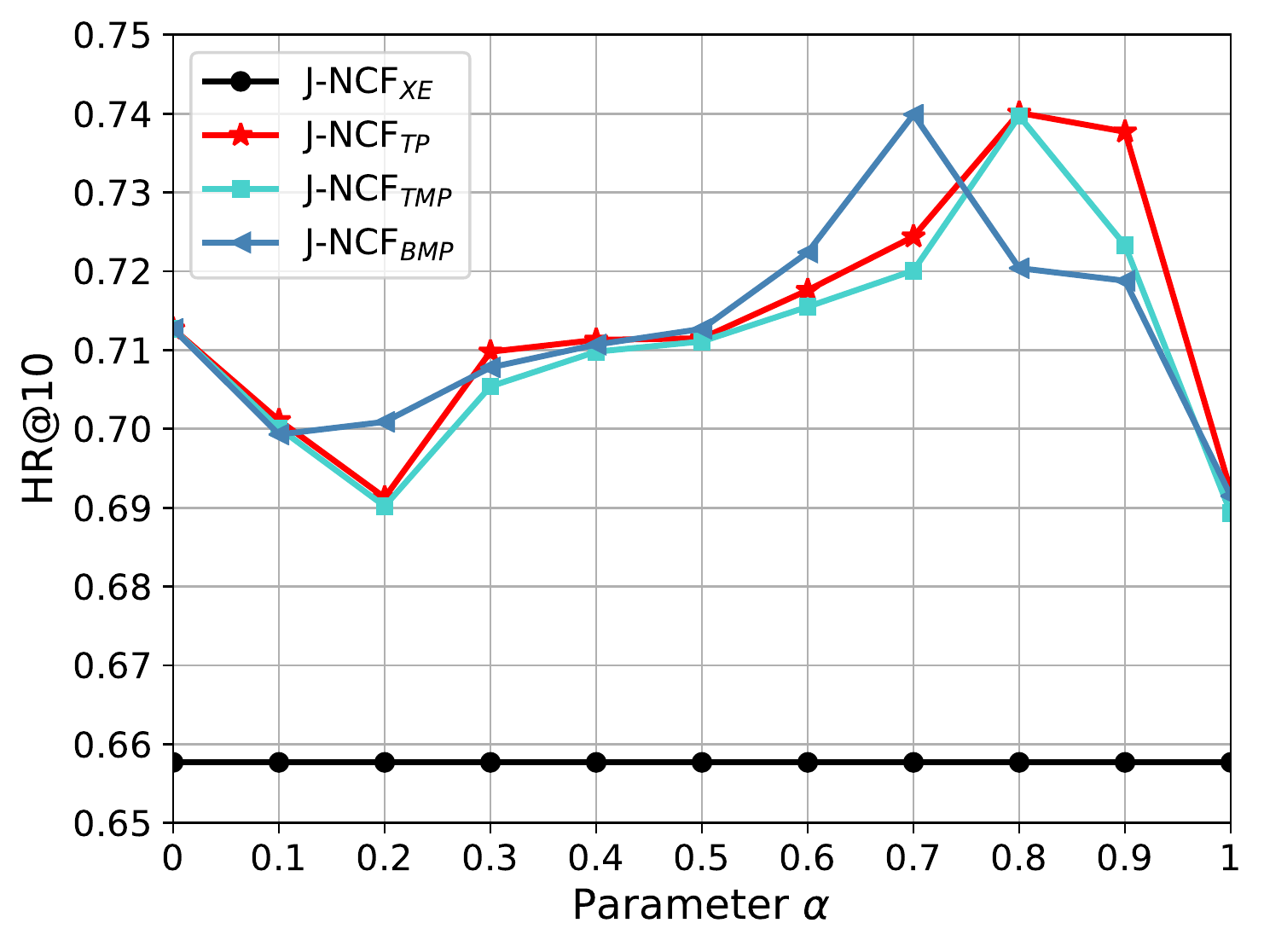}
                \caption{\raggedright Performance in terms of HR@10 on the ML1M dataset.}
              \label{diff-ML1M_hr}
        \end{subfigure}
        ~%space
        \begin{subfigure}[t]{0.49\textwidth}
                \includegraphics[width=\textwidth]{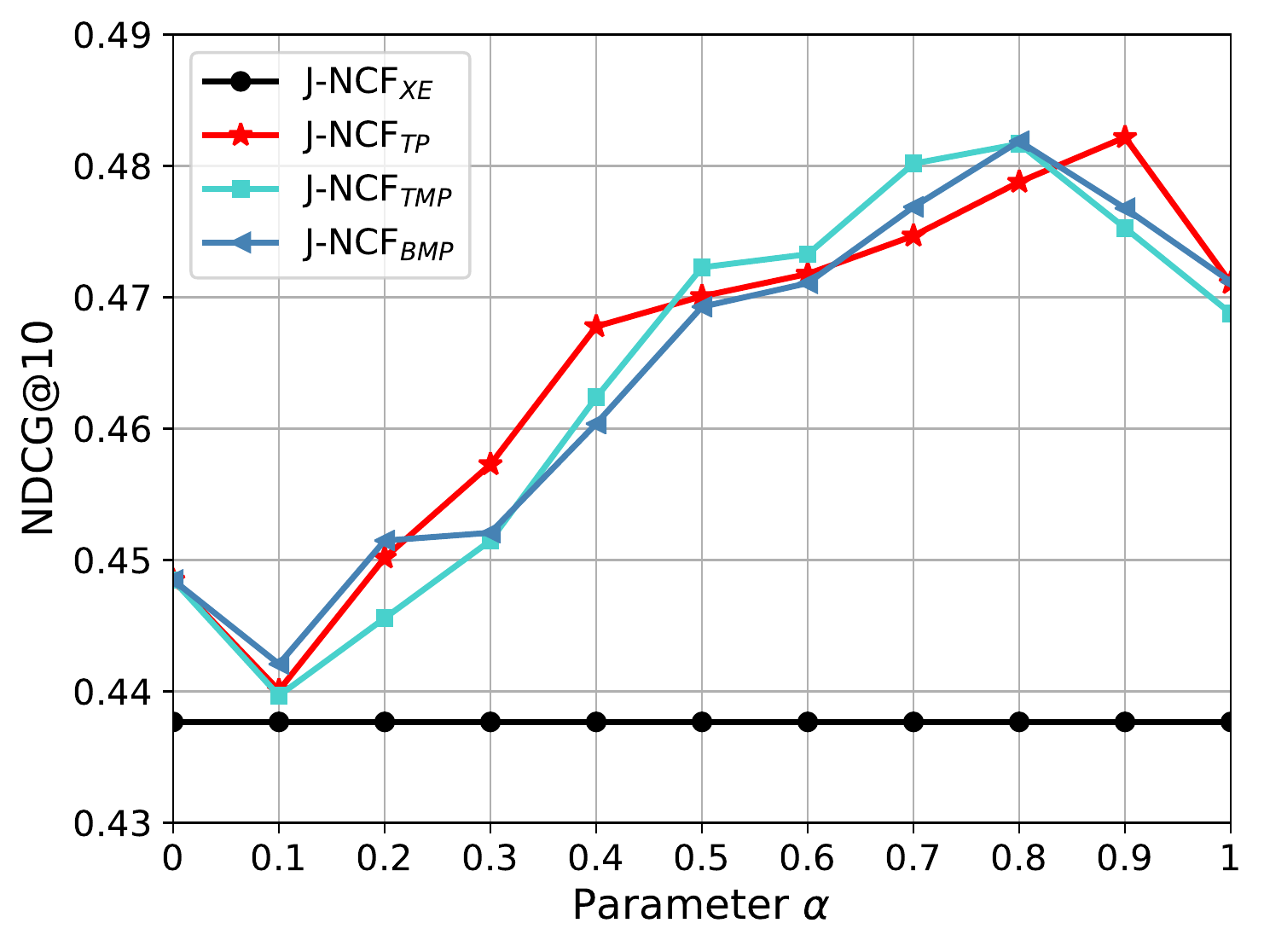}
                \caption{\raggedright Performance in terms of NDCG@10 on the ML1M dataset.}
                \label{diff-ML1M_ndcg}
        \end{subfigure}
        \begin{subfigure}[t]{0.49\textwidth}
                \includegraphics[width=\textwidth]{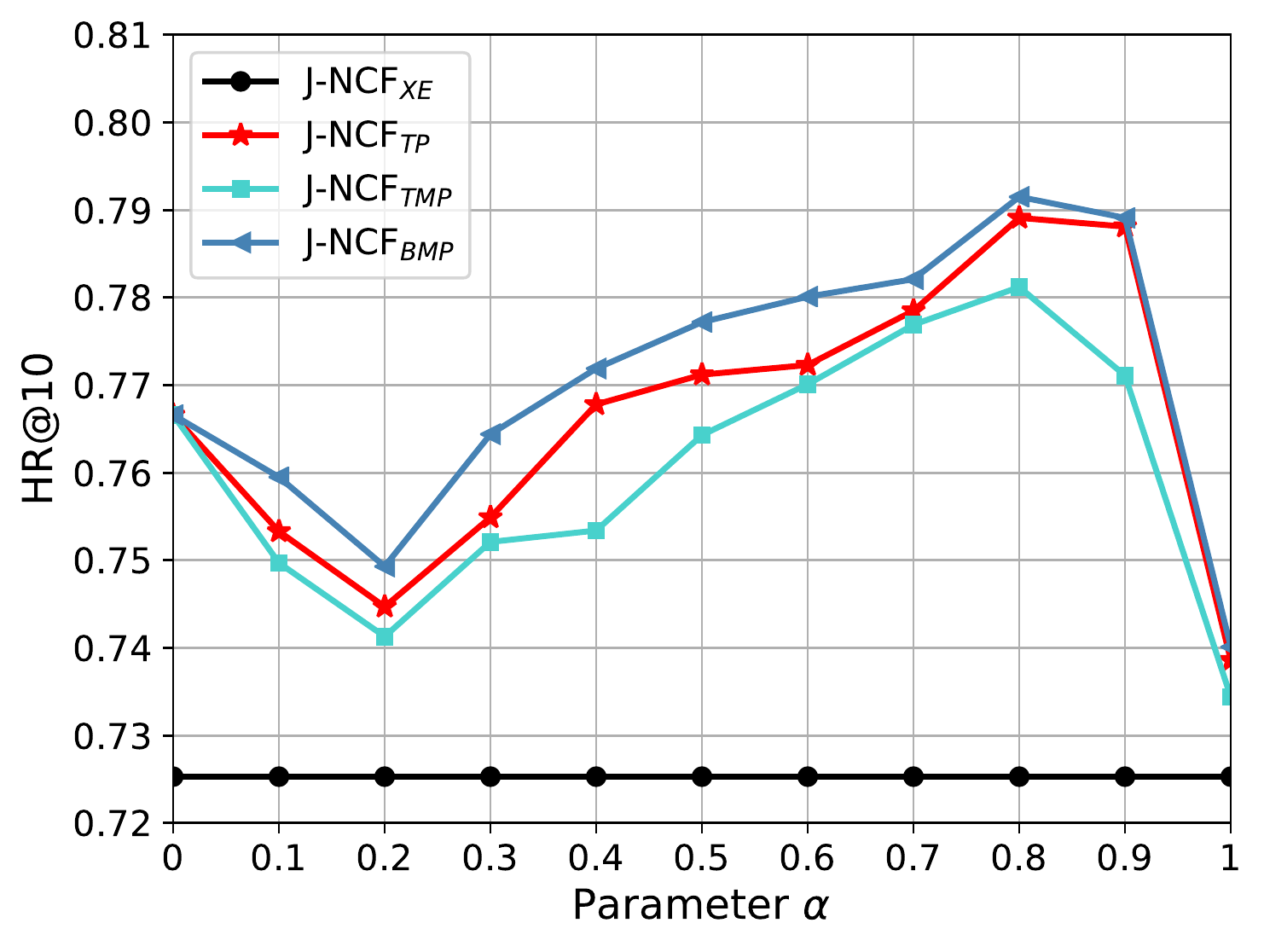}
                \caption{\raggedright Performance in terms of HR@10 on the AMovies dataset.}
                \label{diff-Amovies_hr}
        \end{subfigure}
        ~%space
        \begin{subfigure}[t]{0.49\textwidth}
                \includegraphics[width=\textwidth]{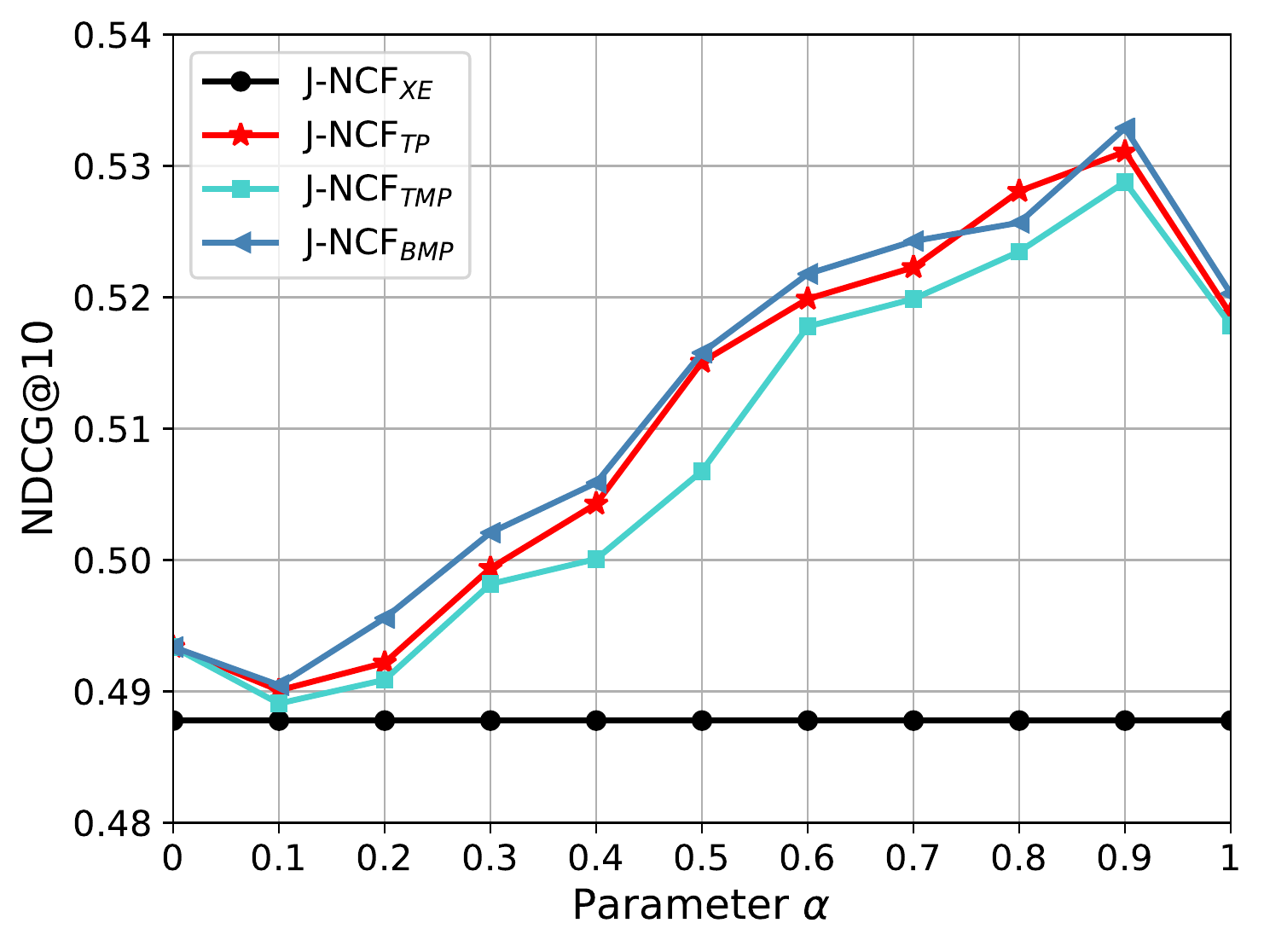}
                \caption{\raggedright Performance in terms of NDCG@10 on the AMovies dataset.}
                \label{diff-Amovies_ndcg}
        \end{subfigure}
        \smallskip
        \caption{Performance of the \ac{J-NCF} models applied with different loss functions where the parameter $\alpha$ in Eq.~\eqref{equation13} ranges from $0$ to $1$ with a step size of $0.1$.}
\label{diff-lossfunction} 
\end{figure*}

\chen{As for the overall performance, we can see that when applied with a list-wise loss function, J-NCF$_{\mathit{XE}}$ has the worst performance among the four models. 
The other three models, which combine pair-wise and point-wise losses, show relatively similar results in terms of HR@10 and NDCG@10. 
When $\alpha=0$, it results in J-NCF$_{\mathit{point}}$. When $\alpha=1$, it leads to \ac{J-NCF}, a model with only corresponding pair-wise loss functions.
It is obvious that solely based on point-wise loss, \ac{J-NCF} has better performance in terms of HR@10 while worse performance regarding NDCG@10 than \ac{J-NCF} with only pair-wise loss. 
This can be explained by the fact that pair-wise loss can help \ac{J-NCF} learn to rank items in right positions.}

\chen{In Fig.~\ref{diff-ML100k_hr}, the performance of all models increases from $\alpha=0.2$ to $\alpha=0.7$ before a short-term decrease and then a dramatic drop after reaching the peak at $\alpha=0.7$.  
The performance of J-NCF$_{\mathit{TP}}$, J-NCF$_{\mathit{TMP}}$ and J-NCF$_{\mathit{BMP}}$ is comparable in terms of HR@10. 
As for NDCG@10, shown in Fig.~\ref{diff-ML100k_ndcg}, J-NCF$_{\mathit{TP}}$ shows better performance than the other two models and achieves the highest point at $\alpha=0.9$.}

\chen{Regarding the performance on the ML1M dataset, similar trends can be found in Fig.~\ref{diff-ML1M_hr} and Fig.~\ref{diff-ML1M_ndcg} as in Fig.~\ref{diff-ML100k_hr} and Fig.~\ref{diff-ML100k_ndcg}, respectively. 
For the AMovies dataset shown in Fig.~\ref{diff-Amovies_hr} and Fig.~\ref{diff-Amovies_ndcg}, J-NCF$_{\mathit{BMP}}$ shows slightly better performance than both J-NCF$_{\mathit{TP}}$ and J-NCF$_{\mathit{TMP}}$ in terms of HR@10, while the performance of J-NCF$_{\mathit{BMP}}$ and J-NCF$_{\mathit{TP}}$ is similar in terms of NDCG@10, which is a little better than that of J-NCF$_{\mathit{TMP}}$.}

\chen{As discussed in \citep{2018lossfunction}, the BPR-max and TOP1-max loss functions have been proposed to overcome vanishing gradients as the number of negative samples increases. 
Since we use a small number of negative samples in our paper, the performance is relatively similar between the three models, J-NCF$_{\mathit{TP}}$, J-NCF$_{\mathit{TMP}}$ and J-NCF$_{\mathit{BMP}}$. 
As BPR-max and TOP1-max losses need additional softmax calculations for all negative samples, we apply the TOP1 pair-wise loss in Eq.~\eqref{equation13} for \ac{J-NCF} in the experiments on which we report below.}

\subsection{Utility of hybrid loss function}
\label{subsectionlossfunction}
\begin{figure*}[!t]
        \begin{subfigure}[t]{0.49\textwidth}
                \includegraphics[width=\textwidth]{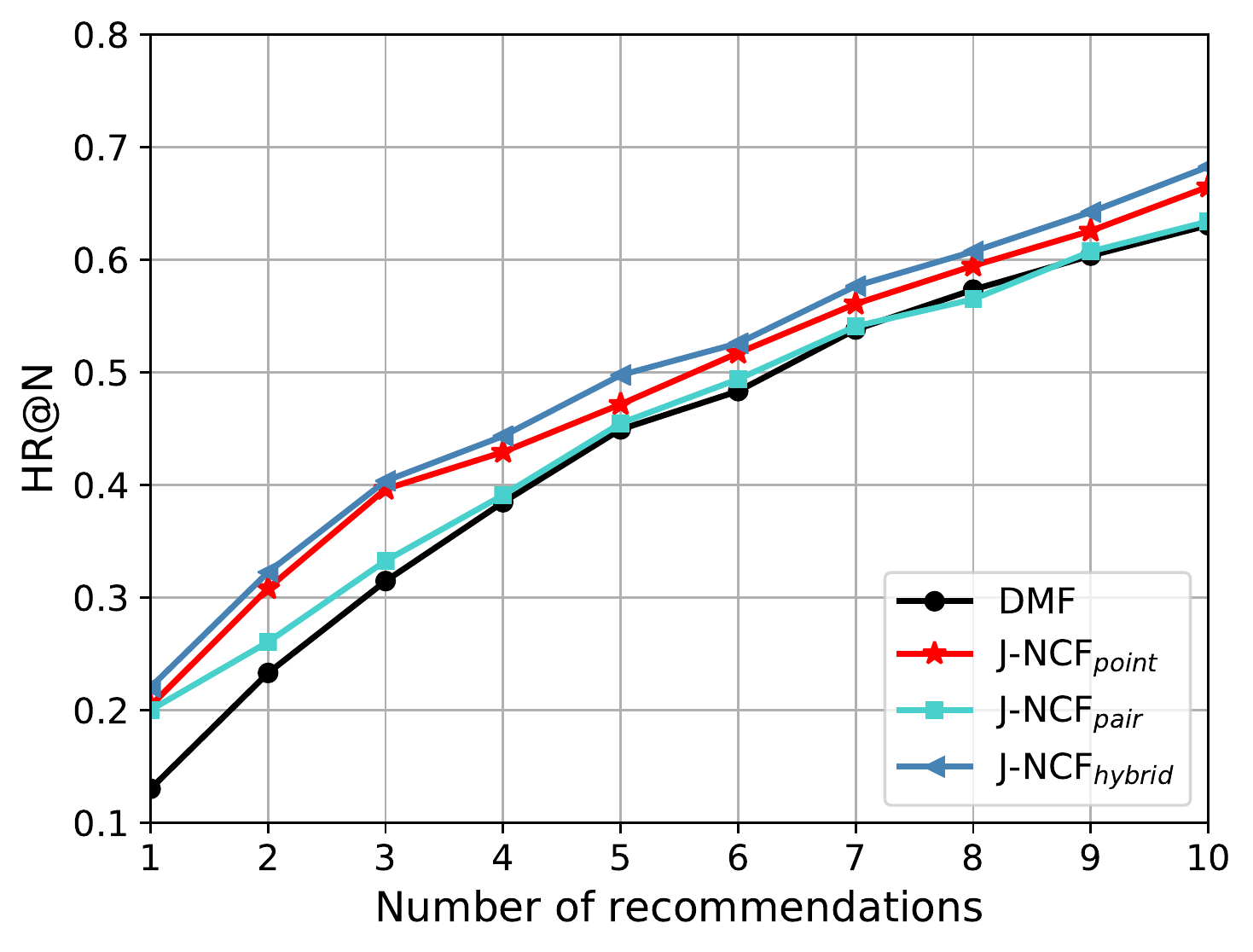}
                \caption{\raggedright Performance in terms of HR@N on the ML100K dataset.}
                \label{ML100k_hr}
        \end{subfigure}
        ~%space
        \begin{subfigure}[t]{0.5\textwidth}
                \includegraphics[width=\textwidth]{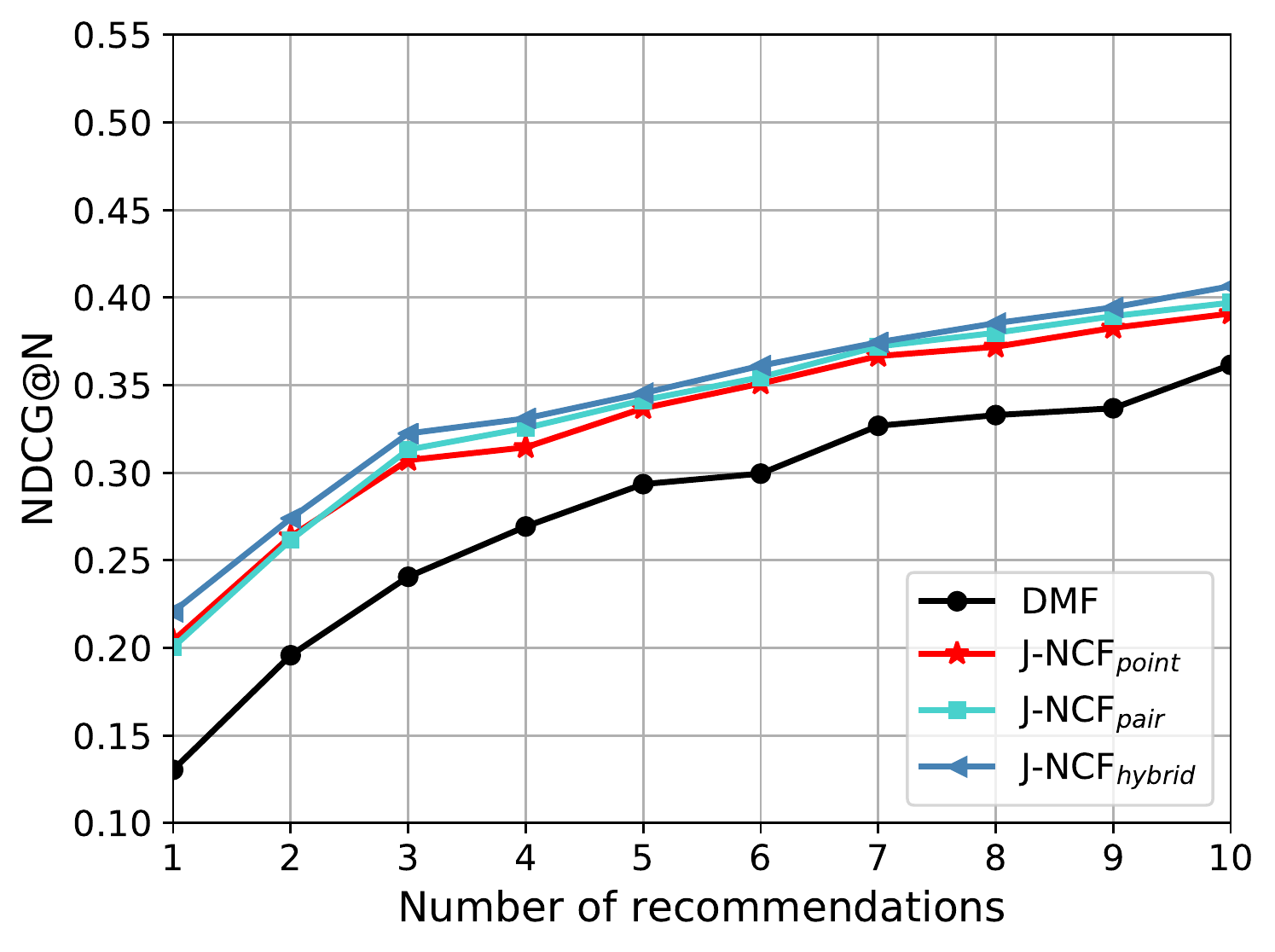}
                \caption{\raggedright Performance in terms of NDCG@N on the ML100K dataset.}
                \label{ML100k_ndcg}
        \end{subfigure}
        \begin{subfigure}[t]{0.49\textwidth}
                \includegraphics[width=\textwidth]{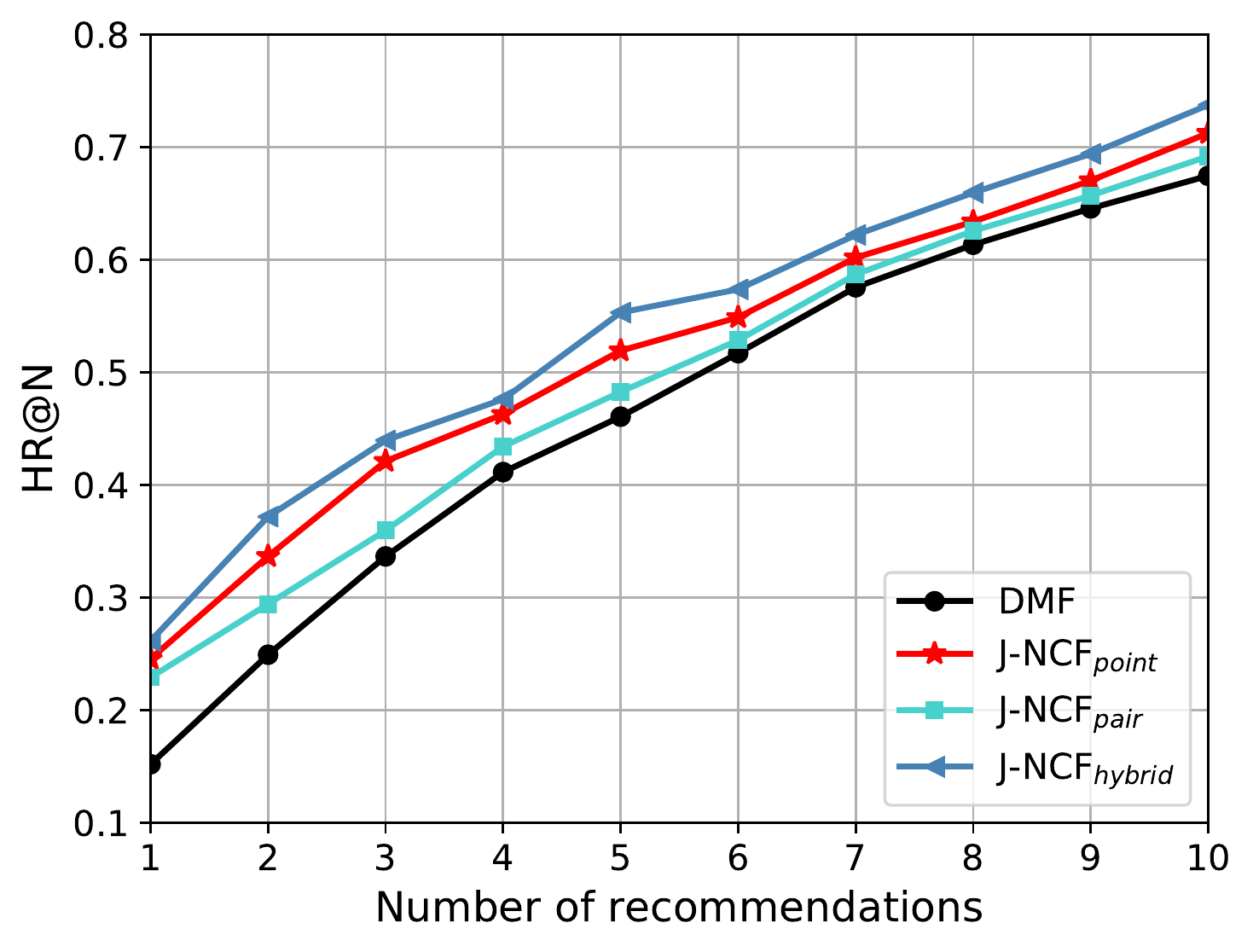}
                \caption{\raggedright Performance in terms of HR@N on the ML1M dataset.}
              \label{ML1M_hr}
        \end{subfigure}
        ~%space
        \begin{subfigure}[t]{0.5\textwidth}
                \includegraphics[width=\textwidth]{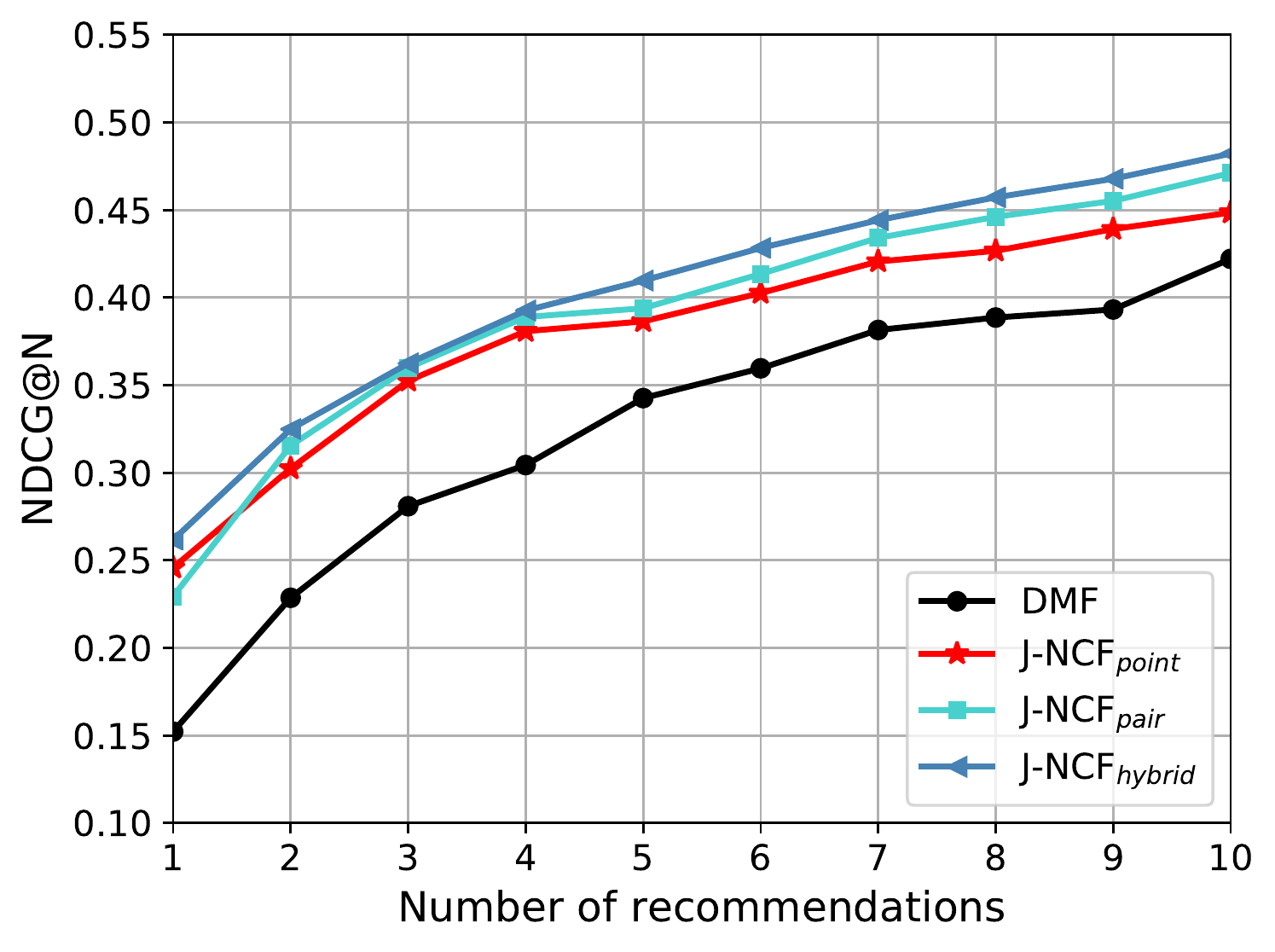}
                \caption{\raggedright Performance in terms of NDCG@N on the ML1M dataset.}
                \label{ML1M_ndcg}
        \end{subfigure}
        \begin{subfigure}[t]{0.49\textwidth}
                \includegraphics[width=\textwidth]{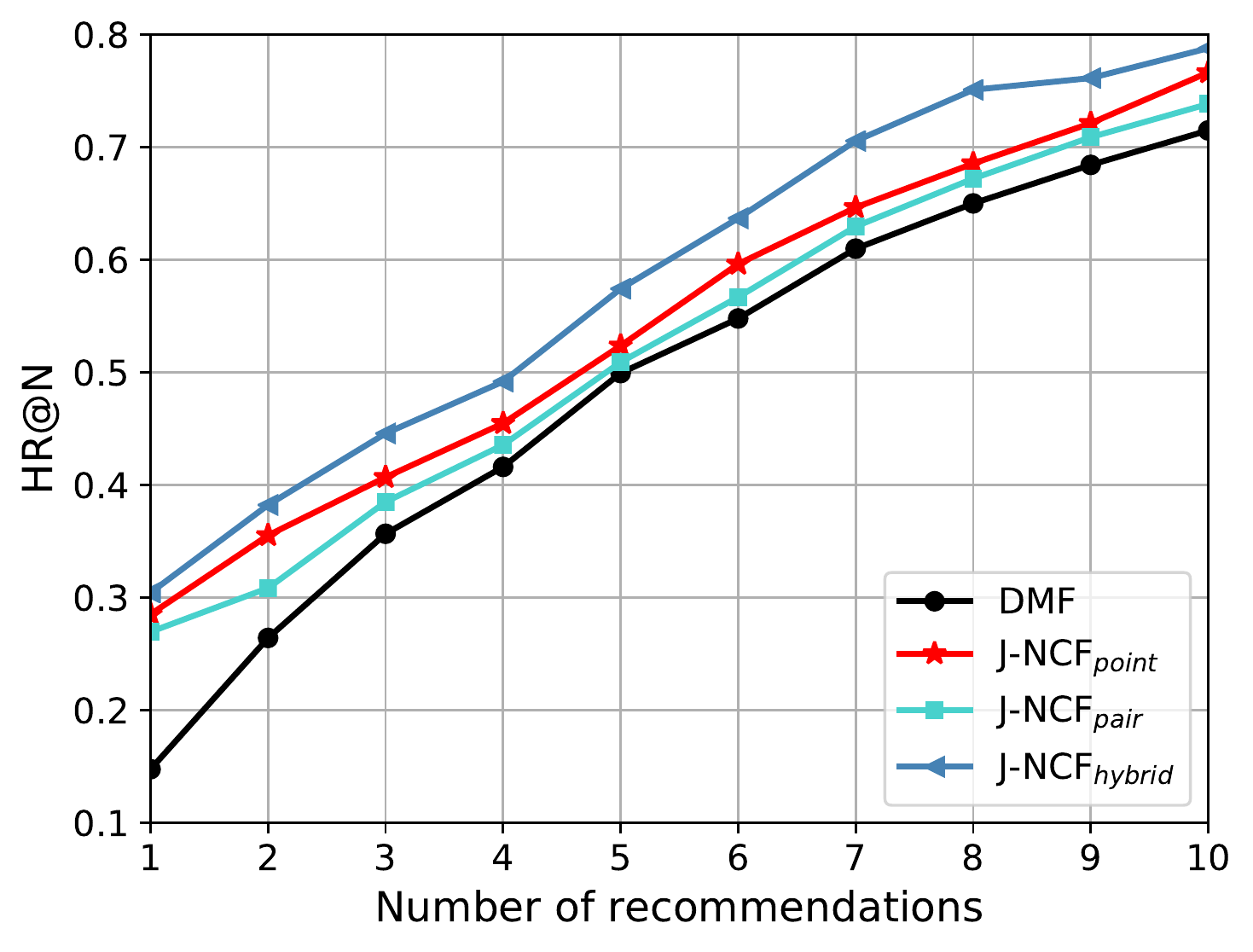}
                \caption{\raggedright Performance in terms of HR@N on the AMovies dataset.}
                \label{Amovies_hr}
        \end{subfigure}
        ~%space
        \begin{subfigure}[t]{0.5\textwidth}
                \includegraphics[width=\textwidth]{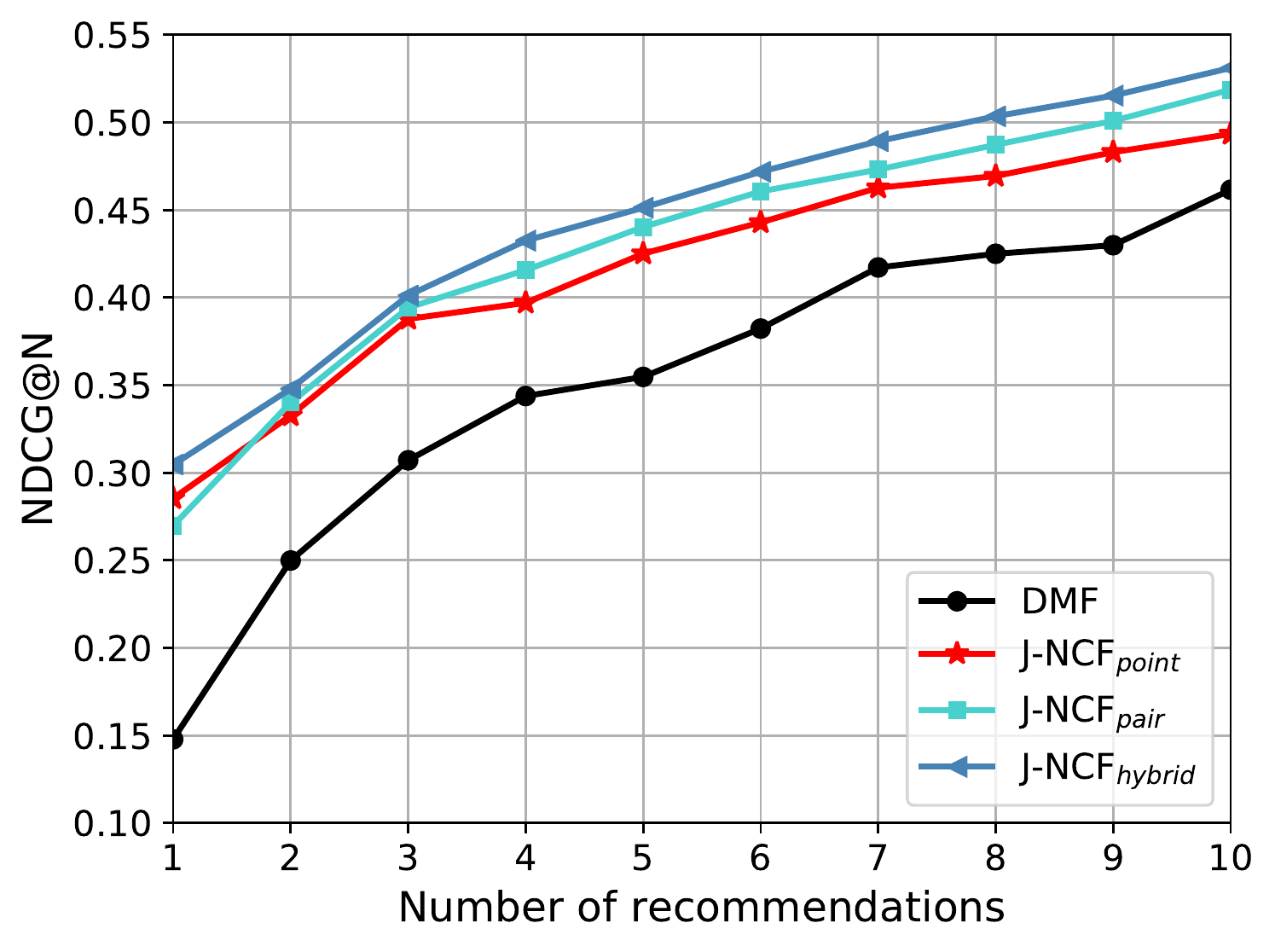}
                \caption{\raggedright Performance in terms of NDCG@N on the AMovies dataset.}
                \label{Amovies_ndcg}
        \end{subfigure}
        \smallskip
        \caption{Performance of Top-N item recommendation where N ranges from 1 to 10. The left and right plots show the  performance in terms of HR@N and NDCG@N, respectively.}
\label{lossfunction}
\end{figure*}

\chen{For \textbf{RQ3}, in order to further investigate the utility of the hybrid loss function (Eq.~\eqref{equation13}), we examine the recommendation performance of the \ac{J-NCF}$_\mathit{c}$ models under different settings, i.e., \ac{J-NCF}$_{\mathit{point}}$ with only point-wise loss based on Eq.~\eqref{equation8} (we incorporate explicit feedback in the same way as Eq.~\eqref{equation14}), \ac{J-NCF}$_{\mathit{pair}}$ with only pair-wise loss based on Eq.~\eqref{equation9}, and \ac{J-NCF}$_{\mathit{hybrid}}$ with our designed loss function from Eq.~\eqref{equation14}. Fig.~\ref{lossfunction} shows the results.}

The overall performance in terms of HR and NDCG increases when the size of the top-N recommended list ranges from 1 to 10, as a large value of $N$ increases the probability of including a user's preferred item in the recommendation list.
\ac{J-NCF}$_{\mathit{hybrid}}$ consistently achieves improvements over \ac{DMF} as well as the two models with a single loss function across positions, which demonstrates the utility of our newly designed loss function. 
Based on the ML100K dataset, \ac{J-NCF}$_{\mathit{hybrid}}$ improves by 2.68\% and 7.61\%, respectively, over \ac{J-NCF}$_{\mathit{point}}$ and \ac{J-NCF}$_{\mathit{pair}}$ in terms of HR@10; improvements of NDCG@10 over \ac{J-NCF}$_{\mathit{point}}$ and \ac{J-NCF}$_{\mathit{pair}}$ are 3.99\% and 2.36\%, respectively.

Comparing \ac{J-NCF}$_{\mathit{point}}$ and \ac{J-NCF}$_{\mathit{pair}}$, we find that \ac{J-NCF}$_{\mathit{point}}$ beats \ac{J-NCF}$_{\mathit{pair}}$ in terms of HR, while \ac{J-NCF}$_{\mathit{pair}}$ shows more competitive performance in terms of NDCG than \ac{J-NCF}$_{\mathit{point}}$.
This confirms the findings in~\citep{BPR2009,EALS2016} that a pair-wise ranking-aware learner has a strong performance for ranking prediction. 
This finding motivates us to incorporate both point-wise loss and pair-wise loss into the hybrid loss function.
Clearly, \ac{J-NCF}$_\mathit{c}$ based models, i.e., \ac{J-NCF}$_{\mathit{point}}$, \ac{J-NCF}$_{\mathit{pair}}$ and \ac{J-NCF}$_{\mathit{hybrid}}$, show a better performance than \ac{DMF}, which also proves that the joint neural structure is effective, i.e., deep interaction modeling can optimize neural matrix factorization and thus improve the recommendation performance.

Comparing the left and right hand sides of Fig.~\ref{lossfunction}, we see that the improvements of \ac{J-NCF}$_{\mathit{hybrid}}$ in terms of NDCG are more significant than those in terms of HR, as indicated by the relative improvements over \ac{DMF} with different sizes of the recommendation list.
In Fig.~\ref{ML100k_hr}, \ac{J-NCF}$_{\mathit{hybrid}}$ shows a 8.78\% improvement over \ac{DMF} in terms of HR at cutoff $N=6$, a 5.91\% improvement at $N=8$ and a 8.24\% improvement at $N=10$ on the ML100K dataset.
In Fig.~\ref{ML100k_ndcg}, the improvements in terms of NDCG at cutoff $N=6$, $N=8$ and $N=10$ are 19.01\%, 15.72\% and 12.42\%, respectively.
\ac{J-NCF}$_\mathit{c}$ with the hybrid loss function cannot only recommend the correct item to a user, but is also competitive in terms of ranking it at the top of the list.

\subsection{Number of of layers in the networks}

In \ac{J-NCF}$_{\mathit{c}}$, we not only learn features of users and items through the DF neural network with multiple hidden layers, but also model user-item interactions with multi-layer perceptrons in the DI network.
Thus it is crucial to see whether \ac{DL} is helpful in our model.
We conduct experiments to examine the performance of \ac{J-NCF}$_\mathit{c}$ with various numbers of layers in the DF and DI networks, respectively. \chen{In addition, we also test the performance of the best baseline model, i.e., \ac{DMF}, with different DF networks.}
The results are shown in Table~\ref{layers}. The $i$ in DF-$i$ and DI-$i$ in Table~\ref{layers} denotes the number of layers in the DF network and DI network of \ac{J-NCF}$_\mathit{c}$, respectively.

\begin{table*}[t]
\captionsetup{justification=justified}
  \centering
  \small
  \caption{Performance of \OurModel{}  and \ac{DMF} with different numbers of layers in terms of HR@10 and NDCG@10. The results produced by the best performing setting on each dataset are boldfaced.}
\label{layers}
%\begin{tabular}{ccccccccccccc}
\begin{tabular}{m{3.2em}<{\raggedright} m{2em}<{\raggedright} m{2.3em}m{2.3em}m{2.3em}m{2.3em}m{2.3em}m{0em}m{2.3em}m{2.3em}m{2.3em}m{2.3em}m{2.3em}}
%{m{13em}<{\centering} m{5em}<{\centering} m{5em}<{\centering}}\hline%{|p{8em}|c|c|c|c|}%{m{7cm}<{\centering}
\toprule
&&\multicolumn{5}{c}{HR@10} && \multicolumn{5}{c}{NDCG@10}\\
\cmidrule{3-7}\cmidrule{9-13}
& &DF-1 & DF-2 & DF-3 & DF-4& DF-5 && DF-1 & DF-2 & DF-3 & DF-4& DF-5 \\
\midrule
\multirow{5}{0.5cm}{ML100K}& DI-1 & .6242\duwen & .6511\duwen & .6713\duwen & .6955\duwen& .7213\duwen &&.3581\duwen & .3721\duwen & .3971\duwen & .4123\duwen& .4313\duwen\\
&DI-2 & .6351\duwen & .6642\duwen & .6829\duwen & .7183\duwen& .7388\duwen && .3694\duwen & .3899\duwen & .4067\duwen  & .4277\duwen& .4426\duwen \\
&DI-3 & .6493\duwen & .6712\duwen & .7144\duwen & .7309\duwen& .7479\duwen && .3811\duwen & .4001\duwen & .4197\duwen & .4388\duwen& .4535\duwen \\
&DI-4 & .6571\duwen & .6832\duwen & .7277\duwen & .7411\duwen& \bf{.7523}\duwen && .3945\duwen & .4183\duwen & .4311\duwen & .4481\duwen& \bf{.4618}\duwen \\
&DI-5 & .6501\duwen & .6799\duwen & .7254\duwen & .7408\duwen& .7501\duwen && .3903\duwen & .4111\duwen & .4287\duwen & .4433\duwen& .4587\duwen \\
&DMF & .6285\duwen & .6309\duwen & .6301\duwen & .6297\duwen& .6298\duwen && .3598\duwen & .3616\duwen & .3614\duwen & .3607\duwen& .3598\duwen \\
\midrule
\multirow{5}{0.5cm}{ML1M}& DI-1 &.6451\duwen & .6671\duwen & .7121\duwen & .7389\duwen& .7619\duwen&&.3622\duwen & .3911\duwen & .4399\duwen & .4893\duwen& .5301\duwen \\
&DI-2 & .6531\duwen & .6999\duwen & .7377\duwen & .7531\duwen& .7814\duwen&& .3889\duwen & .4233\duwen & .4822\duwen & .5211\duwen& .5525\duwen\\
&DI-3 & .6766\duwen & .7198\duwen & .7589\duwen & .7728\duwen& .7929\duwen && .4195\duwen & .4601\duwen & .5177\duwen & .5437\duwen& .5777\duwen \\
&DI-4 & .7134\duwen & .7472\duwen & .7683\duwen & .7834\duwen& \bf{.8088}\duwen && .4581\duwen & .5101\duwen & .5389\duwen & .5663\duwen& \bf{.5906}\duwen \\
&DI-5 & .7099\duwen & .7411\duwen & .7653\duwen & .7821\duwen& .8021\duwen && .4517\duwen & .5078\duwen & .5333\duwen & .5644\duwen& .5878\duwen \\
&DMF & .6673\duwen & .6748\duwen & .6738\duwen & .6722\duwen& .6725\duwen && .3955\duwen & .4221\duwen & .4201\duwen & .4197\duwen& .4199\duwen \\
\midrule
\multirow{5}{0.5cm}{AMovies}& DI-1 & .6611\duwen & .6922\duwen & .7481\duwen & .7911\duwen& .8188\duwen&& .4041\duwen & .4533\duwen & .5004\duwen & .5413\duwen& .5622\duwen \\
&DI-2 & .6872\duwen & .7378\duwen & .7881\duwen & .8101\duwen& .8411\duwen&& .4327\duwen & .4911\duwen & .5311\duwen & .5597\duwen& .5803\duwen\\
&DI-3 & .6989\duwen & .7633\duwen & .8078\duwen& .8378\duwen& .8787\duwen&& .4632\duwen & .5204\duwen & .5501\duwen & .5714\duwen& .6102\duwen\\
&DI-4 & .7414\duwen & .7999\duwen & .8293\duwen& .8612\duwen& \bf{.8893}\duwen&&.5137\duwen & .5461\duwen & .5644\duwen& .5966\duwen& \bf{.6198}\duwen\\
&DI-5 & .7379\duwen & .7922\duwen & .8201\duwen& .8589\duwen& .8821\duwen&&.5111\duwen & .5402\duwen & .5599\duwen& .5934\duwen& .6145\duwen\\
&DMF & .7478\duwen & .7515\duwen & .7491\duwen & .7483\duwen& .7479\duwen && .4551\duwen & .4616\duwen & .4612\duwen & .4603\duwen& .4591\duwen \\
\bottomrule
\end{tabular}
\end{table*}

As shown in Table~\ref{layers}, in terms of HR@10, we can see that with the number of layers increasing, the recommendation performance of J-NCF is improved, which verifies the effectiveness of \ac{DL} techniques for recommender systems. 

%In particular, unlike \ac{DMF}, where adding deeper layers more than 2 seems unuseful (see~\citep[Section 4.4]{DMF2017}), \OurModel{} achieves further improvements when stacking more layers in either the DI or DF network, or both.

Comparing the number of layers in the DI and DF networks, we can find that stacking more layers in the DF network of \OurModel{} seems more helpful than in the DI network in enhancing the recommendation performance. 
For example, based on the ML100K dataset, the improvements of the configuration (DF-3, DI-2) over (DF-2, DI-2) are 2.82\% and 4.31\% in terms of HR@10 and NDCG@10, while the improvements are 1.05\% and 2.62\% for (DF-2, DI-3) over (DF-2, DI-2).
When we stack more than 4 layers in the DI network (e.g., DI-5), the performance of \OurModel{} no longer increases.
However, stacking more layers in the DF network (e.g., DF-5) still seems helpful and the best results produced for each dataset are all based on \OurModel{} with the (DF-5, DI-4) configuration. This may be because deep layers are more helpful in extracting users' as well as items' features and thus enhancing the user-item interactions predictions.
It motivates us to incorporate more auxiliary information for exploring users' and items' features with deep learning techniques in future work.

As for NDCG@10, a similar phenomenon can be found. However, when comparing the scores of HR@10 and NDCG@10 under the same configurations, we can find that deeper layers can lead to more obvious improvements in terms of NDCG@10 than HR@10 on all of the three datasets. The best performance of \ac{J-NCF} with (DF-5, DI-4) outperforms the worst performance of \ac{J-NCF} with (DF-1, DI-1) by 20.52\%, 25.37\% and 34.52\% in terms of HR@10 on the three datasets, respectively. However, the improvements are 28.96\%, 63.05\% and 53.37\% in terms of NDCG@10 on the three datasets.

\chen{As for the baseline model \ac{DMF} shown in the bottom rows in Table~\ref{layers}, when applied with DF-1, \OurModel{} with DI-1 loses to \ac{DMF} on all datasets. Similar results can be found with DF-2, except on ML100K dataset. This can be explained by the fact that the simple concatenation of user's and item's embeddings with only one MLP layer in \OurModel{}  is not efficient for modeling user-item interactions. When applied with more DI layers, \OurModel{} has better performance than \ac{DMF} with the same number of  DF layers. Additionally, we can find that \ac{DMF} achieves the best performance with DF-2 and deeper layers do not seem useful for \ac{DMF} model, which corresponds to the results in~\citep{DMF2017}. However, \OurModel{} achieves further improvements when stacking more layers in either the DI or DF network, or both.}

\subsection{Impact of feedback}
\label{feedback}
In \ac{J-NCF}, we consider different kinds of  user feedback.
On the one hand, we use the interaction matrix as the input of the network with Eq.~\eqref{equation1}, which contains not only implicit feedback but also explicit feedback. On the other hand, our loss function in Eq.~\eqref{equation14} employs a normalized strategy in the form of $Y_{ui} =\frac{y_{ui}}{\mathit{Max}(R_u)}$, where $\mathit{Max}(R_u)$ denotes the largest rating score of user $u$ given to items, to incorporate the explicit feedback.
In order to answer \textbf{RQ5}, we conduct experiments to investigate whether the combination of explicit and implicit feedback works for \ac{J-NCF} with different settings, i.e., \ac{J-NCF}$_{\mathit{ex}}$ with both kinds of feedback in the input and the loss function as well as \ac{J-NCF}$_{\mathit{im}}$ with only implicit feedback by labeling 1 for the interactions and 0 for unknown ratings in the input and the loss function.
Fig.~\ref{feedback} shows the recommendation performance of \ac{J-NCF}$_{\mathit{ex}}$, \ac{J-NCF}$_{\mathit{im}}$, \ac{DMF} and \ac{NCF} across different numbers of training iterations, respectively.
\begin{figure*}[!t]
        \begin{subfigure}[t]{0.49\textwidth}
                \includegraphics[width=\textwidth]{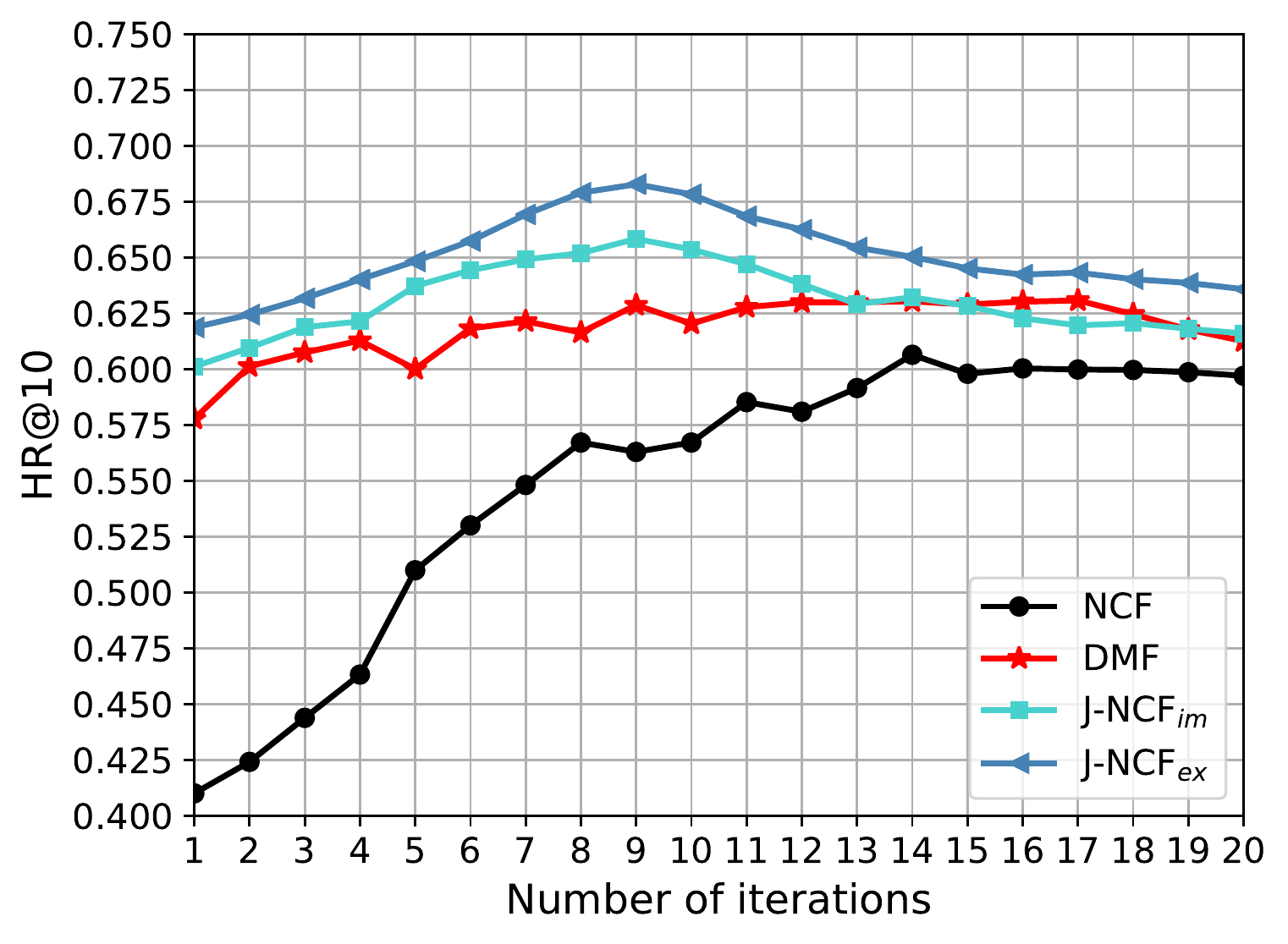}
                \caption{\raggedright Performance in terms of HR@10 on the ML100K datasets.}
                \label{ML100k_feedback_hr}
        \end{subfigure}
        ~%space
        \begin{subfigure}[t]{0.49\textwidth}
                \includegraphics[width=\textwidth]{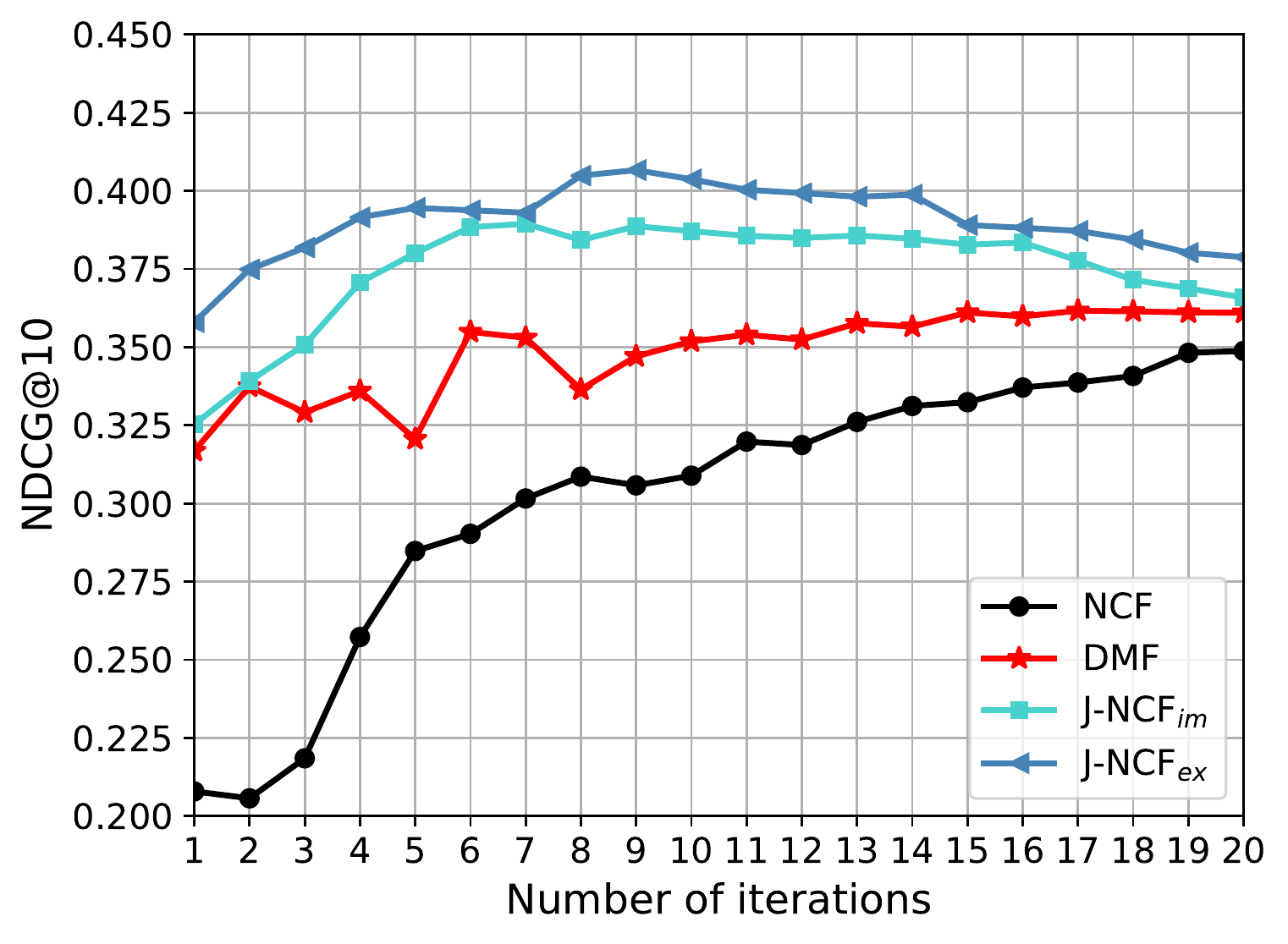}
                \caption{\raggedright Performance in terms of NDCG@10 on the ML100K datasets.}
                \label{ML100k_feedback_ndcg}
        \end{subfigure}
        \begin{subfigure}[t]{0.49\textwidth}
                \includegraphics[width=\textwidth]{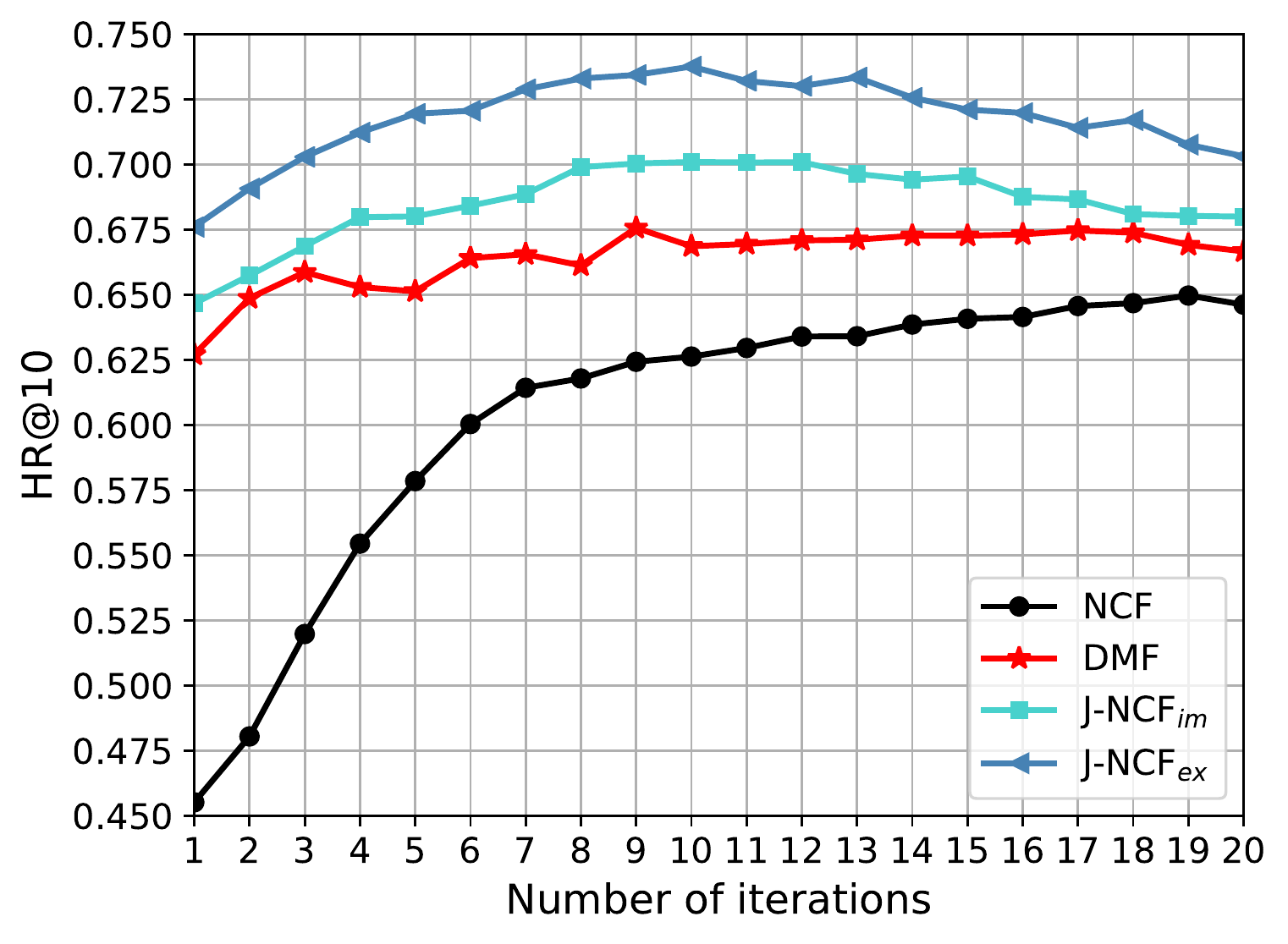}
                \caption{\raggedright Performance in terms of HR@10 on the ML1M datasets.}
                \label{ML1M_feedback_hr}
        \end{subfigure}
        ~%space
        \begin{subfigure}[t]{0.49\textwidth}
                \includegraphics[width=\textwidth]{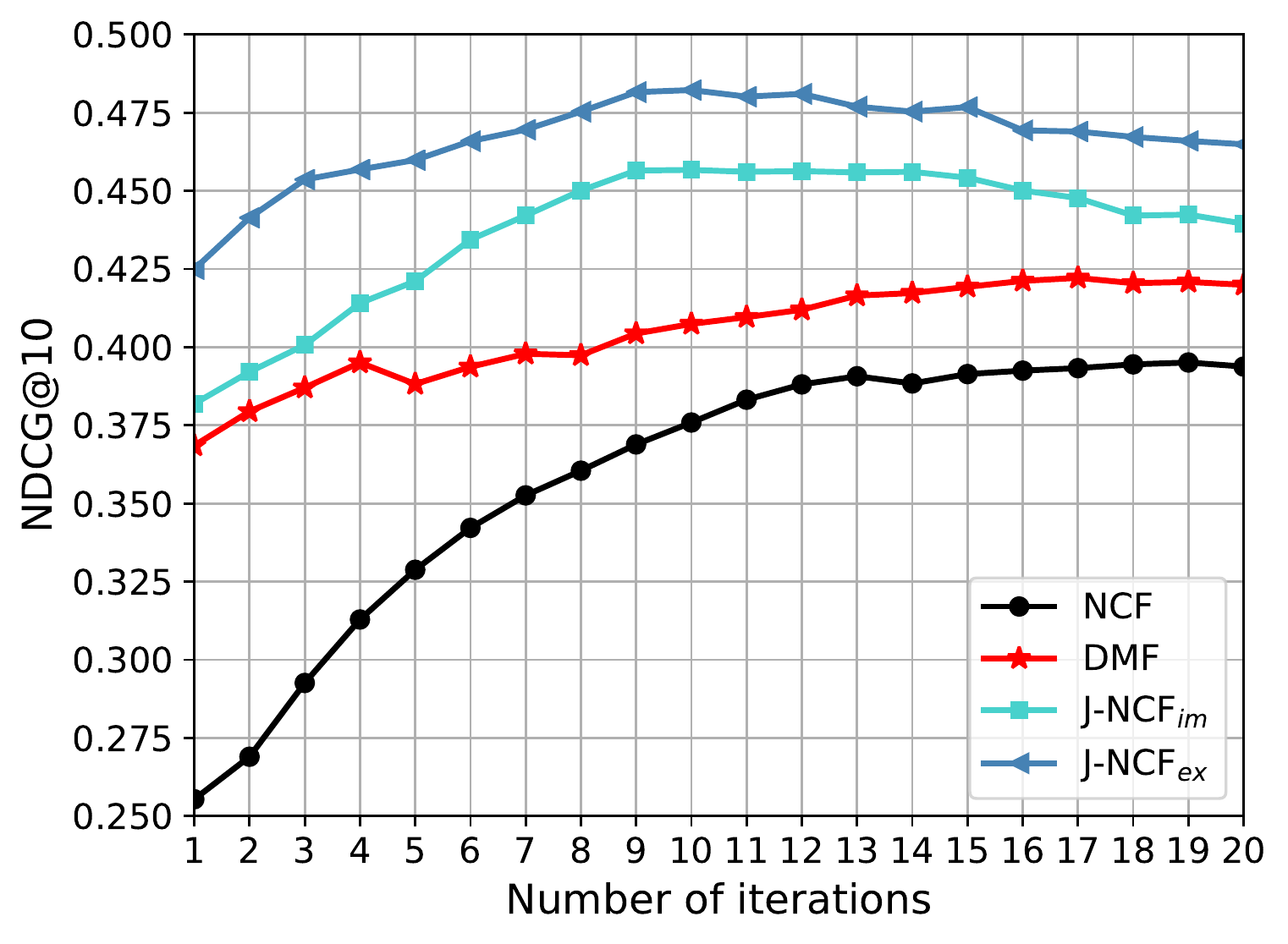}
                \caption{\raggedright Performance in terms of NDCG@10 on the ML1M datasets.}
                \label{ML1M_feedback_ndcg}
        \end{subfigure}
        \begin{subfigure}[t]{0.49\textwidth}
                \includegraphics[width=\textwidth]{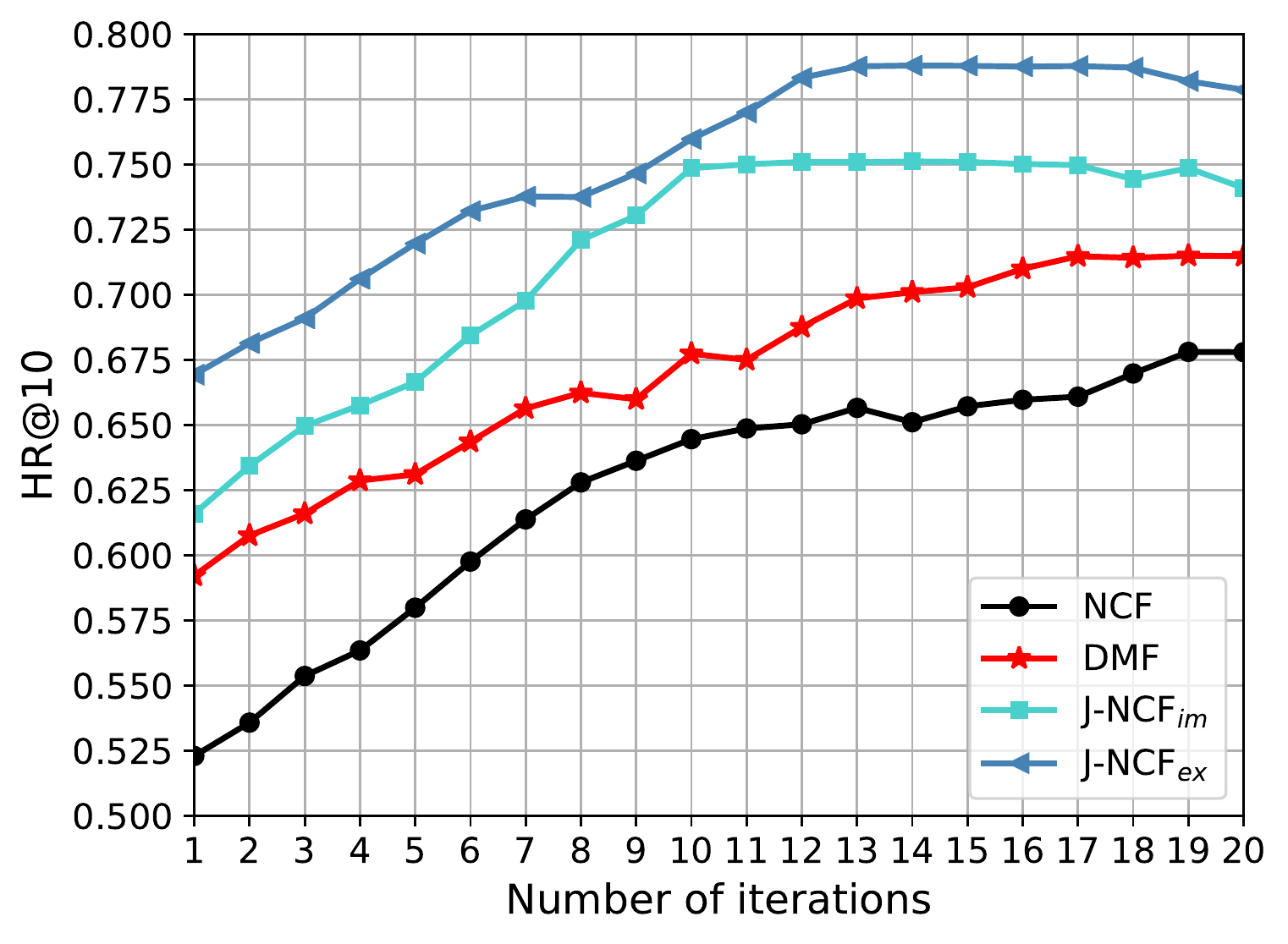}
                \caption{\raggedright Performance in terms of HR@10 on the AMovies datasets.}
                \label{Amovies_feedback_hr}
        \end{subfigure}
        ~%space
        \begin{subfigure}[t]{0.49\textwidth}
                \includegraphics[width=\textwidth]{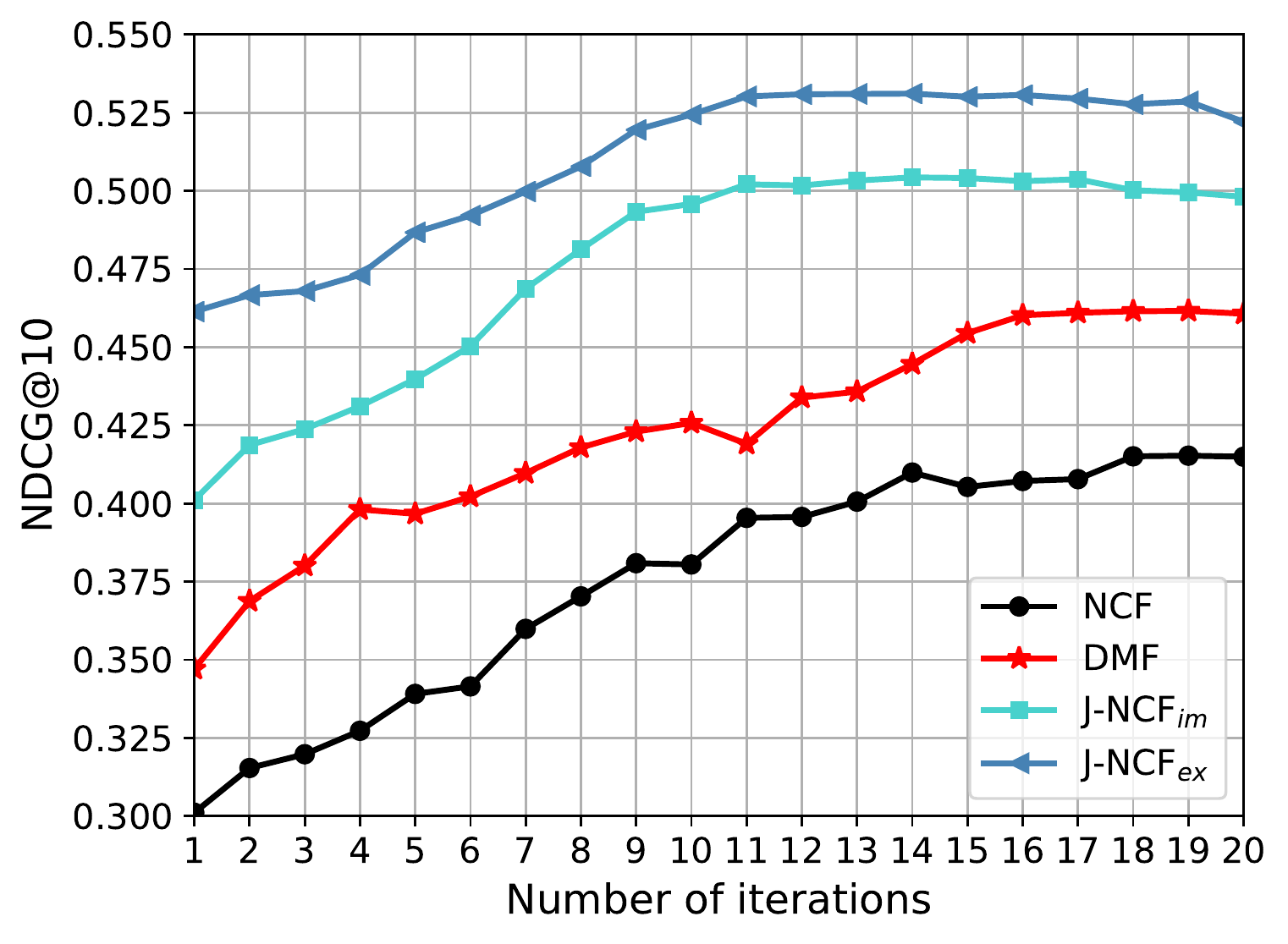}
                \caption{\raggedright Performance in terms of NDCG@10 on the AMovies datasets.}
                \label{Amovies_feedback_ndcg}
        \end{subfigure}
        \smallskip
                \caption{Recommendation performance across different numbers of iterations. The left and right plots show the performance in terms of HR@10 and NDCG@10, respectively.}
\label{feedback}
\end{figure*}

First, from Fig.~\ref{feedback} we can see that \ac{J-NCF}$_{\mathit{ex}}$ with both kinds of feedback achieves a competitive performance across all iterations in terms of HR@10 and NDCG@10 on the three datasets. 
It indicates that the combination of explicit and implicit feedback in the input and the specially designed loss function of \ac{J-NCF} does help to improve the recommendation performance.
Second, as the number of training iterations increases, the recommendation performance of all models is improved and then degraded after reaching a peak.
More iterations may lead to overfitting, which hurts the recommendation performance.
However, comparing \ac{J-NCF} model with the baselines, i.e., \ac{DMF} and \ac{NCF}, we find that \ac{J-NCF} converges to the best performance faster than other models.
For example, on the ML100K dataset, the best result of \ac{J-NCF} is generated after the first 9 effective iterations, while \ac{DMF} and \ac{NCF} need more training iterations to obtain the best results, i.e., 16 and 14 iterations respectively. 
The same phenomenon can be observed on the other two datasets. 
The optimal number of updates needed for \ac{J-NCF}, \ac{DMF} and \ac{NCF} are around 10, 17 and 19 on the ML1M dataset, and 14, 18 and 19 on the AMovies dataset, respectively.
Third, comparing the performance in terms of HR@10 and NDCG@10, we find that \ac{J-NCF}$_{\mathit{ex}}$ shows larger improvements over \ac{J-NCF}$_{\mathit{im}}$ in terms of NDCG@10 than HR@10. For example, the improvements are 3.72\%, 5.22\% and 4.89\% in terms of HR@10, on the ML100K, ML1M and AMovies datasets, respectively, vs.\ improvements of 4.61\%, 5.58\% and 5.31\% in terms of NDCG@10.
This confirms our hypothesis that incorporating both explicit and implicit feedback can improve the ranking precision for recommendation.

\section{Scalability and Sensitivity}
\label{application}

In order to answer \textbf{RQ6} to \textbf{RQ9}, we study the scalability and sensitivity of \ac{J-NCF} as well as the best baseline \ac{DMF} when applied in different settings, i.e., with users with various numbers of ratings in Section~\ref{scalability}, and with datasets with different levels of sparsity in Section~\ref{sensitivity}. 
\chen{In addition, we also investigate the performance of the deep learning-based approaches, i.e., \ac{J-NCF}, \ac{DMF} and {NCF}, when applied with a large and sparse dataset in Section~\ref{largedataset}. 
Moreover, the training and inference time needed for these models on all datasets is discussed in Section~\ref{trainingtime}.}

\subsection{Model scalability with user ratings}
\label{scalability}
\begin{table*}[t]
\captionsetup{justification=justified}
  \centering
  \caption{Recommendation performance across users who are ranked by the number of activities. The results produced by the best performing recommender system in each row are boldfaced. Statistical significance of pairwise differences of \ac{J-NCF}$_{\mathit{m}}$ and \ac{J-NCF}$_{\mathit{c}}$ vs.\ \ac{DMF} is determined by a $t$-test ($^\blacktriangle$/$^\blacktriangledown$ for $\alpha$ = .01, or $^\vartriangle$/$^\triangledown$ for $\alpha$ = .05).}
\label{user_ratings}
\begin{tabular}{ccccccccc}
\toprule
&&\multicolumn{3}{c}{HR@10} && \multicolumn{3}{c}{NDCG@10}\\
\cmidrule{3-5}\cmidrule{7-9}
&&DMF & J-NCF$_{\mathit{m}}$ & J-NCF$_{\mathit{c}}$ &&DMF & J-NCF$_{\mathit{m}}$ & J-NCF$_{\mathit{c}}$ \\
\midrule
\multirow{3}{1.5cm}{ML100K}& 10\% & .7001\duwen & .7400$^\blacktriangle$ & \bf{.8015}$^\blacktriangle$ & &.4358\duwen & .4786$^\blacktriangle$ & \bf{.5001}$^\blacktriangle$ \\
&50\% & .6813\duwen & .7349$^\vartriangle$ & \bf{.7568}$^\blacktriangle$ && .4200\duwen & .4379$^\vartriangle$ & \bf{.4602}$^\blacktriangle$ \\
&90\% & .6279\duwen & .6585$^\vartriangle$ & \bf{.6772}$^\blacktriangle$ && .3813\duwen & .3897$^\vartriangle$ & \bf{.4092}$^\blacktriangle$ \\
\midrule
\multirow{3}{1.5cm}{ML1M}& 10\% & .7548\duwen & .7927$^\blacktriangle$ & \bf{.8511}$^\blacktriangle$ && .5111\duwen & .5417$^\blacktriangle$ & \bf{.5952}$^\blacktriangle$  \\
&50\% & .7211\duwen & .7532$^\blacktriangle$ & \bf{.7982}$^\blacktriangle$  && .4855\duwen & .5266$^\blacktriangle$ & \bf{.5587}$^\blacktriangle$ \\
&90\% & .6601\duwen & .6981$^\blacktriangle$ & \bf{.7277}$^\blacktriangle$ && .4217\duwen & .4432$^\blacktriangle$ & \bf{.4751}$^\blacktriangle$ \\
\midrule
\multirow{3}{1.5cm}{AMovies}& 10\% & .7851\duwen & .8611$^\blacktriangle$ & \bf{.9191}$^\blacktriangle$ && .5349\duwen & .5998$^\blacktriangle$ & \bf{.6611}$^\blacktriangle$ \\
&50\% & .7519\duwen & .7855$^\blacktriangle$ & \bf{.8411}$^\blacktriangle$  && .5033\duwen & .5466$^\blacktriangle$ & \bf{.5821}$^\blacktriangle$ \\
&90\% & .7013\duwen & .7411$^\blacktriangle$ & \bf{.7732}$^\blacktriangle$ && .4597\duwen & .5038$^\blacktriangle$ & \bf{.5301}$^\blacktriangle$ \\
\bottomrule
\end{tabular}
\end{table*}

In Fig.~\ref{distribution}, we have shown that in every dataset most users only have a few ratings, thus it is meaningful to investigate how the performance of \ac{J-NCF} and \ac{DMF} varies with different numbers of user ratings.
Following~\citep{PF2013}, we look at the performance for users of varying degrees of activity, measured by percentile.
For example, in Table~\ref{user_ratings}, we first rank the users according to their numbers of their activities. 10\% shows the mean performance across the bottom 10\% of users, who are least active; the 90\% mark shows the mean performance for all but the top 10\% most active users.

As shown in Table~\ref{user_ratings}, \ac{J-NCF}$_\mathit{c}$ outperforms the best baseline model \ac{DMF} for users across all activity levels, i.e., both the ``inactive'' users who constitute the majority, and the relatively few ``very active'' users who give more ratings. 
In addition, \ac{J-NCF}$_\mathit{c}$ always achieves the best performance in terms of HR@10 and NDCG@10.
In order to test the robustness of \ac{J-NCF} under different settings, i.e., \ac{J-NCF}$_{\mathit{c}}$  and \ac{J-NCF}$_{\mathit{m}}$, we conduct t-tests between the two versions of \ac{J-NCF} with \ac{DMF}, respectively.
Significant improvements against the baseline \ac{DMF} in terms of HR@10 and NDCG@10 are observed for both \ac{J-NCF}$_{\mathit{m}}$ and \ac{J-NCF}$_{\mathit{c}}$ at the $\alpha=.01$ level across all activity levels, except for \ac{J-NCF}$_{\mathit{m}}$ on the ML100K dataset with 50\% and 90\% users, for which we observe significant improvements at the $\alpha=.05$ level in terms of HR@10 and NDCG@10.

Specifically, \ac{J-NCF} shows larger improvements over the \ac{DMF} model for ``inactive'' users than for ``very active'' users.
For example, when incorporating users with more interactions, i.e., from 50\% to 90\%, the improvements change from 11.08\% to 7.85\% in terms of HR@10, and 9.57\% to 7.32\% in terms of NDCG@10 on the ML100K dataset.
This may be because the ``very active'' users have many interactions with the items that have few ratings and collaborative filtering lacks information for recommending items based only on the rating matrix.
This naturally suggest a line of future work in which one would extend \ac{J-NCF} with more auxiliary information, such as content information, to explore more accurate relationships between items.

\todo{To conclude and answer \textbf{RQ6}, the \ac{J-NCF} models can beat the best baseline model for users across all activity levels. \ac{J-NCF}$_\mathit{c}$ shows the best performance in all datasets. In addition, for ``inactive'' users, \ac{J-NCF} shows larger improvements over \ac{DMF} than for ``very active'' users.}

\subsection{Sensitivity to data sparsity}
\label{sensitivity}
\begin{figure*}[!ht]
        \begin{subfigure}[t]{0.49\textwidth}
                \includegraphics[width=\textwidth]{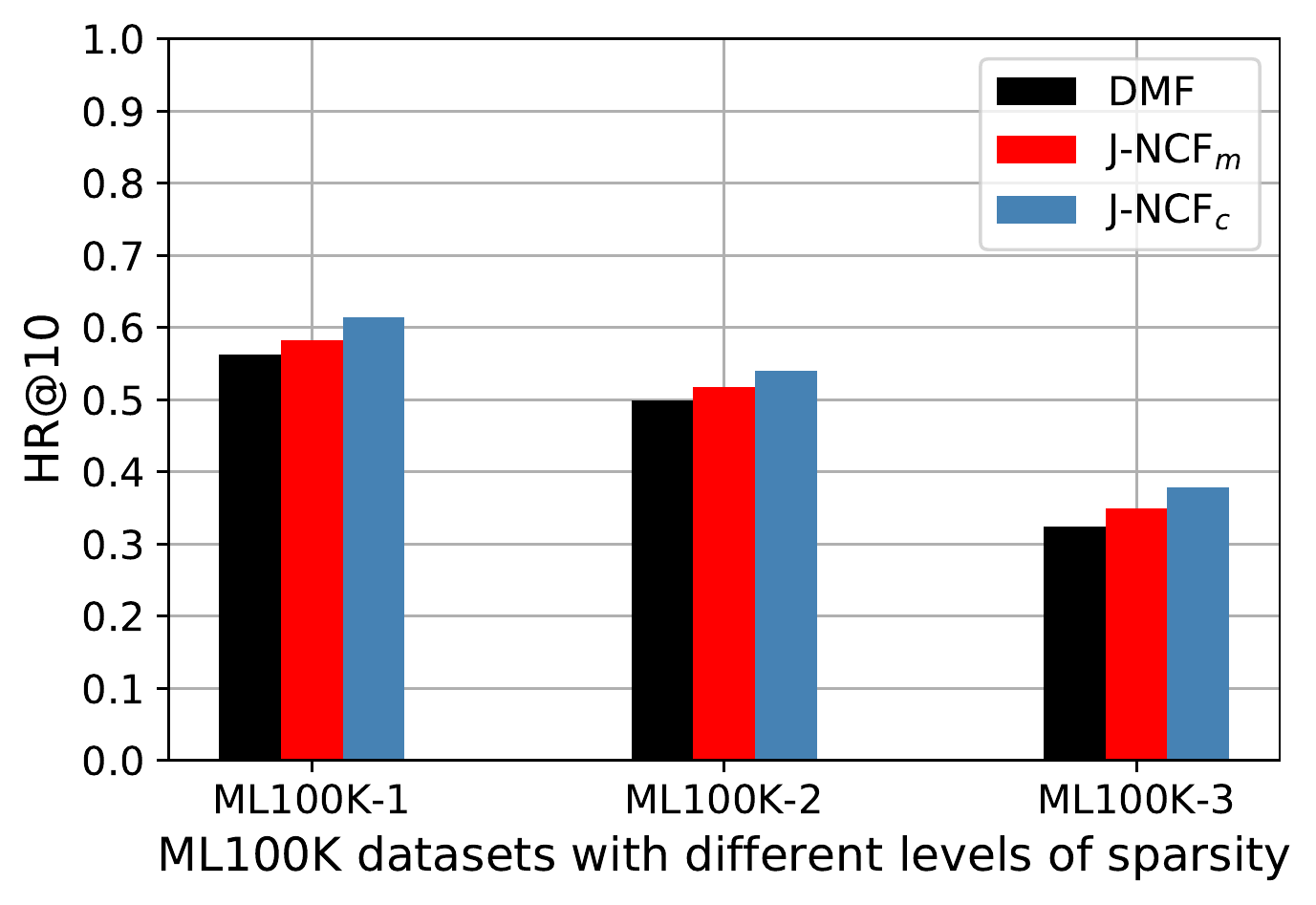}
                \caption{\raggedright Performance in terms of HR@10 on the ML100K datasets.}
                \label{ML100k_data_hr}
        \end{subfigure}
        ~%space
        \begin{subfigure}[t]{0.5\textwidth}
                \includegraphics[width=\textwidth]{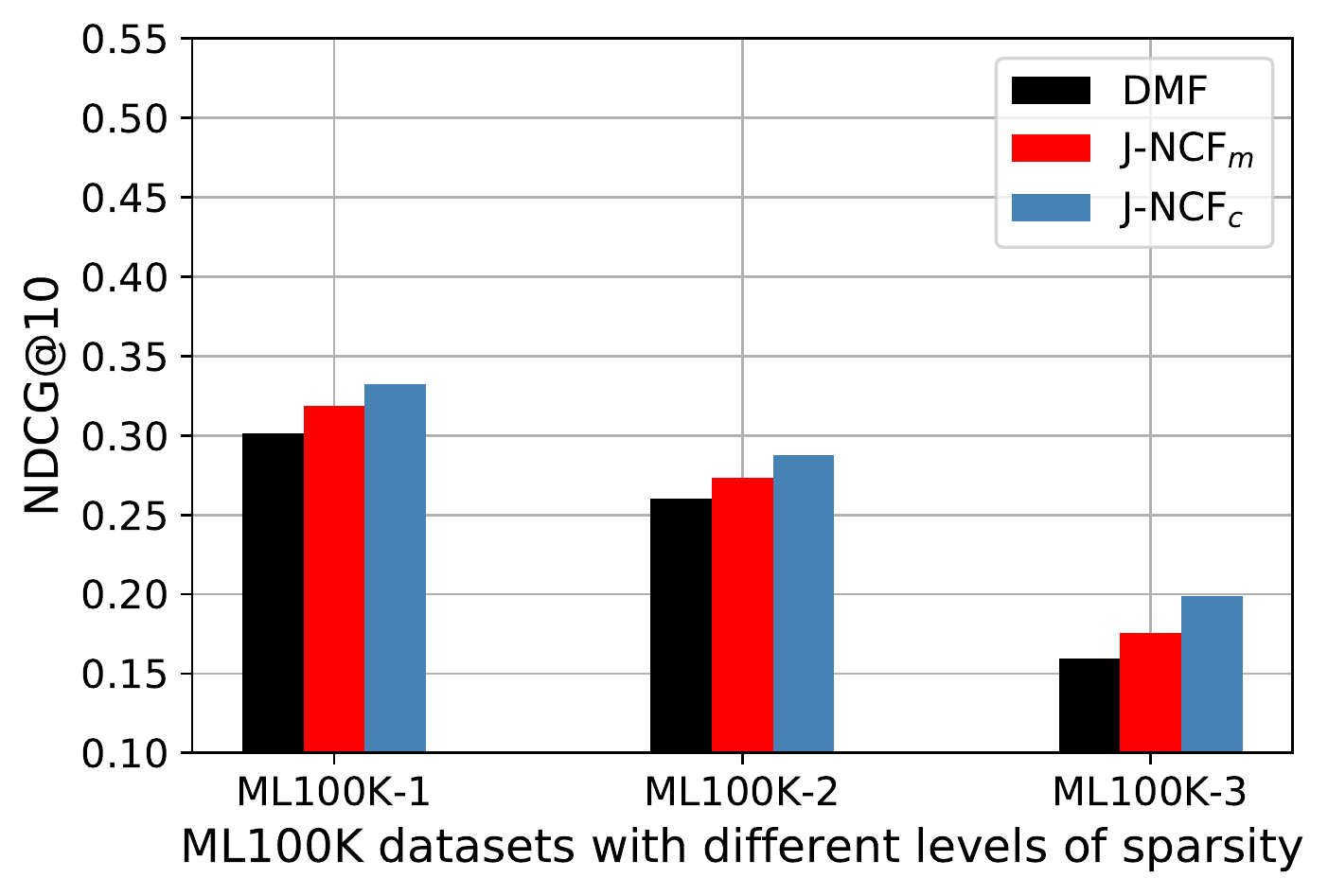}
                \caption{\raggedright Performance in terms of NDCG@10 on the ML100K datasets.}
                \label{ML100k_data_ndcg}
        \end{subfigure}
        \begin{subfigure}[t]{0.49\textwidth}
                \includegraphics[width=\textwidth]{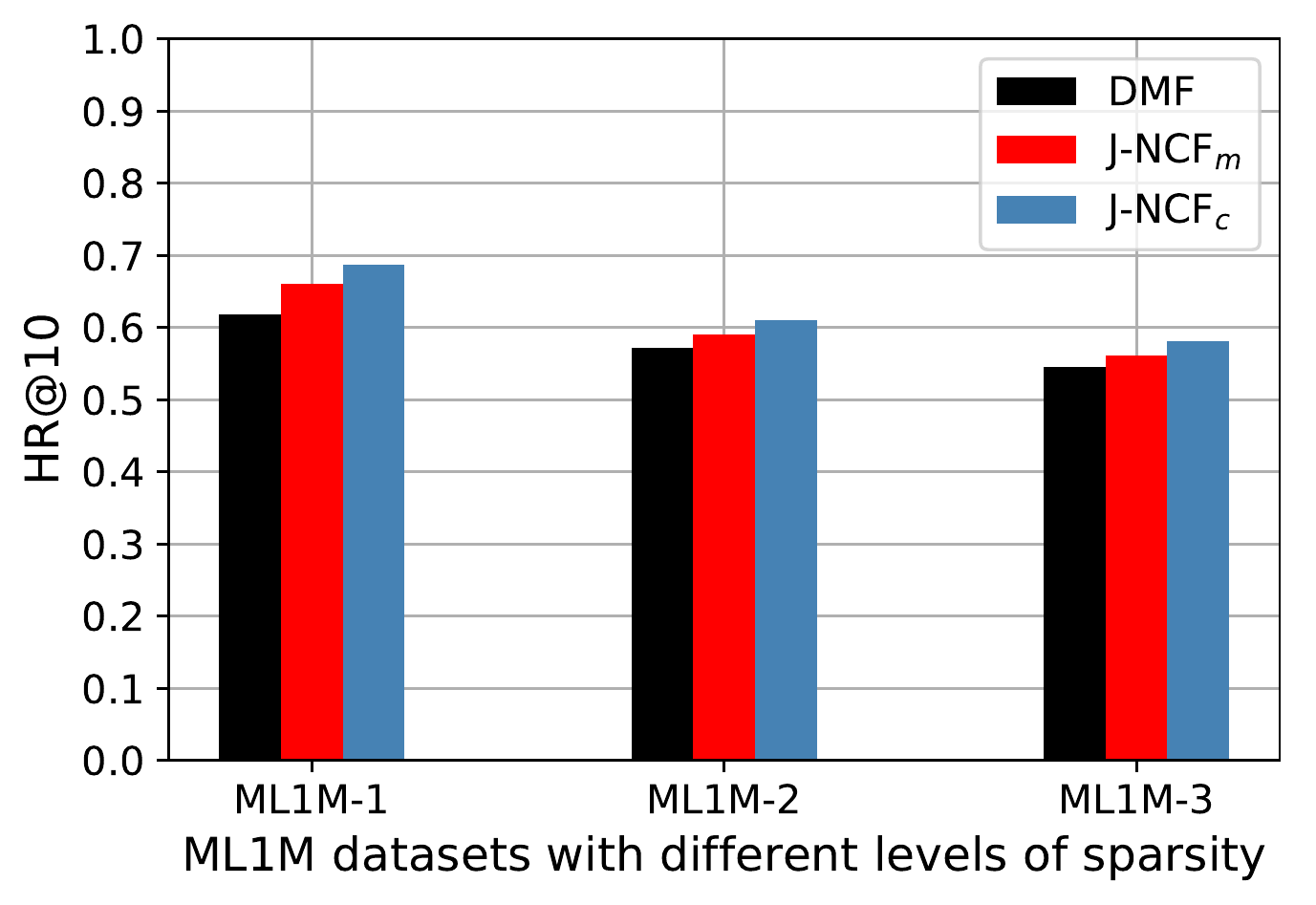}
                \caption{\raggedright Performance in terms of HR@10 on the ML1M datasets.}
                \label{ML1M_data_hr}
        \end{subfigure}
        ~%space
        \begin{subfigure}[t]{0.5\textwidth}
                \includegraphics[width=\textwidth]{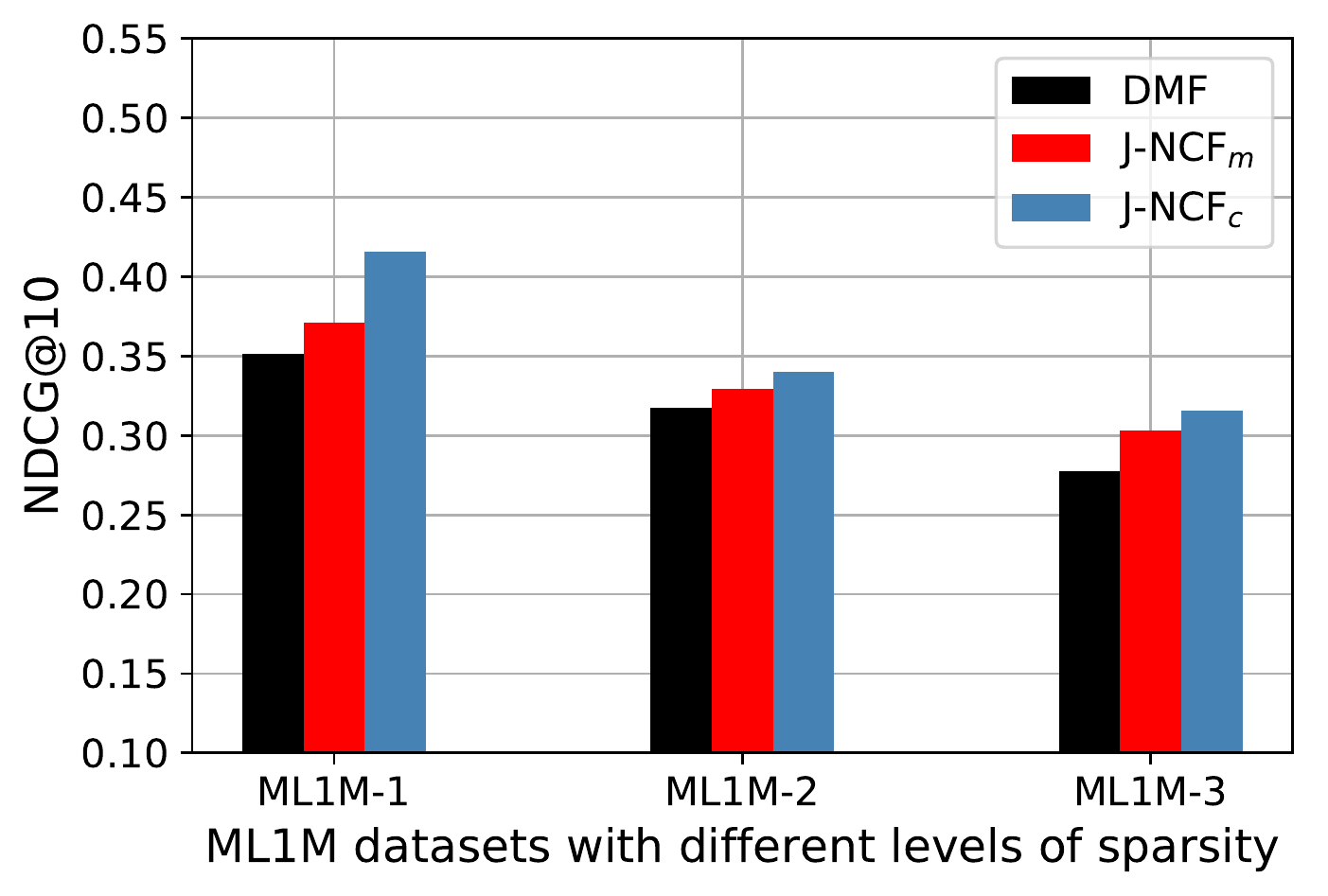}
                \caption{\raggedright Performance in terms of NDCG@10 on the ML1M datasets.}
                \label{ML1M_data_ndcg}
        \end{subfigure}
        \begin{subfigure}[t]{0.49\textwidth}
                \includegraphics[width=\textwidth]{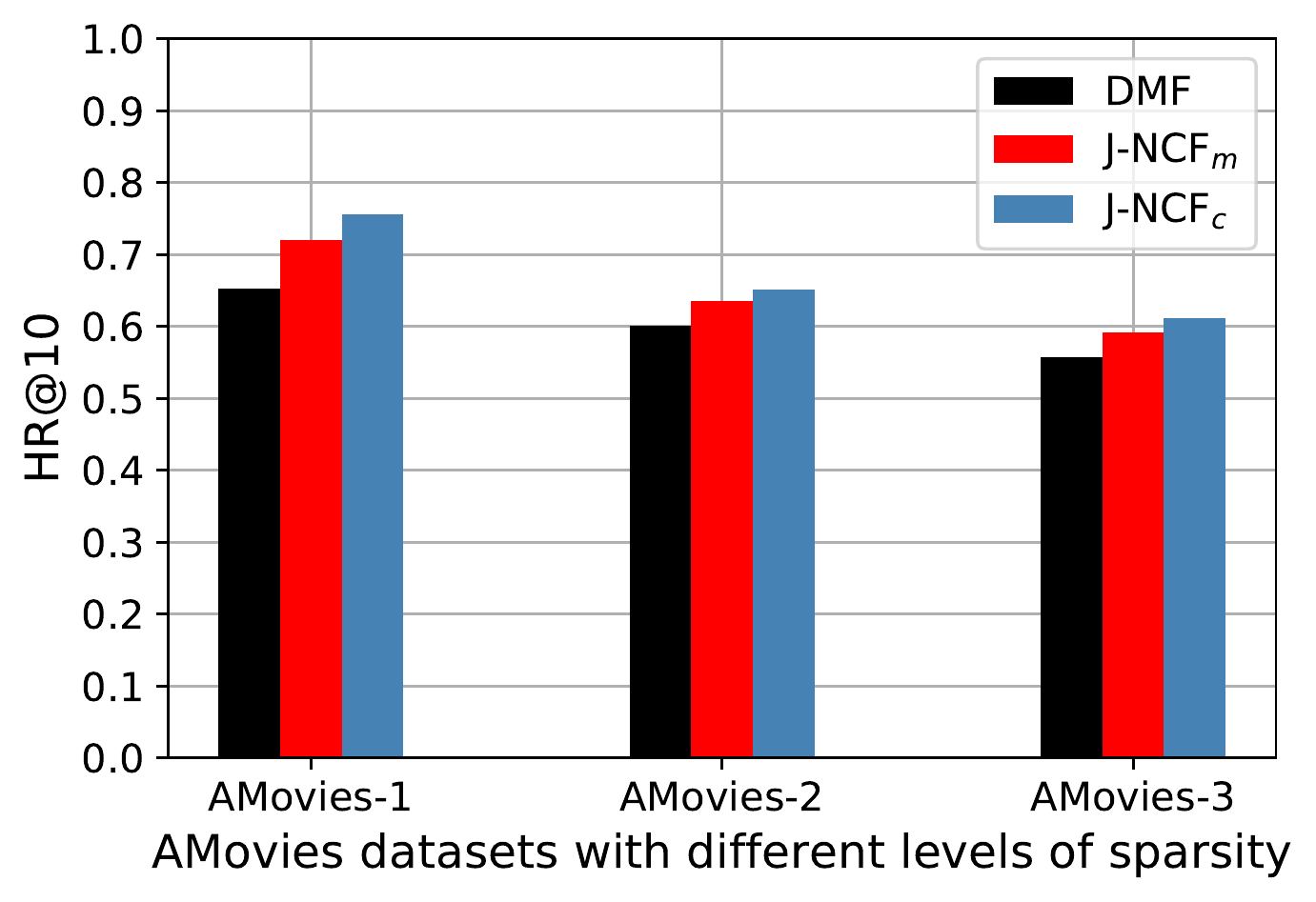}
                \caption{\raggedright Performance in terms of HR@10 on the AMovies datasets.}
                \label{Amovies_data_hr}
        \end{subfigure}
        ~%space1
        \begin{subfigure}[t]{0.5\textwidth}
                \includegraphics[width=\textwidth]{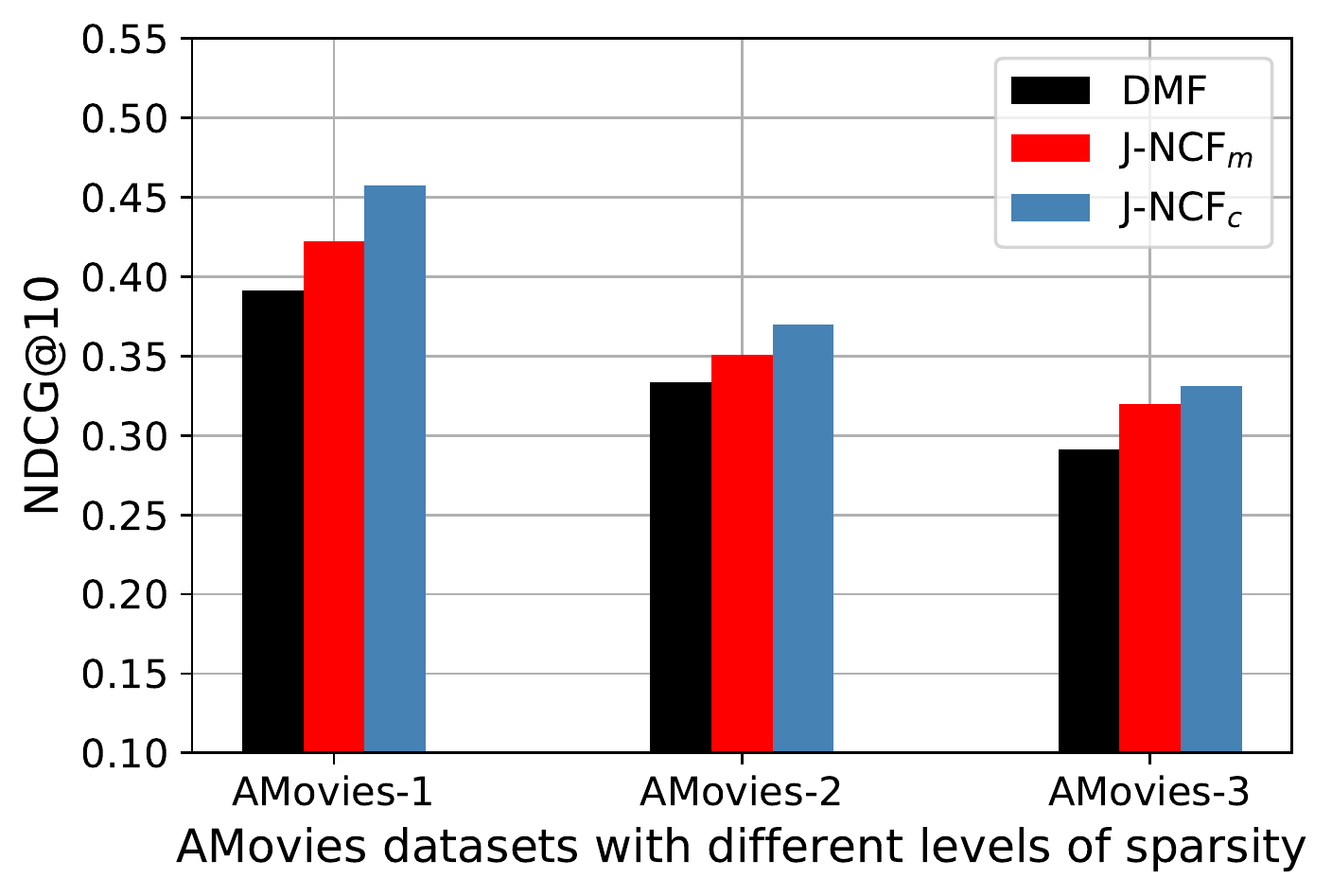}
                \caption{\raggedright Performance in terms of NDCG@10 on the AMovies datasets.}
                \label{Amovies_data_ndcg}
        \end{subfigure}
        \smallskip
                \caption{Recommendation performance across datasets with different levels of sparsity. The left and right plots show the performance in terms of HR@10 and NDCG@10, respectively.}
\label{data_sparsity}
\end{figure*}

To investigate the sensitivity of \ac{J-NCF} to different levels of data sparsity, we examine the recommendation performance on datasets with different levels of sparsity, as presented in Table~\ref{dataset-sparsity}. 
Fig.~\ref{data_sparsity} shows the results.
The overall performance of all models on the AMovies dataset is better than that on the other two datasets. 
%That is to say, the recommendation models achieve better performance when being used on datasets with more users and items.
\chen{That is to say, the recommendation performance may be influenced by the size of a dataset.}
Thus, in order to investigate the model sensitivity across datasets with different degrees of sparsity, it is essential to keep the number of users and items in the same scale for the datasets.

From Fig.~\ref{data_sparsity}, in particular, for the ML100K dataset, the ML1M dataset and the AMovies dataset respectively, we see that the \ac{J-NCF} models outperform the baseline model \ac{DMF} across all sub datasets with different degrees of sparsity in terms of HR@10 and NDCG@10.
In addition, we find that when the density of those datasets goes down, the performance of all models decreases. 
Thus it is interesting to investigate the robustness of \ac{J-NCF} when it is applied to sparse datasets. 
\chen{We find that when applied on small datasets, e.g., subsets of ML100K, our best model, i.e., \ac{J-NCF}$_{\mathit{c}}$, shows higher improvements against \ac{DMF} on sparser datasets. For example, \ac{J-NCF}$_{\mathit{c}}$ achieves 4.91\% and 9.12\% improvements over \ac{DMF} in terms of HR@10 and NDCG@10 on the ML100K-1 subset (Density${}=4.413\%$), while the improvements on the ML100K-3 subset (Density${}=0.630\%$) are 7.77\% and 12.02\% in terms of HR@10 and NDCG@10, respectively.
However, when applied on larger datasets with more users and items, i.e., subsets of ML1M and AMovies, \ac{J-NCF}$_{\mathit{c}}$ shows higher improvements against \ac{DMF} on denser datasets. For instance, \ac{J-NCF}$_{\mathit{c}}$ achieves 11.13\% improvements over \ac{DMF} in terms of HR@10 on the ML1M-1 subset (Density${}=3.7982\%$), while the improvements on the ML1M-3 subset (Density${}=0.7499\%$) are 6.53\% in terms of HR@10. 
These results may indicate that when the dataset becomes larger and sparser, it will be more difficult for models to improve their recommendation performances, which motivates us to conduct a further investigation to answer \textbf{RQ8}; see Section~\ref{largedataset} below. }

In addition, comparing the left and right-hand side plots in Fig.~\ref{data_sparsity}, we find that \ac{J-NCF}$_{\mathit{c}}$ shows a better performance in terms of NDCG@10 than HR@10. For example, the improvements of \ac{J-NCF}$_{\mathit{c}}$ over \ac{DMF} are 9.19\%, 8.28\% and 15.11\% in terms of HR@10 on ML100K-1, ML100K-2 and ML100K-3 datasets, respectively, while the improvements are 10.11\%, 10.65\% and 20.55\% in terms of NDCG@10. This result is consistent with our findings in Section~\ref{subsectionlossfunction}.

\todo{Thus in answer to \textbf{RQ7}, the \ac{J-NCF} models outperform the best baseline model \ac{DMF} across all datasets with different degrees of sparsity in terms of both metrics. 
Specifically, when applied on large datasets, i.e., ML1M and AMovies, \ac{J-NCF}$_{\mathit{c}}$ shows higher improvements against \ac{DMF} on denser datasets. 
In addition, the improvements of \ac{J-NCF}$_{\mathit{c}}$ over \ac{DMF} in terms of NDCG@10 are larger than in terms of HR@10.}

\subsection{Performance with a large and sparse dataset}
\label{largedataset}
\begin{figure*}[!t]
        \begin{subfigure}[t]{0.49\textwidth}
                \includegraphics[width=\textwidth]{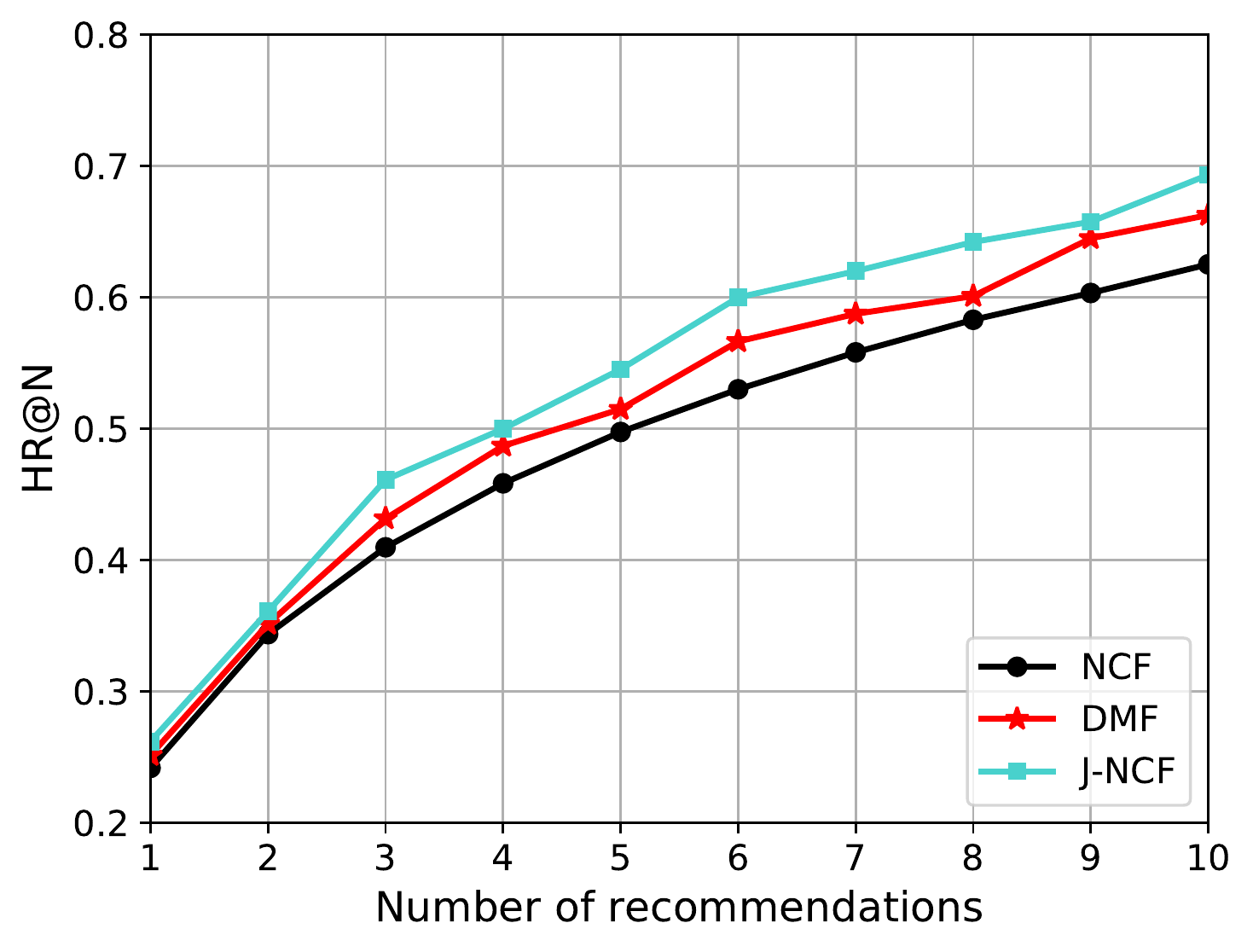}
                \caption{\raggedright Performance in terms of HR@N on AEle dataset.}
                \label{large_hr}
        \end{subfigure}
        ~%space
        \begin{subfigure}[t]{0.5\textwidth}
                \includegraphics[width=\textwidth]{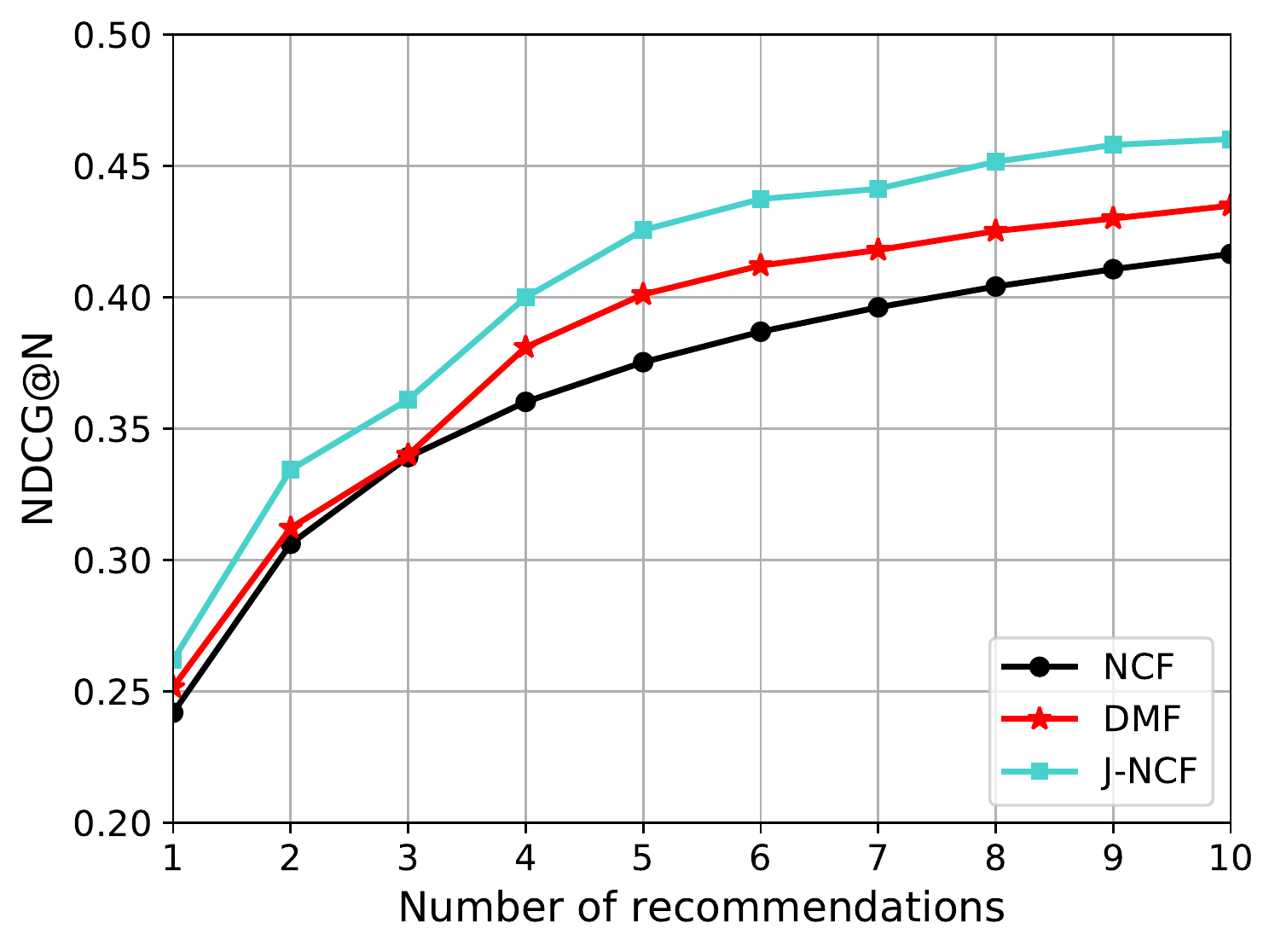}
                \caption{\raggedright Performance in terms of NDCG@N on AEle dataset.}
                \label{large_ndcg}
        \end{subfigure}
        \smallskip
        \caption{Performance of Top-N item recommendation where N ranges from 1 to 10, tested on AEle dataset.}
\label{larger_dataset}
\end{figure*}

\chen{For \textbf{RQ8}, in order to see if our model is able to work well on a large and sparse dataset, we examine our model as well as two baseline models, i.e., \ac{NCF} and \ac{DMF}, on the Amazon Electronic (AEle) dataset, which is larger and sparser than the MovieLens and Amazon Movies datasets. Fig.~\ref{larger_dataset} shows the performance of the three models with different sizes of top-N recommended lists.}

\chen{It is clear that \ac{J-NCF} outperforms \ac{DMF} as well as \ac{NCF} in terms of HR and NDCG across different numbers of recommendations.  With the size of top-N recommended lists ranging from 1 to 10, the overall performances of all models increase, which is consistent with the conclusion in Section~\ref{subsectionlossfunction}. Comparing the results shown in Fig.~\ref{large_hr} and Fig.~\ref{large_ndcg}, the improvements of \ac{J-NCF} over \ac{DMF} in terms of NDCG are more significant than those in terms of HR. For example, when $N=5$ and $N=10$, the improvements of \ac{J-NCF} over \ac{DMF} in terms of HR are 5.88\% and 4.62\%, while the improvements are 6.12\% and 5.82\% in terms of NDCG, respectively. 
To conclude and answer \textbf{RQ8}, \ac{J-NCF} can also work well with large and sparse datasets, especially in ranking items correctly.}

\subsection{Training and inference time}
\label{trainingtime}
\chen{To answer \textbf{RQ9},  we investigate the scalability of \ac{J-NCF} regarding training and inference time in Table~\ref{training_time}. 
As shown in Table~\ref{training_time}, in the ``Training'' part, ``Total time'' denotes the time needed for training the model to the best performance. And the ``Average epoch'' means the average training time for a single epoch in the training process. In the ``Prediction'' part, ``Total time'' denotes the prediction time needed for the whole test set. Since the test set contains the latest interaction of every user, the ``Average ranking'' indicates the time needed for providing a ranked list containing top 10 recommendations for a single user. }
\begin{table*}[t]
\captionsetup{justification=justified}
  \centering
  \caption{Training and prediction time needed for baseline models as well as J-NCF on all datasets.}
\label{training_time}
\begin{tabular}{lrrrrrr}
\toprule
&&\multicolumn{2}{c}{Training} && \multicolumn{2}{c}{Prediction}\\
\cmidrule{3-4}\cmidrule{6-7}
&&Total time(s) & Average epoch(s)&& Total time(s) & Average ranking(s) \\
\midrule
\multirow{3}{1.2cm}{ML100K}& NCF & 46.344 & 1.943 & &1.389 & 0.00147\\
&DMF& 180.017& 9.587 && 1.558& 0.00165 \\
&J-NCF & 116.023  & 10.925 && 1.607 & 0.00170\\
\midrule
\multirow{3}{1.2cm}{ML1M}& NCF & 494.038 & 17.751 & &8.251& 0.00137 \\
&DMF & 5,451.671 & 320.687 && 12.376& 0.00205 \\
&J-NCF & 3,539.059& 340.048  && 13.858 & 0.00229 \\
\midrule
\multirow{3}{1.2cm}{AMovies}& NCF & 977.265 & 25.836 & &25.599 & 0.00170 \\
&DMF & 39,249.657& 2,180.537 && 34.955 & 0.00232 \\
&J-NCF & 31,414.628 & 2,206.084 && 37.818 & 0.00251\\
\midrule
\multirow{3}{1.2cm}{AEle}& NCF & 61,812.187& 326.828 & &2,919.005 & 0.00239 \\
&DMF & 788,138.604 & 43,785.478 &&4,360.187 &0.00357 \\
&J-NCF &723,586.192 & 45,224.137 && 4,775.443& 0.00391\\
\bottomrule
\end{tabular}
\end{table*}

\chen{As we can see in Table~\ref{training_time}, when the size of the dataset becomes larger, the time needed for both training and prediction gets increased significantly for all models. 
\ac{NCF} consistently costs the least time among the three models for both training and prediction processes on all datasets. 
For the training process, the average training time for one epoch of \ac{J-NCF} is slightly higher than \ac{DMF}. 
However, the total training time for \ac{J-NCF} is less than for \ac{DMF}. 
It can be explained by the fact that \ac{J-NCF} needs fewer iterations to obtain the best results than \ac{DMF}, as indicated in Section~\ref{feedback}. 
Thus, \ac{J-NCF} costs less time for training to the best performance than \ac{DMF}. 
For the prediction process, although the total time needed for \ac{J-NCF} and \ac{DMF} is more than \ac{NCF}, the three models cost roughly similar amounts of time for providing a top 10 ranked list for a single user, which is around a few milliseconds.}

\section{Conclusions and Future Work}
\label{Conclusion}
We have proposed a joint neural collaborative filtering model, \ac{J-NCF}, for recommender systems.
\ac{J-NCF} uses a unified deep neural network to tightly couple two important parts in a recommender system, i.e., deep feature learning of users and items, and deep modeling of user-item interactions.
For the user and item feature extraction, we use a deep neural network with matrix factorization and a combination of explicit and implicit feedback as input.
Then we adopt another neural network for modeling user-item interactions using the feature vectors as inputs.
Thus, \ac{J-NCF} enables the two parts to be optimized with each other through a joint training process.
In order to make \ac{J-NCF} fit the top-N recommendation task, we design a new loss function that incorporates information from both pair-wise and point-wise loss.

The experimental results confirm the effectiveness of \ac{J-NCF}. 
In addition, we have also experimentally investigated the performance of \ac{J-NCF} under various settings, e.g., with different loss functions, with varying numbers of layers in the networks, and with using different feedback as inputs. 
The results confirm the effectiveness of our hybrid loss function and demonstrate that \ac{J-NCF} performs better with more layers in the networks and using the combination of implicit and explicit feedback as input. 

In addition, we have investigated the robustness of \ac{J-NCF} with different degrees of data sparsity and different numbers of user ratings.
\ac{J-NCF} outperforms the best baseline model \ac{DMF} for users across all activity levels, especially for ``inactive users'' who constitute the majority of users in the datasets.
As for datasets with different levels of sparsity, in general, \ac{J-NCF} shows more competitive recommendation performance on all datasets than the state-of-the-art baseline model \ac{DMF}. 
\chen{Moreover, we have also tested \ac{J-NCF} model with a large and sparse dataset, i.e., AEle, and the results show that \ac{J-NCF} also outperforms state-of-the-art baseline models on the dataset.}

As to future work, first, we plan to extend \ac{J-NCF} with more auxiliary information~\citep{Jointreview2017,image2017,similarity2016,QAC2016}, such as the content information of items as well as reviews, to get a more informed expression of users as well as items. 
As collaborative filtering usually suffers from limited performance due to the sparsity of user-item interactions~\citep{cold2017}, auxiliary information could be used to boost the performance. 
It would also be interesting to explore heterogeneous information in a knowledge base to improve the quality of recommender systems with deep learning~\citep{knowledgebase2016}. 
Second, we plan to explore the context information of a user in a session with recurrent neural networks to deal with dynamic aspects recommender systems~\citep{SessionRS2017,Hidasi2016,DQAC2016,timeQAC}. 
In addition, an attention mechanism could be applied to \ac{J-NCF}, which can filter out uninformative content and select the most representative items while providing good interpretability~\citep{Attentive2017}.
\chen{Finally, as we have found that \ac{J-NCF} is computationally more expensive than \ac{NCF}, we plan to optimize the structure and implementation details of our model to make it more efficient.}

\begin{acks}
We would like to thank our anonymous reviewers for their helpful comments and valuable suggestions.
\end{acks}

\bibliographystyle{ACM-Reference-Format}
\bibliography{I-NMF}


\begin{thebibliography}{56}

%%% the Web address.

\ifx \showCODEN    \undefined \def \showCODEN     #1{\unskip}     \fi
\ifx \showDOI      \undefined \def \showDOI       #1{#1}\fi
\ifx \showISBNx    \undefined \def \showISBNx     #1{\unskip}     \fi
\ifx \showISBNxiii \undefined \def \showISBNxiii  #1{\unskip}     \fi
\ifx \showISSN     \undefined \def \showISSN      #1{\unskip}     \fi
\ifx \showLCCN     \undefined \def \showLCCN      #1{\unskip}     \fi
\ifx \shownote     \undefined \def \shownote      #1{#1}          \fi
\ifx \showarticletitle \undefined \def \showarticletitle #1{#1}   \fi
\ifx \showURL      \undefined \def \showURL       {\relax}        \fi
% invisible to TeX
\providecommand\bibfield[2]{#2}
\providecommand\bibinfo[2]{#2}
\providecommand\natexlab[1]{#1}
\providecommand\showeprint[2][]{arXiv:#2}

\bibitem[\protect\citeauthoryear{Adeniyi, Wei, and Yang}{Adeniyi
  et~al\mbox{.}}{2016}]%
        {KNN2016}
\bibfield{author}{\bibinfo{person}{David~Adedayo Adeniyi},
  \bibinfo{person}{Zhaoqiang Wei}, {and} \bibinfo{person}{Yongquan Yang}.}
  \bibinfo{year}{2016}\natexlab{}.
\newblock \showarticletitle{Automated web usage data mining and recommendation
  system using K-Nearest Neighbor (KNN) classification method}.
\newblock \bibinfo{journal}{\emph{Applied Computing and Informatics}}
  \bibinfo{volume}{12}, \bibinfo{number}{1} (\bibinfo{year}{2016}),
  \bibinfo{pages}{90--108}.
\newblock


\bibitem[\protect\citeauthoryear{Adomavicius and Tuzhilin}{Adomavicius and
  Tuzhilin}{2005}]%
        {RS-CF2005}
\bibfield{author}{\bibinfo{person}{Gediminas Adomavicius} {and}
  \bibinfo{person}{Alexander Tuzhilin}.} \bibinfo{year}{2005}\natexlab{}.
\newblock \showarticletitle{Toward the Next Generation of Recommender Systems:
  A Survey of the State-of-the-Art and Possible Extensions}.
\newblock \bibinfo{journal}{\emph{IEEE Transactions on Knowledge and Data
  Engineering.}} \bibinfo{volume}{17}, \bibinfo{number}{6}
  (\bibinfo{year}{2005}), \bibinfo{pages}{734--749}.
\newblock


\bibitem[\protect\citeauthoryear{Basiliyos, Charles, and Bernabe}{Basiliyos
  et~al\mbox{.}}{2017}]%
        {DL-RS1}
\bibfield{author}{\bibinfo{person}{Betru Basiliyos, Tilahun},
  \bibinfo{person}{Onana Charles, Awono}, {and} \bibinfo{person}{Batchakui
  Bernabe}.} \bibinfo{year}{2017}\natexlab{}.
\newblock \showarticletitle{Deep Learning Methods on Recommender System: A
  Survey of State-of-the-art.}
\newblock \bibinfo{journal}{\emph{International Journal of Computer
  Applications}} \bibinfo{volume}{162}, \bibinfo{number}{10}
  (\bibinfo{year}{2017}), \bibinfo{pages}{17--22}.
\newblock


\bibitem[\protect\citeauthoryear{Bellogin, Castells, and Cantador}{Bellogin
  et~al\mbox{.}}{2011}]%
        {Precision2011}
\bibfield{author}{\bibinfo{person}{Alejandro Bellogin}, \bibinfo{person}{Pablo
  Castells}, {and} \bibinfo{person}{Ivan Cantador}.}
  \bibinfo{year}{2011}\natexlab{}.
\newblock \showarticletitle{Precision-oriented Evaluation of Recommender
  Systems: An Algorithmic Comparison}. In \bibinfo{booktitle}{\emph{RecSys
  '11}}. \bibinfo{publisher}{ACM}, \bibinfo{pages}{333--336}.
\newblock


\bibitem[\protect\citeauthoryear{Cai and de~Rijke}{Cai and de~Rijke}{2016a}]%
        {similarity2016}
\bibfield{author}{\bibinfo{person}{Fei Cai} {and} \bibinfo{person}{Maarten de
  Rijke}.} \bibinfo{year}{2016}\natexlab{a}.
\newblock \showarticletitle{Learning from homologous queries and semantically
  related terms for query auto completion}.
\newblock \bibinfo{journal}{\emph{Information Processing \& Management}}
  \bibinfo{volume}{52}, \bibinfo{number}{4} (\bibinfo{year}{2016}),
  \bibinfo{pages}{628--643}.
\newblock


\bibitem[\protect\citeauthoryear{Cai and de~Rijke}{Cai and de~Rijke}{2016b}]%
        {QAC2016}
\bibfield{author}{\bibinfo{person}{Fei Cai} {and} \bibinfo{person}{Maarten de
  Rijke}.} \bibinfo{year}{2016}\natexlab{b}.
\newblock \showarticletitle{A Survey of Query Auto Completion in Information
  Retrieval}.
\newblock \bibinfo{journal}{\emph{Foundations and Trends in Information
  Retrieval}} \bibinfo{volume}{10}, \bibinfo{number}{4} (\bibinfo{year}{2016}),
  \bibinfo{pages}{273--363}.
\newblock


\bibitem[\protect\citeauthoryear{Cai, Liang, and de~Rijke}{Cai
  et~al\mbox{.}}{2016a}]%
        {timeQAC}
\bibfield{author}{\bibinfo{person}{Fei Cai}, \bibinfo{person}{Shangsong Liang},
  {and} \bibinfo{person}{Maarten de Rijke}.} \bibinfo{year}{2016}\natexlab{a}.
\newblock \showarticletitle{Prefix-Adaptive and Time-Sensitive Personalized
  Query Auto Completion}.
\newblock \bibinfo{journal}{\emph{IEEE Transactions on Knowledge and Data
  Engineering}} \bibinfo{volume}{28}, \bibinfo{number}{9} (\bibinfo{date}{Sep}
  \bibinfo{year}{2016}), \bibinfo{pages}{2452--2466}.
\newblock


\bibitem[\protect\citeauthoryear{Cai, Reinanda, and de~Rijke}{Cai
  et~al\mbox{.}}{2016b}]%
        {DQAC2016}
\bibfield{author}{\bibinfo{person}{Fei Cai}, \bibinfo{person}{Ridho Reinanda},
  {and} \bibinfo{person}{Maarten de Rijke}.} \bibinfo{year}{2016}\natexlab{b}.
\newblock \showarticletitle{Diversifying Query Auto-Completion}.
\newblock \bibinfo{journal}{\emph{ACM Transactions on Information Systems}}
  \bibinfo{volume}{34}, \bibinfo{number}{4} (\bibinfo{date}{June}
  \bibinfo{year}{2016}), \bibinfo{pages}{25:1--25:33}.
\newblock


\bibitem[\protect\citeauthoryear{Chatzis, Christodoulou, and Andreou}{Chatzis
  et~al\mbox{.}}{2017}]%
        {SessionRS2017}
\bibfield{author}{\bibinfo{person}{Sotirios~P. Chatzis},
  \bibinfo{person}{Panayiotis Christodoulou}, {and} \bibinfo{person}{Andreas~S.
  Andreou}.} \bibinfo{year}{2017}\natexlab{}.
\newblock \showarticletitle{Recurrent Latent Variable Networks for
  Session-Based Recommendation}. In \bibinfo{booktitle}{\emph{DLRS '17}}.
  \bibinfo{pages}{38--45}.
\newblock


\bibitem[\protect\citeauthoryear{Chen, Zhang, He, Nie, Liu, and Chua}{Chen
  et~al\mbox{.}}{2017}]%
        {Attentive2017}
\bibfield{author}{\bibinfo{person}{Jingyuan Chen}, \bibinfo{person}{Hanwang
  Zhang}, \bibinfo{person}{Xiangnan He}, \bibinfo{person}{Liqiang Nie},
  \bibinfo{person}{Wei Liu}, {and} \bibinfo{person}{Tat-Seng Chua}.}
  \bibinfo{year}{2017}\natexlab{}.
\newblock \showarticletitle{Attentive Collaborative Filtering: Multimedia
  Recommendation with Item- and Component-Level Attention}. In
  \bibinfo{booktitle}{\emph{SIGIR '17}}. \bibinfo{publisher}{ACM},
  \bibinfo{pages}{335--344}.
\newblock


\bibitem[\protect\citeauthoryear{Chen, Cai, Chen, and de~Rijke}{Chen
  et~al\mbox{.}}{2018}]%
        {QS2018}
\bibfield{author}{\bibinfo{person}{Wanyu Chen}, \bibinfo{person}{Fei Cai},
  \bibinfo{person}{Honghui Chen}, {and} \bibinfo{person}{Maarten de Rijke}.}
  \bibinfo{year}{2018}\natexlab{}.
\newblock \showarticletitle{Attention-based Hierarchical Neural Query
  Suggestion}. In \bibinfo{booktitle}{\emph{SIGIR '18}}.
  \bibinfo{publisher}{ACM}, \bibinfo{pages}{1093--1096}.
\newblock


\bibitem[\protect\citeauthoryear{Cheng, Koc, Harmsen, Shaked, Chandra, Aradhye,
  Anderson, Corrado, Chai, Ispir, Anil, Haque, Hong, Jain, Liu, and Shah}{Cheng
  et~al\mbox{.}}{2016}]%
        {Wide2017}
\bibfield{author}{\bibinfo{person}{Heng-Tze Cheng}, \bibinfo{person}{Levent
  Koc}, \bibinfo{person}{Jeremiah Harmsen}, \bibinfo{person}{Tal Shaked},
  \bibinfo{person}{Tushar Chandra}, \bibinfo{person}{Hrishi Aradhye},
  \bibinfo{person}{Glen Anderson}, \bibinfo{person}{Greg Corrado},
  \bibinfo{person}{Wei Chai}, \bibinfo{person}{Mustafa Ispir},
  \bibinfo{person}{Rohan Anil}, \bibinfo{person}{Zakaria Haque},
  \bibinfo{person}{Lichan Hong}, \bibinfo{person}{Vihan Jain},
  \bibinfo{person}{Xiaobing Liu}, {and} \bibinfo{person}{Hemal Shah}.}
  \bibinfo{year}{2016}\natexlab{}.
\newblock \showarticletitle{Wide \& Deep Learning for Recommender Systems}. In
  \bibinfo{booktitle}{\emph{DLRS 2016}}. \bibinfo{publisher}{ACM},
  \bibinfo{pages}{7--10}.
\newblock


\bibitem[\protect\citeauthoryear{Cremonesi, Koren, and Turrin}{Cremonesi
  et~al\mbox{.}}{2010}]%
        {Performance2010}
\bibfield{author}{\bibinfo{person}{Paolo Cremonesi}, \bibinfo{person}{Yehuda
  Koren}, {and} \bibinfo{person}{Roberto Turrin}.}
  \bibinfo{year}{2010}\natexlab{}.
\newblock \showarticletitle{Performance of Recommender Algorithms on Top-n
  Recommendation Tasks}. In \bibinfo{booktitle}{\emph{RecSys '10}}.
  \bibinfo{publisher}{ACM}, \bibinfo{pages}{39--46}.
\newblock


\bibitem[\protect\citeauthoryear{Guo, Tang, Ye, Li, and He}{Guo
  et~al\mbox{.}}{2017}]%
        {DeepFM2017}
\bibfield{author}{\bibinfo{person}{Huifeng Guo}, \bibinfo{person}{Ruiming
  Tang}, \bibinfo{person}{Yunming Ye}, \bibinfo{person}{Zhenguo Li}, {and}
  \bibinfo{person}{Xiuqiang He}.} \bibinfo{year}{2017}\natexlab{}.
\newblock \showarticletitle{DeepFM: A Factorization-machine Based Neural
  Network for CTR Prediction}. In \bibinfo{booktitle}{\emph{IJCAI'17}}.
  \bibinfo{publisher}{AAAI Press}, \bibinfo{pages}{1725--1731}.
\newblock


\bibitem[\protect\citeauthoryear{He and Chua}{He and Chua}{2017}]%
        {NFM2017}
\bibfield{author}{\bibinfo{person}{Xiangnan He} {and} \bibinfo{person}{Tat-Seng
  Chua}.} \bibinfo{year}{2017}\natexlab{}.
\newblock \showarticletitle{Neural Factorization Machines for Sparse Predictive
  Analytics}. In \bibinfo{booktitle}{\emph{SIGIR '17}}.
  \bibinfo{publisher}{ACM}, \bibinfo{pages}{355--364}.
\newblock


\bibitem[\protect\citeauthoryear{He, Liao, Zhang, Nie, Hu, and Chua}{He
  et~al\mbox{.}}{2017}]%
        {NCF2017}
\bibfield{author}{\bibinfo{person}{Xiangnan He}, \bibinfo{person}{Lizi Liao},
  \bibinfo{person}{Hanwang Zhang}, \bibinfo{person}{Liqiang Nie},
  \bibinfo{person}{Xia Hu}, {and} \bibinfo{person}{Tat-Seng Chua}.}
  \bibinfo{year}{2017}\natexlab{}.
\newblock \showarticletitle{Neural Collaborative Filtering}. In
  \bibinfo{booktitle}{\emph{WWW '17}}. \bibinfo{publisher}{ACM},
  \bibinfo{pages}{173--182}.
\newblock


\bibitem[\protect\citeauthoryear{He, Zhang, Kan, and Chua}{He
  et~al\mbox{.}}{2016}]%
        {EALS2016}
\bibfield{author}{\bibinfo{person}{Xiangnan He}, \bibinfo{person}{Hanwang
  Zhang}, \bibinfo{person}{Min-Yen Kan}, {and} \bibinfo{person}{Tat-Seng
  Chua}.} \bibinfo{year}{2016}\natexlab{}.
\newblock \showarticletitle{Fast Matrix Factorization for Online Recommendation
  with Implicit Feedback}. In \bibinfo{booktitle}{\emph{SIGIR '16}}.
  \bibinfo{publisher}{ACM}, \bibinfo{pages}{549--558}.
\newblock


\bibitem[\protect\citeauthoryear{Herlocker, Konstan, Terveen, and
  Riedl}{Herlocker et~al\mbox{.}}{2004}]%
        {TOIS2004}
\bibfield{author}{\bibinfo{person}{Jonathan~L. Herlocker},
  \bibinfo{person}{Joseph~A. Konstan}, \bibinfo{person}{Loren~G. Terveen},
  {and} \bibinfo{person}{John~T. Riedl}.} \bibinfo{year}{2004}\natexlab{}.
\newblock \showarticletitle{Evaluating Collaborative Filtering Recommender
  Systems}.
\newblock \bibinfo{journal}{\emph{ACM Transactions on Information Systems}}
  \bibinfo{volume}{22}, \bibinfo{number}{1} (\bibinfo{year}{2004}),
  \bibinfo{pages}{5--53}.
\newblock


\bibitem[\protect\citeauthoryear{Hidasi and Karatzoglou}{Hidasi and
  Karatzoglou}{2018}]%
        {2018lossfunction}
\bibfield{author}{\bibinfo{person}{Bal\'{a}zs Hidasi} {and}
  \bibinfo{person}{Alexandros Karatzoglou}.} \bibinfo{year}{2018}\natexlab{}.
\newblock \showarticletitle{Recurrent Neural Networks with Top-k Gains for
  Session-based Recommendations}. In \bibinfo{booktitle}{\emph{CIKM '18}}.
  \bibinfo{publisher}{ACM}, \bibinfo{pages}{843--852}.
\newblock


\bibitem[\protect\citeauthoryear{Hidasi, Karatzoglou, Baltrunas, and
  Tikk}{Hidasi et~al\mbox{.}}{2016a}]%
        {2016session-based}
\bibfield{author}{\bibinfo{person}{Balazs Hidasi}, \bibinfo{person}{Alexandros
  Karatzoglou}, \bibinfo{person}{Linas Baltrunas}, {and}
  \bibinfo{person}{Domonkos Tikk}.} \bibinfo{year}{2016}\natexlab{a}.
\newblock \showarticletitle{Session-based Recommendations with Recurrent Neural
  Networks}. In \bibinfo{booktitle}{\emph{ICLR '16}}.
\newblock


\bibitem[\protect\citeauthoryear{Hidasi, Quadrana, Karatzoglou, and
  Tikk}{Hidasi et~al\mbox{.}}{2016b}]%
        {Hidasi2016}
\bibfield{author}{\bibinfo{person}{Bal\'{a}zs Hidasi}, \bibinfo{person}{Massimo
  Quadrana}, \bibinfo{person}{Alexandros Karatzoglou}, {and}
  \bibinfo{person}{Domonkos Tikk}.} \bibinfo{year}{2016}\natexlab{b}.
\newblock \showarticletitle{Parallel Recurrent Neural Network Architectures for
  Feature-rich Session-based Recommendations}. In
  \bibinfo{booktitle}{\emph{RecSys '16}}. \bibinfo{publisher}{ACM},
  \bibinfo{pages}{241--248}.
\newblock


\bibitem[\protect\citeauthoryear{Hong-Jian, Xinyu, Jianbing, Shujian, and
  Jiajun}{Hong-Jian et~al\mbox{.}}{2017}]%
        {DMF2017}
\bibfield{author}{\bibinfo{person}{Xue Hong-Jian}, \bibinfo{person}{Dai Xinyu},
  \bibinfo{person}{Zhang Jianbing}, \bibinfo{person}{Huang Shujian}, {and}
  \bibinfo{person}{Chen Jiajun}.} \bibinfo{year}{2017}\natexlab{}.
\newblock \showarticletitle{Deep Matrix Factorization Models for Recommender
  Systems}. In \bibinfo{booktitle}{\emph{IJCAI '17}}.
  \bibinfo{pages}{3203--3209}.
\newblock


\bibitem[\protect\citeauthoryear{Huang, He, Gao, Deng, Acero, and Heck}{Huang
  et~al\mbox{.}}{2013}]%
        {Huang:2013}
\bibfield{author}{\bibinfo{person}{Po-Sen Huang}, \bibinfo{person}{Xiaodong
  He}, \bibinfo{person}{Jianfeng Gao}, \bibinfo{person}{Li Deng},
  \bibinfo{person}{Alex Acero}, {and} \bibinfo{person}{Larry Heck}.}
  \bibinfo{year}{2013}\natexlab{}.
\newblock \showarticletitle{Learning Deep Structured Semantic Models for Web
  Search Using Clickthrough Data}. In \bibinfo{booktitle}{\emph{CIKM '13}}.
  \bibinfo{publisher}{ACM}, \bibinfo{pages}{2333--2338}.
\newblock


\bibitem[\protect\citeauthoryear{Kabbur, Ning, and Karypis}{Kabbur
  et~al\mbox{.}}{2013}]%
        {FISM2013}
\bibfield{author}{\bibinfo{person}{Santosh Kabbur}, \bibinfo{person}{Xia Ning},
  {and} \bibinfo{person}{George Karypis}.} \bibinfo{year}{2013}\natexlab{}.
\newblock \showarticletitle{FISM: Factored item similarity models for Top-N
  recommender systems}. In \bibinfo{booktitle}{\emph{KDD '13}}.
  \bibinfo{publisher}{ACM}, \bibinfo{pages}{659--667}.
\newblock


\bibitem[\protect\citeauthoryear{Kim, Park, Oh, Lee, and Yu}{Kim
  et~al\mbox{.}}{2016}]%
        {CNN2016}
\bibfield{author}{\bibinfo{person}{Donghyun Kim}, \bibinfo{person}{Chanyoung
  Park}, \bibinfo{person}{Jinoh Oh}, \bibinfo{person}{Sungyoung Lee}, {and}
  \bibinfo{person}{Hwanjo Yu}.} \bibinfo{year}{2016}\natexlab{}.
\newblock \showarticletitle{Convolutional Matrix Factorization for Document
  Context-Aware Recommendation}. In \bibinfo{booktitle}{\emph{RecSys '16}}.
  \bibinfo{pages}{233--240}.
\newblock


\bibitem[\protect\citeauthoryear{Kingma and Ba}{Kingma and Ba}{2014}]%
        {Adam2014}
\bibfield{author}{\bibinfo{person}{Diederik Kingma} {and}
  \bibinfo{person}{Jimmy Ba}.} \bibinfo{year}{2014}\natexlab{}.
\newblock \showarticletitle{Adam: A Method for Stochastic Optimization}.
\newblock \bibinfo{journal}{\emph{arXiv preprint arXiv:1412.6980}}
  (\bibinfo{year}{2014}).
\newblock


\bibitem[\protect\citeauthoryear{Koren}{Koren}{2008}]%
        {SVD++2008}
\bibfield{author}{\bibinfo{person}{Yehuda Koren}.}
  \bibinfo{year}{2008}\natexlab{}.
\newblock \showarticletitle{Factorization Meets the Neighborhood: A
  Multifaceted Collaborative Filtering Model}. In \bibinfo{booktitle}{\emph{KDD
  '08}}. \bibinfo{publisher}{ACM}, \bibinfo{pages}{426--434}.
\newblock


\bibitem[\protect\citeauthoryear{Koren, Bell, and Volinsky}{Koren
  et~al\mbox{.}}{2009}]%
        {LFM2009}
\bibfield{author}{\bibinfo{person}{Yehuda Koren}, \bibinfo{person}{Robert
  Bell}, {and} \bibinfo{person}{Chris Volinsky}.}
  \bibinfo{year}{2009}\natexlab{}.
\newblock \showarticletitle{Matrix Factorization Techniques for Recommender
  Systems}.
\newblock \bibinfo{journal}{\emph{Computer}} \bibinfo{volume}{42},
  \bibinfo{number}{8} (\bibinfo{year}{2009}), \bibinfo{pages}{30--37}.
\newblock


\bibitem[\protect\citeauthoryear{Li, Kawale, and Fu}{Li et~al\mbox{.}}{2015}]%
        {DAE2015}
\bibfield{author}{\bibinfo{person}{Sheng Li}, \bibinfo{person}{Jaya Kawale},
  {and} \bibinfo{person}{Yun Fu}.} \bibinfo{year}{2015}\natexlab{}.
\newblock \showarticletitle{Deep Collaborative Filtering via Marginalized
  Denoising Auto-encoder}. In \bibinfo{booktitle}{\emph{CIKM '15}}.
  \bibinfo{publisher}{ACM}, \bibinfo{pages}{811--820}.
\newblock


\bibitem[\protect\citeauthoryear{Lian, Zhang, Xie, and Sun}{Lian
  et~al\mbox{.}}{2017}]%
        {CCCFNet2017}
\bibfield{author}{\bibinfo{person}{Jianxun Lian}, \bibinfo{person}{Fuzheng
  Zhang}, \bibinfo{person}{Xing Xie}, {and} \bibinfo{person}{Guangzhong Sun}.}
  \bibinfo{year}{2017}\natexlab{}.
\newblock \showarticletitle{CCCFNet: A Content-Boosted Collaborative Filtering
  Neural Network for Cross Domain Recommender Systems}. In
  \bibinfo{booktitle}{\emph{WWW '17}}. \bibinfo{publisher}{ACM},
  \bibinfo{pages}{817--818}.
\newblock


\bibitem[\protect\citeauthoryear{Linden, Smith, and York}{Linden
  et~al\mbox{.}}{2003}]%
        {neighborRS2003}
\bibfield{author}{\bibinfo{person}{Greg Linden}, \bibinfo{person}{Brent Smith},
  {and} \bibinfo{person}{Jeremy York}.} \bibinfo{year}{2003}\natexlab{}.
\newblock \showarticletitle{Amazon.com Recommendations: Item-to-Item
  Collaborative Filtering}.
\newblock \bibinfo{journal}{\emph{IEEE Internet Computing}}
  \bibinfo{volume}{7}, \bibinfo{number}{1} (\bibinfo{year}{2003}),
  \bibinfo{pages}{76--80}.
\newblock


\bibitem[\protect\citeauthoryear{Liu and Wu}{Liu and Wu}{2017}]%
        {DL-RS2}
\bibfield{author}{\bibinfo{person}{Juntao Liu} {and} \bibinfo{person}{Caihua
  Wu}.} \bibinfo{year}{2017}\natexlab{}.
\newblock \showarticletitle{Deep Learning Based Recommendation: A Survey}. In
  \bibinfo{booktitle}{\emph{ICISA '17}}. \bibinfo{pages}{451--458}.
\newblock


\bibitem[\protect\citeauthoryear{Liu, Ouyang, Rong, and Xiong}{Liu
  et~al\mbox{.}}{2015}]%
        {CRBM2105}
\bibfield{author}{\bibinfo{person}{Xiaomeng Liu}, \bibinfo{person}{Yuanxin
  Ouyang}, \bibinfo{person}{Wenge Rong}, {and} \bibinfo{person}{Zhang Xiong}.}
  \bibinfo{year}{2015}\natexlab{}.
\newblock \showarticletitle{Item Category Aware Conditional Restricted
  Boltzmann Machine Based Recommendation}. In \bibinfo{booktitle}{\emph{ICONIP
  '15}}. \bibinfo{pages}{609--616}.
\newblock


\bibitem[\protect\citeauthoryear{Onal, Zhang, Altingovde, Rahman, Karagoz,
  Braylan, Dang, Chang, Kim, McNamara, Angert, Banner, Khetan, McDonnell,
  Nguyen, Xu, Wallace, de~Rijke, and Lease}{Onal et~al\mbox{.}}{2018}]%
        {onal-neural-2018}
\bibfield{author}{\bibinfo{person}{Kezban~Dilek Onal}, \bibinfo{person}{Ye
  Zhang}, \bibinfo{person}{Ismail~Sengor Altingovde},
  \bibinfo{person}{Md~Mustafizur Rahman}, \bibinfo{person}{Pinar Karagoz},
  \bibinfo{person}{Alex Braylan}, \bibinfo{person}{Brandon Dang},
  \bibinfo{person}{Heng-Lu Chang}, \bibinfo{person}{Henna Kim},
  \bibinfo{person}{Quinten McNamara}, \bibinfo{person}{Aaron Angert},
  \bibinfo{person}{Edward Banner}, \bibinfo{person}{Vivek Khetan},
  \bibinfo{person}{Tyler McDonnell}, \bibinfo{person}{An~Thanh Nguyen},
  \bibinfo{person}{Dan Xu}, \bibinfo{person}{Byron~C. Wallace},
  \bibinfo{person}{Maarten de Rijke}, {and} \bibinfo{person}{Matthew Lease}.}
  \bibinfo{year}{2018}\natexlab{}.
\newblock \showarticletitle{Neural information retrieval: At the end of the
  early years}.
\newblock \bibinfo{journal}{\emph{Information Retrieval Journal}}
  \bibinfo{volume}{21}, \bibinfo{number}{2--3} (\bibinfo{date}{June}
  \bibinfo{year}{2018}), \bibinfo{pages}{111--182}.
\newblock


\bibitem[\protect\citeauthoryear{Paterek}{Paterek}{2007}]%
        {improving2007}
\bibfield{author}{\bibinfo{person}{Arkadiusz Paterek}.}
  \bibinfo{year}{2007}\natexlab{}.
\newblock \showarticletitle{Improving regularized singular value decomposition
  for collaborative filtering}. In \bibinfo{booktitle}{\emph{KDD '07}}.
  \bibinfo{publisher}{ACM}.
\newblock


\bibitem[\protect\citeauthoryear{Prem, Hofman, and Blei}{Prem
  et~al\mbox{.}}{2013}]%
        {PF2013}
\bibfield{author}{\bibinfo{person}{Gopalan Prem}, \bibinfo{person}{Jake~M.
  Hofman}, {and} \bibinfo{person}{David~M. Blei}.}
  \bibinfo{year}{2013}\natexlab{}.
\newblock \showarticletitle{Scalable Recommendation with Poisson
  Factorization}.
\newblock \bibinfo{journal}{\emph{arXiv preprint arXiv:1311.1704}}
  (\bibinfo{year}{2013}).
\newblock


\bibitem[\protect\citeauthoryear{Rendle, Freudenthaler, Gantner, and
  Schmidt-Thieme}{Rendle et~al\mbox{.}}{2009}]%
        {BPR2009}
\bibfield{author}{\bibinfo{person}{Steffen Rendle}, \bibinfo{person}{Christoph
  Freudenthaler}, \bibinfo{person}{Zeno Gantner}, {and} \bibinfo{person}{Lars
  Schmidt-Thieme}.} \bibinfo{year}{2009}\natexlab{}.
\newblock \showarticletitle{BPR: Bayesian Personalized Ranking from Implicit
  Feedback}. In \bibinfo{booktitle}{\emph{UAI '09}}. \bibinfo{pages}{452--461}.
\newblock


\bibitem[\protect\citeauthoryear{Salakhutdinov and Mnih}{Salakhutdinov and
  Mnih}{2007}]%
        {2007PMF}
\bibfield{author}{\bibinfo{person}{Ruslan Salakhutdinov} {and}
  \bibinfo{person}{Andriy Mnih}.} \bibinfo{year}{2007}\natexlab{}.
\newblock \showarticletitle{Probabilistic Matrix Factorization}. In
  \bibinfo{booktitle}{\emph{NIPS'07}}. \bibinfo{publisher}{Curran Associates
  Inc.}, \bibinfo{pages}{1257--1264}.
\newblock


\bibitem[\protect\citeauthoryear{Salakhutdinov, Mnih, and Hinton}{Salakhutdinov
  et~al\mbox{.}}{2007}]%
        {RBM2007}
\bibfield{author}{\bibinfo{person}{Ruslan Salakhutdinov},
  \bibinfo{person}{Andriy Mnih}, {and} \bibinfo{person}{Geoffrey Hinton}.}
  \bibinfo{year}{2007}\natexlab{}.
\newblock \showarticletitle{Restricted Boltzmann Machines for Collaborative
  Filtering}. In \bibinfo{booktitle}{\emph{ICML '07}}.
  \bibinfo{pages}{791--798}.
\newblock


\bibitem[\protect\citeauthoryear{Sarwar, Karypis, Konstan, and Riedl}{Sarwar
  et~al\mbox{.}}{2001}]%
        {Item2001}
\bibfield{author}{\bibinfo{person}{Badrul Sarwar}, \bibinfo{person}{George
  Karypis}, \bibinfo{person}{Joseph Konstan}, {and} \bibinfo{person}{John
  Riedl}.} \bibinfo{year}{2001}\natexlab{}.
\newblock \showarticletitle{Item-based Collaborative Filtering Recommendation
  Algorithms}. In \bibinfo{booktitle}{\emph{WWW '01}}.
  \bibinfo{publisher}{ACM}, \bibinfo{pages}{285--295}.
\newblock


\bibitem[\protect\citeauthoryear{Sarwar, Karypis, Konstan, and Riedl}{Sarwar
  et~al\mbox{.}}{2000}]%
        {Application2000}
\bibfield{author}{\bibinfo{person}{Badrul~Munir Sarwar},
  \bibinfo{person}{George Karypis}, \bibinfo{person}{Joseph~A. Konstan}, {and}
  \bibinfo{person}{John~Thomas Riedl}.} \bibinfo{year}{2000}\natexlab{}.
\newblock \showarticletitle{Application of Dimensionality Reduction in
  Recommender System--A Case Study}. In \bibinfo{booktitle}{\emph{ACM WebKDD
  Workshop}}. \bibinfo{publisher}{ACM}.
\newblock


\bibitem[\protect\citeauthoryear{Sedhain, Menon, Sanner, and Xie}{Sedhain
  et~al\mbox{.}}{2015}]%
        {Autorec2016}
\bibfield{author}{\bibinfo{person}{Suvash Sedhain}, \bibinfo{person}{Aditya
  Menon}, \bibinfo{person}{Scott Sanner}, {and} \bibinfo{person}{Lexing Xie}.}
  \bibinfo{year}{2015}\natexlab{}.
\newblock \showarticletitle{AutoRec: Autoencoders Meet Collaborative
  Filtering}. In \bibinfo{booktitle}{\emph{WWW '15}}. \bibinfo{publisher}{ACM},
  \bibinfo{pages}{111--112}.
\newblock


\bibitem[\protect\citeauthoryear{Shi, Zhao, and Shen}{Shi
  et~al\mbox{.}}{2017}]%
        {cold2017}
\bibfield{author}{\bibinfo{person}{Lei Shi}, \bibinfo{person}{Wayne~Xin Zhao},
  {and} \bibinfo{person}{Yi-Dong Shen}.} \bibinfo{year}{2017}\natexlab{}.
\newblock \showarticletitle{Local Representative-Based Matrix Factorization for
  Cold-Start Recommendation}.
\newblock \bibinfo{journal}{\emph{ACM Transaction on Information Systems}}
  \bibinfo{volume}{36}, \bibinfo{number}{2} (\bibinfo{date}{Aug.}
  \bibinfo{year}{2017}), \bibinfo{pages}{22:1--22:28}.
\newblock


\bibitem[\protect\citeauthoryear{Su and Khoshgoftaar}{Su and
  Khoshgoftaar}{2009}]%
        {Cf-survey2009}
\bibfield{author}{\bibinfo{person}{Xiaoyuan Su} {and} \bibinfo{person}{Taghi~M.
  Khoshgoftaar}.} \bibinfo{year}{2009}\natexlab{}.
\newblock \showarticletitle{A Survey of Collaborative Filtering Techniques}.
\newblock \bibinfo{journal}{\emph{Advances in Artificial Intelligence}}
  \bibinfo{volume}{2009} (\bibinfo{year}{2009}), \bibinfo{pages}{Article 4}.
\newblock


\bibitem[\protect\citeauthoryear{Trapit, David, and Andrew}{Trapit
  et~al\mbox{.}}{2016}]%
        {RNN2016}
\bibfield{author}{\bibinfo{person}{Bansal Trapit}, \bibinfo{person}{Belanger
  David}, {and} \bibinfo{person}{McCallum Andrew}.}
  \bibinfo{year}{2016}\natexlab{}.
\newblock \showarticletitle{Ask the GRU: Multi-task Learning for Deep Text
  Recommendations}. In \bibinfo{booktitle}{\emph{RecSys '16}}.
  \bibinfo{pages}{107--114}.
\newblock


\bibitem[\protect\citeauthoryear{Truyen, Phung, and Venkatesh}{Truyen
  et~al\mbox{.}}{2009}]%
        {ORBM2009}
\bibfield{author}{\bibinfo{person}{Tran~The Truyen}, \bibinfo{person}{Dinh~Q.
  Phung}, {and} \bibinfo{person}{Svetha Venkatesh}.}
  \bibinfo{year}{2009}\natexlab{}.
\newblock \showarticletitle{Ordinal Boltzmann Machines for Collaborative
  Filtering}. In \bibinfo{booktitle}{\emph{UAI '09}}.
  \bibinfo{pages}{548--556.}
\newblock


\bibitem[\protect\citeauthoryear{van~den Oord, Dieleman, and Schrauwen}{van~den
  Oord et~al\mbox{.}}{2013}]%
        {DLcontentRS}
\bibfield{author}{\bibinfo{person}{Aaron van~den Oord}, \bibinfo{person}{Sander
  Dieleman}, {and} \bibinfo{person}{Benjamin Schrauwen}.}
  \bibinfo{year}{2013}\natexlab{}.
\newblock \showarticletitle{Deep Content-based Music Recommendation}. In
  \bibinfo{booktitle}{\emph{NIPS '13}}. \bibinfo{pages}{2643--2651}.
\newblock


\bibitem[\protect\citeauthoryear{Wang and Blei}{Wang and Blei}{2011}]%
        {CTR2011}
\bibfield{author}{\bibinfo{person}{Chong Wang} {and} \bibinfo{person}{David~M.
  Blei}.} \bibinfo{year}{2011}\natexlab{}.
\newblock \showarticletitle{Collaborative Topic Modeling for Recommending
  Scientific Articles}. In \bibinfo{booktitle}{\emph{KDD '11}}.
  \bibinfo{pages}{448--456}.
\newblock


\bibitem[\protect\citeauthoryear{Wang, Wang, and Yeung}{Wang
  et~al\mbox{.}}{2015}]%
        {CDL2015}
\bibfield{author}{\bibinfo{person}{Hao Wang}, \bibinfo{person}{Naiyan Wang},
  {and} \bibinfo{person}{Dit-Yan Yeung}.} \bibinfo{year}{2015}\natexlab{}.
\newblock \showarticletitle{Collaborative Deep Learning for Recommender
  Systems}. In \bibinfo{booktitle}{\emph{KDD '15}}. \bibinfo{publisher}{ACM},
  \bibinfo{pages}{1235--1244}.
\newblock


\bibitem[\protect\citeauthoryear{Wang, Wang, Tang, Shu, Ranganath, and
  Liu}{Wang et~al\mbox{.}}{2017}]%
        {image2017}
\bibfield{author}{\bibinfo{person}{Suhang Wang}, \bibinfo{person}{Yilin Wang},
  \bibinfo{person}{Jiliang Tang}, \bibinfo{person}{Kai Shu},
  \bibinfo{person}{Suhas Ranganath}, {and} \bibinfo{person}{Huan Liu}.}
  \bibinfo{year}{2017}\natexlab{}.
\newblock \showarticletitle{What Your Images Reveal: Exploiting Visual Contents
  for Point-of-Interest Recommendation}. In \bibinfo{booktitle}{\emph{WWW
  '17}}. \bibinfo{pages}{391--400}.
\newblock


\bibitem[\protect\citeauthoryear{Wang, He, Wang, Feng, and Chua}{Wang
  et~al\mbox{.}}{2019}]%
        {NGCF19}
\bibfield{author}{\bibinfo{person}{Xiang Wang}, \bibinfo{person}{Xiangnan He},
  \bibinfo{person}{Meng Wang}, \bibinfo{person}{Fuli Feng}, {and}
  \bibinfo{person}{Tat-Seng Chua}.} \bibinfo{year}{2019}\natexlab{}.
\newblock \showarticletitle{Neural Graph Collaborative Filtering}. In
  \bibinfo{booktitle}{\emph{{SIGIR '19}}}. \bibinfo{publisher}{ACM}.
\newblock


\bibitem[\protect\citeauthoryear{Wu, DuBois, Zheng, and Ester}{Wu
  et~al\mbox{.}}{2016}]%
        {CDAE2016}
\bibfield{author}{\bibinfo{person}{Yao Wu}, \bibinfo{person}{Christopher
  DuBois}, \bibinfo{person}{Alice~X. Zheng}, {and} \bibinfo{person}{Martin
  Ester}.} \bibinfo{year}{2016}\natexlab{}.
\newblock \showarticletitle{Collaborative Denoising Auto-Encoders for Top-N
  Recommender Systems}. In \bibinfo{booktitle}{\emph{WSDM '16}}.
  \bibinfo{publisher}{ACM}, \bibinfo{pages}{153--162}.
\newblock


\bibitem[\protect\citeauthoryear{Zhang, Yuan, Lian, Xie, and Ma}{Zhang
  et~al\mbox{.}}{2016}]%
        {knowledgebase2016}
\bibfield{author}{\bibinfo{person}{Fuzheng Zhang},
  \bibinfo{person}{Nicholas~Jing Yuan}, \bibinfo{person}{Defu Lian},
  \bibinfo{person}{Xing Xie}, {and} \bibinfo{person}{Wei-Ying Ma}.}
  \bibinfo{year}{2016}\natexlab{}.
\newblock \showarticletitle{Collaborative Knowledge Base Embedding for
  Recommender Systems}. In \bibinfo{booktitle}{\emph{KDD '16}}.
  \bibinfo{publisher}{ACM}, \bibinfo{pages}{353--362}.
\newblock


\bibitem[\protect\citeauthoryear{Zhang, Yao, and Sun}{Zhang
  et~al\mbox{.}}{2017}]%
        {RSsurvey2017}
\bibfield{author}{\bibinfo{person}{Shuai Zhang}, \bibinfo{person}{Lina Yao},
  {and} \bibinfo{person}{Aixin Sun}.} \bibinfo{year}{2017}\natexlab{}.
\newblock \showarticletitle{Deep Learning based Recommender System: {A} Survey
  and New Perspectives}.
\newblock \bibinfo{journal}{\emph{arXiv preprint arXiv:1707.07435}}
  (\bibinfo{year}{2017}).
\newblock


\bibitem[\protect\citeauthoryear{Zheng, Noroozi, and S.~Yu.}{Zheng
  et~al\mbox{.}}{2017}]%
        {Jointreview2017}
\bibfield{author}{\bibinfo{person}{Lei Zheng}, \bibinfo{person}{Vahid Noroozi},
  {and} \bibinfo{person}{Philip S.~Yu.}} \bibinfo{year}{2017}\natexlab{}.
\newblock \showarticletitle{Joint Deep Modeling of Users and Items Using
  Reviews for Recommendation}. In \bibinfo{booktitle}{\emph{WSDM '17}}.
  \bibinfo{publisher}{ACM}, \bibinfo{pages}{425--434}.
\newblock


\bibitem[\protect\citeauthoryear{Zheng, Tang, Ding, and Zhou}{Zheng
  et~al\mbox{.}}{2016}]%
        {NAuto2016}
\bibfield{author}{\bibinfo{person}{Yin Zheng}, \bibinfo{person}{Bangsheng
  Tang}, \bibinfo{person}{Wenkui Ding}, {and} \bibinfo{person}{Hanning Zhou}.}
  \bibinfo{year}{2016}\natexlab{}.
\newblock \showarticletitle{A Neural Autoregressive Approach to Collaborative
  Filtering}. In \bibinfo{booktitle}{\emph{ICML'16}}.
  \bibinfo{pages}{764--773}.
\newblock


\end{thebibliography}
%\bibliography{sample-bibliography}

\end{document}